\documentclass[twocolumn,numberedappendix]{openjournal}
\usepackage{natbib}
\usepackage{graphicx}
\usepackage{graphicx,amsmath,amssymb,amstext}
\usepackage{amsbsy,amsfonts,amsthm,color}
\usepackage{caption}
\usepackage{subcaption}
\usepackage{cancel}
\usepackage{xcolor}
\usepackage{graphicx}
\usepackage[colorlinks,linkcolor=blue,citecolor=blue,urlcolor=blue ]{hyperref}
\usepackage[title]{appendix}
 
\usepackage{siunitx}

\usepackage{tikz}
\usetikzlibrary{shapes,arrows,shadows,fit}
\usetikzlibrary{positioning}
\usetikzlibrary{bayesnet}
\definecolor{ourgreen}{HTML}{06D6A0}
\definecolor{ourblue}{HTML}{118AB2}
\usepackage{hyperref}
\hypersetup{colorlinks,linkcolor=ourblue,citecolor=ourblue,urlcolor=ourblue}

\newcommand{\subf}[1]{#1}   %% \newcommand{\subf}[1]{\textit{#1}}

\newcommand{\imagunit}{\ensuremath{\textrm{i}\mkern1mu}}

\newcommand{\half}{\frac{1}{2}}

\newcommand{\transpose}{\ensuremath{{\textrm{T}}}}

\newcommand{\sfA}{\ensuremath{{\sf{A}}}}
\newcommand{\sfB}{\ensuremath{{\sf{B}}}}
\newcommand{\sfC}{{\ensuremath{\sf{C}}}}
\newcommand{\sfD}{{\ensuremath{\sf{D}}}}

\newcommand{\sfG}{\ensuremath{{\sf{G}}}}
\newcommand{\sfJ}{{\ensuremath{\sf{J}}}}
\newcommand{\sfK}{\ensuremath{{\sf{K}}}}
\newcommand{\sfL}{{\ensuremath{\sf{L}}}}
\newcommand{\sfM}{\ensuremath{{\sf{M}}}}
\newcommand{\sfN}{{\ensuremath{\sf{N}}}}

\newcommand{\sfR}{\ensuremath{{\sf{R}}}}
\newcommand{\sfU}{\ensuremath{{\sf{U}}}}
\newcommand{\sfY}{\ensuremath{{\sf{Y}}}}
\newcommand{\sfZ}{\ensuremath{{\sf{Z}}}}

\newcommand{\fata}{\ensuremath{\boldsymbol{a}}}
\newcommand{\fatd}{\ensuremath{\boldsymbol{d}}}
\newcommand{\fatn}{\ensuremath{\boldsymbol{n}}}
\newcommand{\fatp}{\ensuremath{\boldsymbol{p}}}
\newcommand{\fats}{\ensuremath{\boldsymbol{s}}}
\newcommand{\fatx}{\ensuremath{\boldsymbol{x}}}
\newcommand{\faty}{\ensuremath{\boldsymbol{y}}}

\newcommand{\calP}{\ensuremath{\mathcal{P}}}
\newcommand{\calB}{\ensuremath{\mathcal{B}}}
\newcommand{\calO}{\ensuremath{\mathcal{O}}}

\newcommand{\calL}{\ensuremath{\mathcal{L}}}

\newcommand{\calG}{\ensuremath{\mathcal{G}}}

\newcommand{\mathd}{\ensuremath{\mathrm{d}}}
\newcommand{\both}{\ensuremath{\boldsymbol{\theta}}}

\newcommand{\almanac}{\textsc{Almanac}}

\newcommand{\lpot}{\ensuremath{{\Tilde{\Psi}}}} 
\newcommand{\nlp}{\ensuremath{{\psi}}} 

\DeclareMathOperator{\diag}{diag}
\DeclareMathOperator{\Tr}{Tr}
\DeclareMathOperator{\Rot}{Rot}
\DeclareMathOperator{\Flip}{Flip}
\DeclareMathOperator{\Var}{Var}

\newcommand{\determinant}[1]{\ensuremath{{ | #1 | }}}

\definecolor{ork}{rgb}{0.9,0.1,0.3}
\definecolor{grbl}{rgb}{0.3,0.6,0.7}
\definecolor{bleu}{rgb}{0,0.5,0.6}

\usepackage{soul}

\usepackage{xcolor}

\begin{document}

\journalinfo{The Open Journal of Astrophysics}
\shorttitle{Almanac: MCMC-Based Signal Extraction on the Sphere}

\title{Almanac: MCMC-based signal extraction of power spectra and maps on the sphere}

\author{Elena Sellentin$^{1,2}$}
\author{Arthur Loureiro$^{3,4,5,6}$}
\author{Lorne Whiteway$^6$}
\author{Javier Silva Lafaurie$^{1,2}$}
\author{Sreekumar T. Balan$^{6,7}$}
\author{Malak Olamaie$^8$}
\author{Andrew H. Jaffe$^4$}
\author{Alan F. Heavens$^4$}

\affiliation{$^1$ Mathematical Institute, Leiden University, Snellius Gebouw, Niels Bohrweg 1, NL-2333 CA Leiden, The Netherlands}
\affiliation{$^2$ Leiden Observatory, Leiden University, Oort Gebouw, Niels Bohrweg 2, NL-2333 CA Leiden, The Netherlands}
\affiliation{$^3$ The Oskar Klein Centre, Department of Physics, Stockholm University, AlbaNova University Centre, SE-106 91 Stockholm, Sweden}
\affiliation{$^4$ Astrophysics Group and Imperial Centre for Inference and Cosmology (ICIC), Blackett Laboratory, Imperial College London, Prince Consort Road, London SW7 2AZ, UK}
\affiliation{$^5$ Institute for Astronomy, University of Edinburgh, Royal Observatory, Blackford Hill, Edinburgh EH9 3HJ, UK}
\affiliation{$^6$ Department of Physics and Astronomy, University College London, Gower Street, London WC1E 6BT, UK}
\affiliation{$^7$ Proportunity, GG 405, Metal Box Factory, 30 Great Guildford St, London, SE1 0HS, UK}
\affiliation{$^8$ School of Science, Technology and Health, York St John University, Lord Mayor's Walk, York, YO31 7EX, UK}

\email{sellentin@strw.leidenuniv.nl}
\email{arthur.loureiro@fysik.su.se}
\email{lorne.whiteway@star.ucl.ac.uk}
\email{silvalafaurie@strw.leidenuniv.nl}
\email{sree@proportunity.com}
\email{m.olamaie@yorksj.ac.uk}
\email{a.jaffe@imperial.ac.uk}
\email{a.heavens@imperial.ac.uk}

%%%%%%%%%%%%%%%%%%%%%%%%%%%%%%%%%%%%%%%%%%%%%%%%%%%%%%%%%%%%%%%%%%%%%%%%
%                               ABSTRACT
%%%%%%%%%%%%%%%%%%%%%%%%%%%%%%%%%%%%%%%%%%%%%%%%%%%%%%%%%%%%%%%%%%%%%%%%

\begin{abstract}
Inference in cosmology often starts with noisy observations of random fields on the celestial sphere, such as maps of the microwave background radiation, continuous maps of cosmic structure in different wavelengths, or maps of point tracers of the cosmological fields. \textsc{Almanac} uses Hamiltonian Monte Carlo sampling to infer the underlying all-sky noiseless maps of cosmic structures, in multiple redshift bins, together with their auto- and cross-power spectra.  It can sample many millions of parameters, handling the highly variable signal-to-noise of typical cosmological signals, and it provides science-ready posterior data products. In the case of spin-weight 2 fields, \textsc{Almanac} infers $E$- and $B$-mode power spectra and parity-violating $EB$ power, and, by sampling the full posteriors rather than point estimates, it avoids the problem of $EB$-leakage. For theories with no $B$-mode signal, inferred non-zero $B$-mode power may be a useful diagnostic of systematic errors or an indication of new physics. \textsc{Almanac}'s aim is to characterise the statistical properties of the maps, with outputs that are completely independent of the cosmological model, beyond an assumption of statistical isotropy.  Inference of parameters of any particular cosmological model follows in a separate analysis stage. We demonstrate our signal extraction on a CMB-like experiment. 

\end{abstract}

\maketitle

\section{Introduction}
There is strong observational evidence that our Universe has always been permeated by random structures. Astronomical data show that the matter in our Universe is distributed as a random field; this field can be observed across a wide range of the electromagnetic spectrum, using continuous tracers (such as 21cm radiation from neutral hydrogen), or local point tracers (such as galaxies), or by probing via weak gravitational lensing (where cosmic shear fields are sampled by the shapes of galaxy images in tomographic bins). Other random fields, such as the cosmic microwave background radiation (CMB), also contain information about the cosmological model. Thus fields on the sky are a ubiquitous source of information, and methods that can extract as much information as possible from fields are particularly valuable.

Many cosmological experiments have relied on the analysis of angular two-point statistics (such as the angular power spectrum): CMB \citep{2009-WMAP-Cls,2020-PlanckCls}, photometric \citep{2007-Padmanabhan-Cls}, radio \citep{2004-Blake-Radio-Cls}, and even spectroscopic galaxy surveys \citep{2019-Loureiro-BOSS}. There are two routes to the measurement of two-point functions from measurements of cosmic fields. First, one can define a two-point (i.e., quadratic) estimator, which uses the fact that two-point functions are ensemble averages over multiples of pairs of field values. Second, further assuming ergodicity and relying on the statistical homogeneity and isotropy of the field leads to a family of estimators based on taking the harmonic transform of noisy and incomplete measurements of the field. These \textit{pseudo-harmonic coefficients} are multiplied together pairwise with an appropriate coupling matrix, and an additive bias is removed \citep{1973-Peebles}. At the level of the estimator, this approach does not rely on specifying a particular underlying distribution for the cosmological signal (although once we require error bars on our estimator we must supply a sampling distribution whereupon we lose that seeming advantage\footnote{If we further require a minimum variance quadratic estimator, we are led to a formalism mathematically equivalent to the Bayesian approach \citep{bjk98}.}).

These latter steps ensure that the estimator remains unbiased. This is particularly straightforward for all-sky measurements with uniform noise; here the coupling matrix just specifies taking the average of the pseudo-harmonic coefficients at a given degree $\ell$, and the additive bias corresponds to the spectrum of the noise. However, point estimates of two-point statistics will not in general capture the full range of information available in cosmological surveys \citep{2021-Harnois-Beyond2pt,2021-LeclercqHeavens,2021-Porth,2020-Naidoo,2022-Sarma}, either because of non-gaussianity, or because of non-optimality of the statistics. 

In addition to limitations in information extraction, quadratic estimators of the two-point function have other well-known drawbacks such as a) complicated modelling for the effects of the survey geometry on the estimator's covariance matrices \citep{2001-Cooray,2019-Wu-Cov,2019-Harnois-Cov,2022-Upham-PClCov} and b) difficulties in mitigating or detecting leakage between $E$- and $B$-modes \citep{2003Bunn-EBLeak,2008-Bunn-EBLeak,2010-Kim,2019-Liu}. The latter is a concern for both CMB polarization and cosmic shear studies. 

Alternately, we can take a Bayesian approach from the start. If we assume a statistically isotropic and gaussian random field on the sky (or, equivalently, a statistically homogeneous three-dimensional field  measured through various two-dimensional quantities), we are led to straightforward likelihoods and priors \citep{bjk98}. The gaussian prior on the cosmological fields is a maximum-entropy assumption: they are the least constraining distributions under the assumption of given mean and covariance \citep{jaynes03}. The original implementations of this idea marginalised over the 
field values analytically, giving a likelihood function gaussian in the data, with covariance matrices given by the sum of noise and signal contributions, the latter itself a linear function of the desired power spectrum parameters. However, this algorithm is naturally dominated by matrix manipulations with operation counts scaling as the cube of the number of observed pixels (or spherical harmonic modes), infeasible for current mega-pixel experiments.

Because of these properties, hierarchical modeling via field-level inference (FLI) has gained significant popularity in recent years; its objective is the optimal inference of the information in the large-scale structure as observed by galaxy surveys \citep{2010-Jasche,2013-Jasche-BORG1,2015-Schneider,2015-Leclercq,2016-Lavaux,2017-Bohm-FLI,2019-Jasche-BORG2,2019-Taylor-DELFI,2020-Porqueres-LyAlpha,2021-Jeffrey,2022-Sarma,2022-Andrews-BORG-fnl} or cosmic shear \citep{2016-Alsing, 2017-Alsing,2021-Porqueres-WL-1,2022-Porqueres-2, Fiedorowicz2022, 2022AlmanacWL, Porqueres2023, ZhouDodelson2023}. In this form, FLI explores a range of information not necessarily encapsulated by point estimates of two-point functions; it hierarchically performs inference directly on cosmological fields (such as the matter density and the shear fields), on their statistics, and on cosmological parameters. Using a toy model, \cite{2021-LeclercqHeavens} compared the FLI approach with the traditional two-point function approach and with simulation-based inference, showing that FLI can be more accurate and more precise than either.\footnote{Note however that one can find more informative statistics that preserve much of the information content of the fields, for example by using machine learning methods  \citep[e.g.,][]{2018-Charnock,2021-Jeffrey,2021-Makinen}.}  The advantage of field-level inference is that by linking the (almost raw) field data to the model parameters through a Bayesian Hierarchical Model (BHM; in practice this means sampling a likelihood function that is complicated but that correctly represents the data model), essentially all of the statistical information is used, including all orders of correlations.  For the gaussian fields of the CMB, this approach was introduced and developed with Gibbs sampling for temperature fields by \citet{2004-Wandelt,2004-Eriksen} and later extended to polarisation fields by \citet{2007-Larson-CMB-Pol-Gibbs} \citep[see also][]{2007-Eriksen,2009-Jewell,2016-Racine}.

For late-time non-gaussian fields, field-level inference can take one of two forms. The first explores a data model that starts with the initial density field and forward models using assumptions about gravity and cosmology; this is the approach of the \textsc{BORG} framework \citep{2013-Jasche-BORG1, 2019-Jasche-BORG2, 2021-Porqueres-WL-1}. An alternative, which we use here, is to sample the fields jointly with their two-point statistics, without reference to a cosmological model (beyond some mild assumptions of isotropy).  The analysis is then cosmology-independent and does not have to be repeated if the cosmological model changes (provided symmetries such as statistical isotropy are still respected). The analysis does not test cosmological models directly, but the resulting samples can then be used in a further stage of inference to test cosmology.   We develop the idea presented in \cite{Taylor,2016-Alsing,2017-Alsing}: simultaneously infer cosmic fields and their angular power spectra for data distributed on the sphere. Future surveys will cover large sky fractions, so a curved-sky approach is needed to avoid potential biases to the inferred cosmological parameters from the flat-sky approximation \citep{2017-Lemos-FlatSky,2017-Kitching-FlatSky,2017-Kilbinger-FlatSky,2021-Matthewson-FlatSky}. 

The maximum-entropy properties of our Bayesian approach do not force higher-order moments to zero. Rather, where the signal-to-noise is sufficiently high, the resulting samples of the underlying field will retain non-gaussian information and can be used to probe non-gaussianity in a further inference stage if desired (although this should be treated with care and, ideally, estimated in a self-consistent workflow).

The algorithm, called \almanac{}, has a generic structure that allows it to analyse multiple fields, both spin-weight 0 and spin-weight 2.  It can therefore handle the vast majority of cosmological observations. It generalises the ideas presented in \citet{Taylor} from the CMB to multiple fields, and generalises cosmic shear inference first presented in \cite{2016-Alsing} to the curved sky while also reaching the challenging scales relevant for weak lensing analysis, $\ell_{\rm max} \sim \calO(10^4)$. This implies extremely high dimensional posteriors, with $\calO(10^7)$ parameters. There are significant challenges in sampling such a high-dimensional parameter space; the methods most commonly used are Gibbs sampling (see, e.g., \cite{2022-Colombo} and references therein) and Hamiltonian Monte Carlo (HMC). HMC has advantages for highly correlated distributions; a disadvantage is that it requires derivatives of the target function (but here this is not a difficulty as we can compute these analytically). \almanac{} therefore uses a HMC sampler.

This paper is a companion to \cite{2022AlmanacWL}{}; the companion paper develops \almanac{} further for correlated fields, and specialises to the spin-weight 2 fields of cosmic shear, with emphasis on the tomographic analysis of different redshift slices.  

In Section~\ref{sec:methods} we outline the mathematical formalism for representing statistically isotropic intensity and polarization fields on the sphere. In Section~\ref{sec:datamodel} we outline the data model that generates our BHM. In Section~\ref{sec:HMCsampler} we give a brief review of Hamiltonian Monte Carlo, and in Section~\ref{sec:HMCtuning} we show how we optimize the sampler for our application. In Section~\ref{sec:results} we apply this to representative CMB-like problems, and in Section~\ref{sec:paramresults} discuss how we monitor its convergence. We conclude in Section~\ref{sec:discussion}, and we discuss various mathematical details in a series of appendices.

%%%%%%%%%%%%%%%%%%%%%%%%%%%%%%%%%%%%%%%%%%%%%%%%%%%%%%%%%%%%%%%%%%%%%%%%
%                       FIELDS ON THE SPHERE
%%%%%%%%%%%%%%%%%%%%%%%%%%%%%%%%%%%%%%%%%%%%%%%%%%%%%%%%%%%%%%%%%%%%%%%%
\section{Cosmological Fields on the Sphere}
\label{sec:methods}

This section presents notation for both CMB and weak lensing on the sphere; one aim is to make clear which quantities from these two observation techniques correspond to each other. We denote positions on the sphere via angular polar coordinates $\theta$ (colatitude) and $\phi$ (longitude), and we introduce the unit vector $\hat{n} = (\theta, \phi)$.

We give a brief introduction to spin-weight functions; for more information see \cite{2005-Castro}, and see also Appendix~\ref{apx:SpinWeightFunctions} for an equivalent but coordinate-free treatment. A function ${}_s f$ of integral spin-weight $s$ on the sphere is a complex-valued function (of both position and local orthonormal basis) that transforms as
\begin{equation}
    {}_s f(\hat{n}) \rightarrow e^{-\imagunit{}\alpha s}{}_s f(\hat{n})
\end{equation}
under a rotation by the angle $\alpha$ in the tangent plane at $\hat{n}$. This is made precise in Eq.~\eqref{Eq:apx:spinWeightFunctionDefRotation}. Note that the complex conjugate function will have spin-weight $-s$. The coordinates $\theta, \phi$ give us a standard orthonormal basis on the sphere; if we refer only to this basis then spin-weight $s$ functions become usual complex functions on the sphere (except possibly at the poles). Such functions will have particular behaviour near the poles: if non-zero there, then to first order they must be multiples of $\exp(-\imagunit{} s \phi)$ at the north pole and $\exp(\imagunit{} s \phi)$ at the south pole. Functions of spin-weight zero have the transformation properties of a usual (complex) scalar function. The name `spin-weight $s$' is abbreviated by some authors (including the companion paper \cite{2022AlmanacWL}{}) to `spin-$s$'.

The \textit{spin-weight spherical harmonics} form a convenient basis for representing spin-weight functions on the sphere. Begin by defining a spin-weight raising operator ($\eth$) and a spin-weight lowering operator ($\bar{\eth}$) via \citep{1966-NewmanPenrose,1967-Goldberg}

\begin{align}
\label{eqn:eth}
    \eth & \equiv -\sin^{\rm s}\theta\left(\frac{\partial}{\partial\theta} + \imagunit{} \frac{1}{\sin\theta}\frac{\partial}{\partial\phi}\right)\sin^{\rm -s}\theta, \\
    \bar{\eth} & \equiv - \sin^{\rm -s}\theta\left(\frac{\partial}{\partial\theta} - \imagunit{}\frac{1}{\sin\theta}\frac{\partial}{\partial\phi}\right)\sin^{\rm s}\theta.
\end{align}
The spin-weight spherical harmonics $_s Y_{\ell m}$ of spin-weight $s$, degree $\ell$ and order $m$ are then defined inductively to be the usual spherical harmonics when $s=0$, to be zero when $|s| > \ell$, and otherwise to satisfy

\begin{align}
    \eth \, _s Y_{\ell m} & = \sqrt{(\ell-s)(\ell+s+1)} \enspace _{s+1} Y_{\ell m}, \\
    \bar{\eth} \, _s Y_{\ell m} & = - \sqrt{(\ell+s)(\ell-s+1)} \enspace _{s-1} Y_{\ell m}.
\end{align}

The full \almanac{} data model currently handles an arbitrary combination of spin-weight 0 and spin-weight 2 fields. It could easily be extended, particularly to the spin-weight 4 fields needed to represent gravitational wave backgrounds \citep{ConneelyJaffe,2022-Renzini}. 

The forward spin-weight spherical harmonic transform is 
\begin{equation}
a_{\ell m} = \int {\rm d}\Omega_{\hat{n}}\, {}_{s}f(\hat{n})\, {}_{s}Y_{\ell m}^*(\hat{n}) \, ,
\end{equation}
with inverse
\begin{equation}
    {}_{s}f(\hat{n}) = \sum_{\ell m} a_{\ell m}\ {}_{s}Y_{\ell m }(\hat{n}) \, .
    \label{spherics}
\end{equation}

In all cases the minimum value of $\ell$ is 2: for spin-weight $s=\pm 2$ fields this is required since $\ell \ge |s|$, and for spin-weight $0$ fields this is a consequence of the usual subtraction of the monopole and dipole.

Examples of spin-weight 0 fields are the CMB temperature (also called intensity), the number density of galaxy positions in a galaxy clustering analysis, and the weak lensing magnification field estimated, e.g., by galaxy sizes or fluxes \citep{2015-AlsingMag,2020-Thiele,2022-Mahony}.

Examples of spin-weight $\pm 2$ fields are CMB polarization and weak lensing shear. The linear CMB polarization is described by the Stokes parameters $Q$ and $U$, while weak lensing shear is described by shear components $\gamma_{1}$ and $\gamma_{2}$; geometrically, $Q$ and $\gamma_1$ are analogues, as are $U$ and $\gamma_2$. These (real) components may be combined to yield the (complex) polarization $P = Q + \imagunit U$ and the (complex) shear $\gamma = \gamma_1 + \imagunit \gamma_2$. In both cases the spin-weight of the resulting complex quantity will be $-2$ or $+2$, depending on the choice of coordinate system; see Appendix~\ref{apx:SpinWeightFunctions} for more information. For example, the CMB polarization $P$ is defined by authors such as \citet{ChallinorChon, LewChalTu} to be of spin-weight $-2$ and by authors such as \citet{2003Bunn-EBLeak,1997-Zaldarriaga} to be of spin-weight $+2$. A spin-weight $-2$ field may be converted to spin-weight $+2$ by taking the complex conjugate, i.e., by changing the sign of $U$ (respectively $\gamma_2$), and therefore \almanac{} (and the remainder of this paper) deals only with spin-weight $+2$ (together with spin-weight $0$) fields.

A spin-weight $2$ field ${}_{2}f$ with harmonic coefficients $a_{\ell m}$ can be transformed into a pair of spin-weight $2$ $E$- and $B$-fields with harmonic coefficients

\begin{equation}
    \label{eq:eandb}
    \begin{aligned}
    E_{\ell m} & = -(1/2) \ (a_{\ell m} + (-1)^m (a_{\ell,-m})^*) \quad \textrm{and} \\
    B_{\ell m} & = (\imagunit{}/2) \ (a_{\ell m} + (-1)^{m+1} (a_{\ell,-m})^*)
    \end{aligned}
\end{equation}
so that ${}_{2}f = -(E + \imagunit{} B)$. The sign convention here ($H=1$ in the terminology of \cite{2005PhRvD..71h3008L}) is that of \textsc{HEALPix} \citep{1999-Healpix}. $E$ and $B$ can be written as the second derivative of spin-weight $0$ potential fields $\lpot{}_E$ and $\lpot{}_B$:

\begin{equation}
    E = \eth \eth \lpot{}_E \qquad \textrm{and} \qquad B = \eth \eth \lpot{}_B,
    \label{gradcurl}
\end{equation}
where crucially both potential fields are real-valued (this uses the symmetry ${}_{s} Y_{\ell m}^* = (-1)^{s+m} {}_{-s}Y_{\ell,-m}$)\footnote{Here we correct a minor typographical error in \cite{2005-Castro}.}. As a consequence, $E$ will be curl-free (`electric') and $\imagunit{} B$ will be divergence-free (`magnetic'). In contamination-free weak lensing, $\lpot{}_B$ is expected to be very close to zero \citep[e.g.,][]{BartelSchneid}. In contrast, $\lpot{}_B$ for the CMB is expected to be small but non-zero, as it contains signatures of gravitational waves and gravitational lensing.

Collecting all pixels of all fields in a joint data vector $\fatd$, all spherical harmonic coefficients in a vector $\fata$, and the corresponding spherical harmonics of the appropriate spin-weight in a matrix $\sfY$, we can write Eq.~\eqref{spherics} as
\begin{equation}
    \fatd = \sfR\sfY\fata + {\fatn},
    \label{eq:alm_to_map}
\end{equation}
where we have additionally included a beam-smearing matrix $\sfR$ (introduced in Sect.~\ref{ssec:posterior} for the CMB only; for galaxy surveys and cosmic shear, this factor is absent) and noise $\fatn$. We usually assume uncorrelated additive gaussian noise, $\fatn \sim \calG(0,\sfN)$, with a diagonal noise matrix $\sfN$ in pixel space. Galaxy shot noise, on the other hand, is usually modelled as Poisson.

%%%%%%%%%%%%%%%%%%%%%%%%%%%%%%%%%%%%%%%%%%%%%%%%%%%%%%%%%%%%%%%%%%%%%%%%
%                       ALMANAC DATA MODEL
%%%%%%%%%%%%%%%%%%%%%%%%%%%%%%%%%%%%%%%%%%%%%%%%%%%%%%%%%%%%%%%%%%%%%%%%
\section{Almanac data model}
\label{sec:datamodel}

\subsection{Bayesian Hierarchical Modelling}\label{SSec:BHM}

\begin{figure}
    \begin{center}
    \large{\begin{tikzpicture}[node distance = 1.4cm, auto]

	    \pgfdeclarelayer{background}
	    \pgfdeclarelayer{foreground}
	    \pgfsetlayers{background,main,foreground}

	    \tikzstyle{prob}=[draw, thick, text centered, rounded corners, minimum height=1em, minimum width=1em, fill=ourblue!60]
	    \tikzstyle{var}=[draw, thick, text centered, circle, minimum height=1em, minimum width=1em, fill=white]

        % nodes
        \node[prob](prior){$\pi(\sfC)$};
        \node[var](Cls)[below of= prior]{$\sfC$};
        \node[prob](Palms)[below of= Cls]{$\calG(\fata | \sfC)$};
        \node[var](alms)[below of= Palms]{$\fata$};
        \node[prob](like)[below of= alms]{$\calL(\fatd | \fata, \sfN)$};
        \node[var](noise)[left of= alms]{$\sfN$};
        \node[var](data)[below of= like]{$\fatd$};
        
        %paths between nodes
        \path [draw, line width=0.7pt, arrows={-latex}] (prior) -- (Cls);
        \path [draw, line width=0.7pt, arrows={-latex}] (Cls) -- (Palms);
        \path [draw, line width=0.7pt, arrows={-latex}] (Palms) -- (alms);
        \path [draw, line width=0.7pt, arrows={-latex}] (alms) -- (like);
        \path [draw, line width=0.7pt, arrows={-latex}] (noise) -- (like);
        \path [draw, line width=0.7pt, arrows={-latex}] (like) -- (data);

    \end{tikzpicture}}
    \end{center}
    \caption{
   Directed acyclic graph showing \almanac{}'s Bayesian hierarchical model for the inference of maps and angular power spectra. The angular power spectra $\sfC$ are drawn from a prior distribution $\pi$, spherical harmonic coefficients $\fata$ for the fields are then drawn from a gaussian prior $\calG$, and the fields are then transformed to configuration space, where noise $\fatn$ (drawn from a likelihood $\calL$ with noise covariance $\sfN$) is added to yield the data $\fatd$. The beam (for the CMB) is not shown.}
    \label{Fig:DAG}
\end{figure}
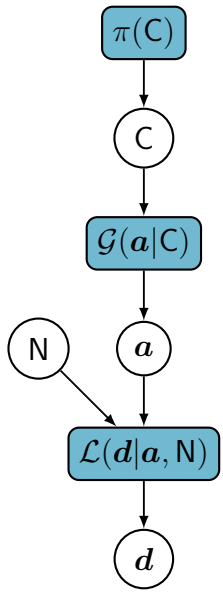

We briefly review the Bayesian forward modelling used in \almanac{}. Our hierarchical data model is a full-sphere curved-sky spin-weight 0 and spin-weight 2 extension of the approach taken taken by \cite{2007-Larson-CMB-Pol-Gibbs}, \cite{Taylor}, and \cite{2016-Alsing, 2017-Alsing}. The model allows both convolution with an observational beam and additive noise in pixel space. We infer all power spectra and underlying maps simultaneously. The basic model is described by a directed acyclic graph, as shown in slightly simplified form in Fig.~\ref{Fig:DAG}.

In the most general case we have a set of data maps; each map will be either spin-weight 0 (i.e., a scalar field, assumed to be real-valued) or spin-weight 2 (split into two real-valued fields, one for the real part and one for the imaginary part of the map). Section~\ref{sec:methods} lists examples of what the maps could represent.

The different fields will be denoted by a superscript field index $i$, and we let $n$ denote the number of fields. In the CMB case, the field index may be used for data at different frequencies (but this is usually not required since the primordial temperature and polarization signals are independent of frequency). For cosmic shear in particular, a tomographic analysis is usually done, and in this case the field index represents the tomographic bin. Note that in this case, the fields in different tomographic bins are highly correlated; this challenging case is treated in detail in the companion paper \citep{2022AlmanacWL}{}.

We work with a sphere that has been pixelised using \textsc{HEALPix} \citep{1999-Healpix}, with resolution $n_{\rm side}$. We denote all data points (pixels) of a single field $i$ as $\fatd^{i}$, and the total data vector $\fatd$ (introduced in Eq~\ref{eq:alm_to_map}) is the stack of these individual fields:
\begin{equation}
    \fatd^{\transpose{}} = (\fatd^{1,T}, \fatd^{2,T}, ... , \fatd^{n,T} ) \, .
\end{equation}

These data maps are represented by their spherical harmonic coefficients. Working with spherical harmonics has the advantage that statistical isotropy is very simply expressed in this basis. Moreover, when transforming to spherical harmonic space no artificial padding needs to be introduced (in comparison to a flat Fourier transform, where padding is needed to suppress boundary effects). We use \textsc{libsharp2} \citep{2013-Libsharp,2018-Libsharp2} to perform the spherical harmonic transforms.

In the spin-weight 0 case with field index $i$, let $\fata^i$ be the spherical harmonic coefficients $S_{\ell m}$ of the field. In the spin-weight 2 case, for which there are two fields (the real part $\fatd^i$ and the imaginary part $\fatd^{i+1}$), let $\fata^i$ be the $E$-mode coefficients $E_{\ell m}$ and $\fata^{i+1}$ the $B$-mode coefficients $B_{\ell m}$. The total spherical harmonic coefficient vector $\fata$ (introduced in Eq~\ref{eq:alm_to_map}) is the stack of these individual vectors:
\begin{equation}
    \fata^{\transpose{}} = (\fata^{1,T}, \fata^{2,T},...,\fata^{n,T}) \, .
\end{equation}
See Appendix~\ref{apx:Preliminaries} for further details of the structure of $\fata$.

The spherical harmonics themselves are stored in a matrix $\sfY$, whose ordering is chosen to align with the ordering of $\fata$ and $\fatd$; smilar remarks hold for the beam-smearing matrix $\sfR$.

In both the CMB and late-time cases, the fields (and therefore harmonic coefficients) will in general be correlated, but under the assumption of statistical isotropy on the sky, there will be no correlations between different values of $\ell$ and $m$ and moreover the amplitude of the correlation will only depend on $\ell$. Hence for each $\ell$ we have a symmetric positive definite matrix $\sfC_{\ell}$ (with entries indexed by field indices $i$ and $j$) satisfying:
\begin{equation}
    \langle \fata^i_{\ell m} \, \fata^{j*}_{\ell'm'} \rangle = \sfC^{ij}_{\ell}\delta_{\ell \ell'}\delta_{m m'} \, .
    \label{eq:deltas}
\end{equation}
We denote the union of all the $\sfC_{\ell}$ as $\sfC$. For example, in the case of a pure weak lensing experiment with $N_{\rm b}$ tomographic bins, each $\sfC_{\ell}$ is a dense $2N_{\rm b} \times 2N_{\rm b}$ symmetric positive definite matrix.

\begin{figure*}%
    \centering
    \subfloat[\centering ]{{\includegraphics[width=.45\textwidth]{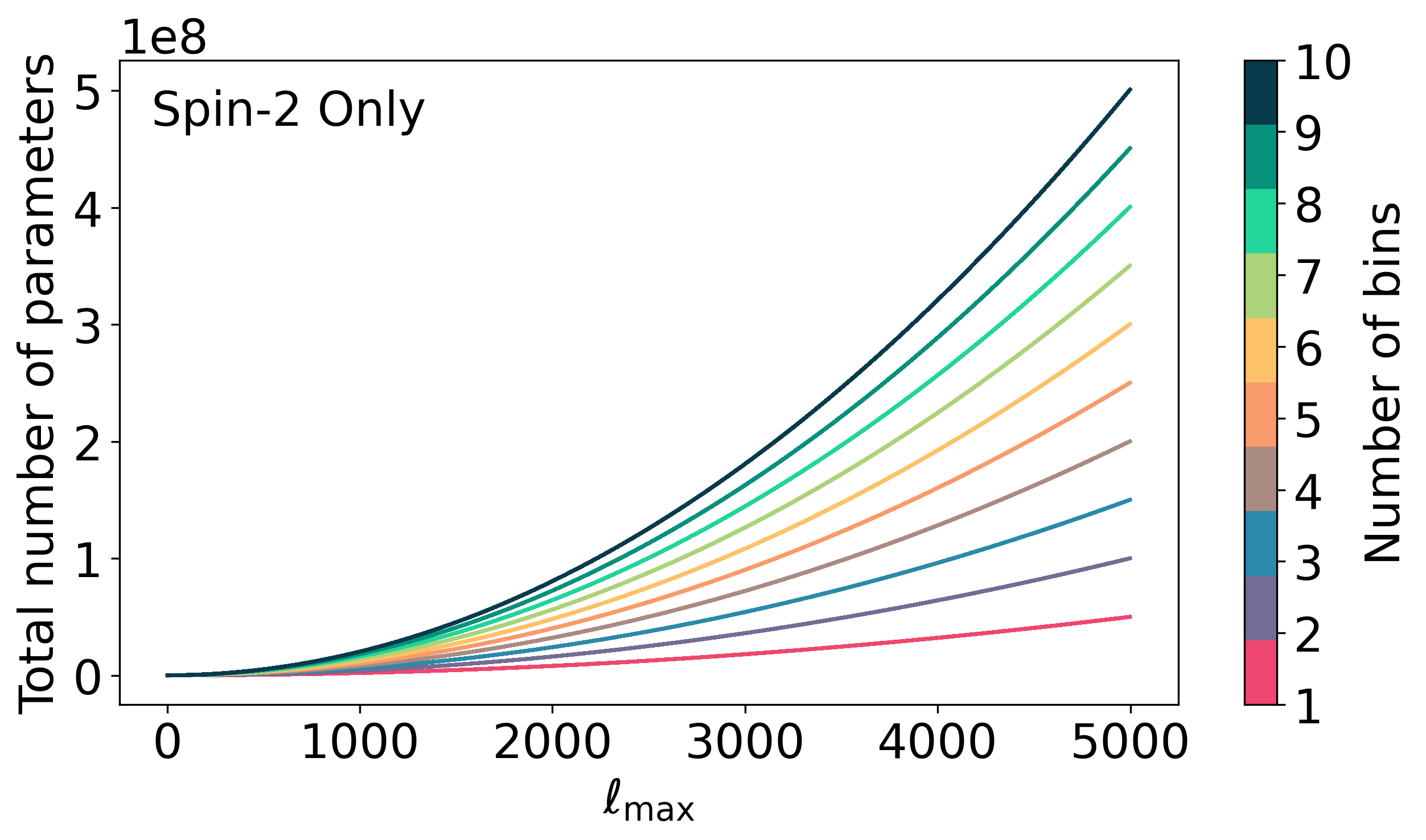} }}%
    \qquad
    \subfloat[\centering ]{{\includegraphics[width=.45\textwidth]{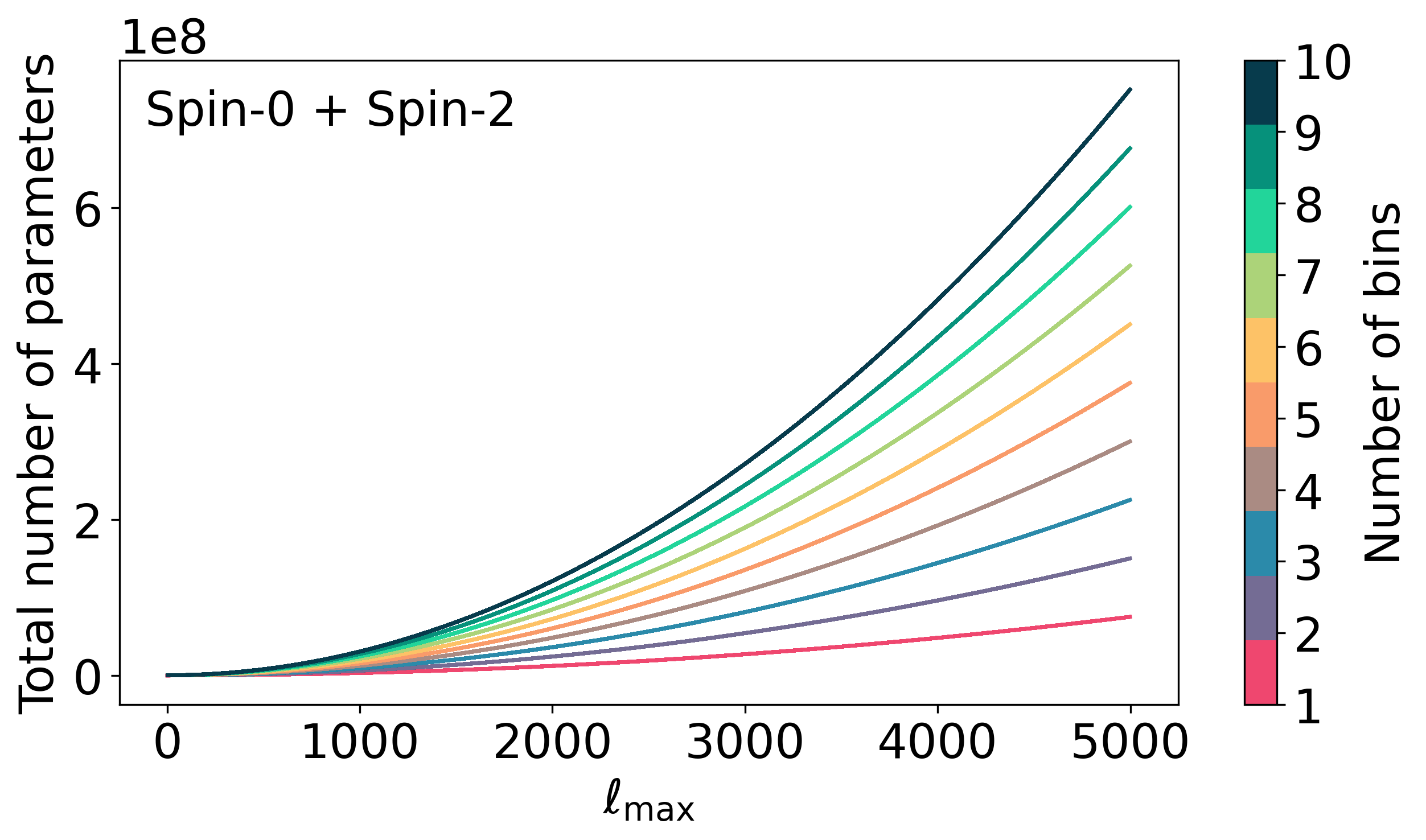} }}%
    \caption{Number of free parameters in a Euclid-like experiment that are jointly estimated by \almanac{}: for a given $\ell_{\rm max}$, all power spectra $C_\ell$ up to $\ell_{\rm max}$ are estimated. Additionally, all $a_{\ell m}$-modes up to $\ell_{\rm max}$ of the underlying (noise-free) field are inferred. The colours indicate the number of redshift bins. This figure assumes both $E$ and $B$-modes are estimated for spin-weight 2 fields (subplot a), and that there is only a single tracer of the spin-weight 0 field in subplot (b), as in galaxy clustering. For a CMB experiment, the spin-weight 0 field would be observed at different frequencies, hence contributing multiple fields. As can be seen, several hundred millions of parameters are to be jointly inferred, implying a requirement for a sampler that can handle this dimensionality very efficiently.}
    \label{Fig:Numpar_euclid}%
    \vspace{8pt}
\end{figure*}

\begin{figure}%
    \centering
    \includegraphics[width=.45\textwidth]{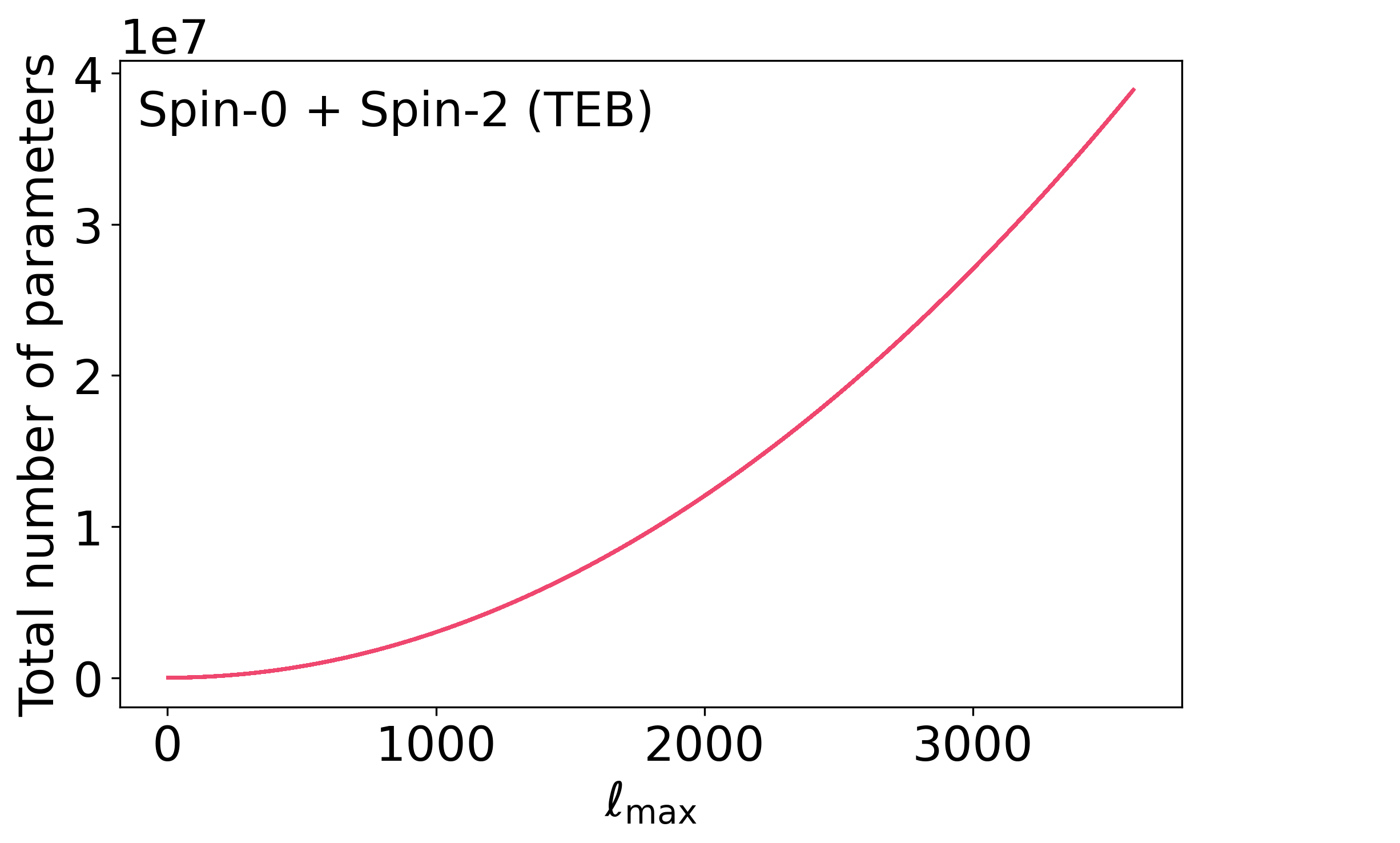}   
    \caption{Same as Fig.~\ref{Fig:Numpar_euclid} but for a Planck-like CMB experiment measuring a $TQU$ field, but typically observed in a number of different wavelengths. The primary CMB temperature and polarization maps, together with their power spectra then yield several ten millions of parameters to be jointly inferred, independently of the number of frequencies.}%
    \label{Fig:Numpar_planck}%
\end{figure}

We show the number of free parameters in \almanac{} as a function of the number of redshift tomographic bins and maximum multipole for a Euclid-like weak lensing only (spin-weight 2 only) and 3$\times$2pt analysis (spin-weight 0 plus spin-weight 2) analysis in Fig.~\ref{Fig:Numpar_euclid}. For a complete Euclid analysis (10 tomographic bins and $\ell_{\rm max} = 5000$), the \almanac{} data model would require around $7\times 10^8$ free parameters. By contrast, a Planck-like CMB analysis would require around $4\times 10^7$ free parameters (an order of magnitude fewer), as shown in Fig.~\ref{Fig:Numpar_planck}. 

We further expect the cross-power spectra $\sfC_{\ell}^{E_iB_j}$ and $\sfC_\ell^{S_iB_j}$ to vanish since $B$ has the opposite parity to $E$ and $T$. The $B$ modes can be constrained to zero in \almanac{} if desired, as there are some circumstances when they should vanish. However, computing them incurs little computational overhead and may be helpful since any observed $EB$ or $TB$ power can indicate contamination, or even new physics \citep{birefringence}.  
More generally we can force to zero all cross-spectra  (between any pairs of different $T$, $E$, or $B$ fields within the same tomographic bin, or any pair of fields in different bins) --- i.e., we can force the matrix $\sfC$ to be diagonal, at the risk of missing important systematics or new physics.

%%%%%%%%%%%%%%%%%%%%%%%%%%%%%%%%%%%%%%%%%%%%%%%%%%%%%%%%%%%%%%%%%%%%%%%%
%                       POSTERIOR & PRIOR
%%%%%%%%%%%%%%%%%%%%%%%%%%%%%%%%%%%%%%%%%%%%%%%%%%%%%%%%%%%%%%%%%%%%%%%%
\subsection{Posterior and Prior}
\label{ssec:posterior}

By Bayes' Theorem the posterior distribution $\calP$ of the full sky maps $\fata$ (in harmonic space) and their power spectra $\sfC$, given the data $\fatd$ and noise covariance $\sfN$, is
\begin{equation}
    \calP(\sfC, \fata | \fatd, \sfN) \propto \calL(\fatd | \fata, \sfN)\calG(\fata | \sfC)\pi(\sfC)\, .
    \label{Eq:Cond_Posterior}
\end{equation}

The first factor in the posterior is the likelihood of the data given the maps we wish to infer, and given the noise covariance:
\begin{equation}
    \calL(\fatd | \fata, \sfN) \propto \exp \left[-\half (\fatd - \sfR\sfY\fata)^{\transpose{}}\sfN^{-1}(\fatd - \sfR\sfY\fata) \right] \, .
    \label{Eq:Likelihood}
\end{equation}
For the CMB, $\sfR$ models the effect of the beam (beam smearing is a convolution in pixel space and hence a multiplication in harmonic space). For a galaxy or cosmic shear survey $\sfR$ is the identity and will be ignored.

The second factor in the posterior is the prior on the spherical harmonic coefficients of the underlying map given the power spectrum; it is taken to be gaussian\footnote{A detail is that for $m \neq 0$ we split $\fata$ into real and imaginary parts, independent of each other, and each with covariance $\sfC/2$. See Appendix~\ref{apx:Preliminaries} for more information.}:

\begin{equation}
    \calG(\fata|\sfC) = \frac{1}{\sqrt{\determinant{2\pi \sfC}}} \exp\left( -\half \fata^{\transpose{}} \sfC^{-1}\fata \right).
    \label{adraw}
\end{equation}
Here, we are taking a theory-agnostic approach, so we assume we have no preferred physical model (which might, for example, impart a particular non-gaussianity due to nonlinear evolution from gaussian initial conditions).  As a consequence, we choose a prior that is minimally informative, and the maximum entropy distribution for given mean and covariance is a gaussian \citep{jaynes03}.  This has the advantage that the maps and power spectra that we sample are independent of cosmology and based only on symmetry properties. 

Furthermore, the gaussian {\it prior} choice does not imply that the ({\it posterior} sample) maps that are created by the sampler are gaussian: for non-gaussian data, the accepted maps (and derived posterior statistics such as the mean) can inherit the non-gaussianity of the data, in particular for highly constraining data.

The third factor in the posterior is a prior on the angular power spectra. We choose a power of the determinant of the $p \times p$ covariance matrix:
\begin{equation}
    \pi(\sfC) = \determinant{\sfC}^q\, .
    \label{Eq:Prior}
\end{equation}
We use $q=0$ in our standard implementation.  In contrast, the Jeffreys prior corresponds to $q = -(p+1)/2$ as is discussed in \cite{SH}. An alternative, frequency-matching prior is given in \citet{2022-Percival}. Note that for subsequent cosmological parameter inference, the choice of prior here is essentially cosmetic, i.e., it impacts how we display our estimates of the spectra. For any further cosmological use, we (typically) have a deterministic model for the spectra $\sfC_\ell$ determined by the vector of cosmological parameters ${\both}$ (including any relevant ancillary parameters, e.g., describing foregrounds or the galaxy distribution). This induces a delta function prior, $\pi( \sfC_\ell | \both ) = \delta_\mathrm{D}[\sfC_\ell - \sfC_\ell(\both)]$, and the $\sfC$ prior cancels out, to be replaced by a prior on the cosmological parameters.

In summary, our model uses as free parameters all power spectra modes and the spherical harmonic coefficients for both the $E$- and $B$-modes of the data in question, and/or a scalar mode, depending on application.

When parts of the sky are masked out, the field values of the masked pixels can therefore be inferred: the observed pixels will constrain the spherical harmonic mode of low $\ell$, and their synthesis will thus yield posterior values for credible field configurations under the mask. This enables \almanac{} to represent the posterior covariance between spherical harmonic modes that are mixed by the presence of the mask and anisotropic noise, and, in particular, to provide a Bayesian solution to the $EB$-leakage problem -- there will be strong posterior dependence between modes that are subject to leakage, which will be evident as covariance between and amongst both map (spherical harmonic coefficient) and power spectrum parameters. $EB$ leakage as such does not occur unless the posterior samples are reduced to point estimates.

%%%%%%%%%%%%%%%%%%%%%%%%%%%%%%%%%%%%%%%%%%%%%%%%%%%%%%%%%%%%%%%%%%%%%%%%
%                           HMC THEORY
%%%%%%%%%%%%%%%%%%%%%%%%%%%%%%%%%%%%%%%%%%%%%%%%%%%%%%%%%%%%%%%%%%%%%%%%
\section{Hamiltonian Monte Carlo}
\label{sec:HMCsampler}

In this section we briefly review the Hamiltonian Monte Carlo sampler. 
Readers familiar with this sampler and its tunable parameters can skip this section. 

In HMC \citep{2001-Hanson-HMC,2007-Hajian-HMC,2011-Neal} we view the negative logarithm of the posterior density as a potential energy

\begin{equation}
    \nlp{}(\faty) = - \ln \calP(\faty)
    \label{Eq:Potential_Generic1}
\end{equation}
where $\faty$ is a point in parameter space. We then augment the parameter space with momentum variables bearing a gaussian probability distribution (whose negative logarithm can be viewed as kinetic energy). The negative logarithm of the combined probability distribution on this augmented space (i.e., \textit{phase space}) may then be viewed as the total energy of a dynamical system. Dynamical flow on this system preserves energy and phase space volume; in probability terms, this corresponds to preserving detailed balance. The dynamical flow does not fully explore phase space (as it cannot leave the subspace corresponding to a fixed total energy); we therefore periodically stop the flow and resample the momentum (this step also preserves detailed balance). In practice the calculation of the dynamical flow is done in discrete steps using a leapfrog algorithm that preserves phase space volume but that only approximately preserves total energy; a Metropolis-Hastings decision \citep{1953-Metropolis, 1970-Hastings} then corrects for the energy error. We finally marginalise by discarding the momentum variables, yielding a set of samples of the original distribution. 

In our case, we jointly sample spherical harmonic coefficients and power spectra (or reparametrizations of these - as discussed in Sect.~\ref{sec:HMCtuning}, we need not directly sample the externally visible parameters $\fata$ and $\sfC$). We gather all parameters to be sampled in the vector $\faty$.

Let $\fatp$ be the momenta conjugate to $\faty$. Because a momentum is a velocity times a mass, HMC requires a positive definite mass matrix $\sfM$ to be set. A good choice for $\sfM$ is crucial for obtaining a highly efficient sampler. If $d = \mathrm{dim}(\faty)$, then the symmetric $\sfM$ introduces $d(d+1)/2$ tunable parameters; HMC tuning is hence a formidable challenge when $d$ is large. Tuning of the \almanac{} sampler is presented in Sect.~\ref{sec:HMCtuning}.

For now we simply assume that the mass matrix $\sfM$ has been set. The Hamiltonian then is
\begin{equation}\label{eq:Hamiltonian}
    H(\faty,\fatp) = \nlp{}(\faty) + \half\fatp^{\transpose{}}\sfM^{-1} \fatp + {\rm constant},
\end{equation}
and the target posterior of $\faty$ is $\exp(-H)$, marginalised over the auxiliary momenta. 

The HMC algorithm begins at a random starting point $\faty_0$, which we assume for now has been chosen; see Sect.~\ref{sec:HMCtuning} for the exact method by which we choose a starting point.

HMC then augments this starting point with an initial momentum vector $\fatp_0$ drawn from a gaussian distribution with mean zero and covariance matrix $\sfM$.
The position and momentum are then updated by solving the Hamiltonian system
\begin{equation}
\begin{aligned}
\frac{\mathd \faty}{\mathd t} & = \frac{\partial H}{\partial \fatp} = \sfM^{-1} \fatp,\\
\frac{\mathd \fatp}{\mathd t} & = - \frac{\partial H}{\partial \faty} = - \nabla_{\faty} \nlp{}(\faty).
\end{aligned}
\label{eom}
\end{equation}
Hamiltonian flow preserves energy and phase space volume and together these properties yield the detailed balance necessary for HMC to converge to the target distribution. In practice we can solve the equations of motion Eq.~\eqref{eom} only approximately, with discretized time. It is important here to use a symplectic integrator as such integrators preserve phase space volume and, although not exactly preserving energy, at least have bounded energy error. To this end, we choose the leapfrog integrator given by 
\begin{equation}
    \begin{aligned}
\fatp(t + \epsilon/2) & = \fatp(t) - \epsilon \nabla_{\faty} \nlp{}[\faty(t)]/2,\\
\faty(t + \epsilon) & = \faty(t) + \epsilon \sfM^{-1} \fatp(t + \epsilon/2),\\
\fatp(t + \epsilon) & = \fatp(t + \epsilon/2) - \epsilon \nabla_{\faty} \nlp{}[ \faty(t+ \epsilon) ]/2;
    \end{aligned}
    \label{LF}
\end{equation}
here $t$ is discretized time and $\epsilon$ is the discrete increment in $t$. The gradients of the potential have closed-form expressions in our science case. The three steps of Eq.~\eqref{LF} make one leapfrog move. A trajectory of $r$ position updates is built up by repeating the leapfrog move $r$ times.

If $\sfM$ is diagonal then it is easier to work with transformed momenta $\tilde{\fatp} = \sfM^{-1/2} \fatp$ (in fact such a transformation is possible for arbitrary positive definite $\sfM$ via a Cholesky decomposition). The kinetic energy term in $H$ simplifies to $\tilde{\fatp}^{\transpose{}} \tilde{\fatp}/2$, i.e., $\tilde{\fatp}$ is distributed with unit covariance, and the leapfrog equations simplify to

\begin{equation}
    \begin{aligned}
\tilde{\fatp}(t + \epsilon/2) & = \tilde{\fatp}(t) - \fats \odot \nabla_{\faty} \nlp{}[\faty(t)]/2,\\
\faty(t + \epsilon) & = \faty(t) + \fats \odot \tilde{\fatp}(t + \epsilon/2),\\
\tilde{\fatp}(t + \epsilon) & = \tilde{\fatp}(t + \epsilon/2) - \fats \odot \nabla_{\faty} \nlp{}[ \faty(t+ \epsilon) ]/2.
    \end{aligned}
    \label{LFprime}
\end{equation}
Here $\fats = \epsilon \diag(\sfM^{-1/2})$ is the vector of \textit{step sizes} and $\odot$ denotes element-wise multiplication; see \cite{2011-Neal}.

We now discard $\sfM$ and instead deal exclusively with the step sizes $\fats$. When optimising the $\fats$ and the number $r$ of leapfrog moves per trajectory one must balance a high acceptance rate (small step size) with numerical efficiency (small $r$) and good mixing (large distances traversed in short time). Small step size and $r$ (at the same time) causes the sampler to traverse only short distances; large step sizes leads to poor energy conservation; large $r$ is computationally expensive. Some tuning is therefore needed; our tuning is presented in Sect.~\ref{sec:HMCtuning}.

If $\nlp{}$ is gaussian with zero off-diagonal correlation then it is ideal to use its inverse covariance matrix as the mass matrix $\sfM$, as this leads to a matching of the scale of the position and momentum variables in each component. For more general posteriors one can approximate this ideal by using step sizes $\fats = (\diag \textrm{Hessian(\nlp{}}))^{-1/2}$; these quantities may be calculated analytically at some typical point, or else estimated using samples of $\nlp{}$.

Our aim is to have a numerically efficient sampler, rather than a perfect solution to the Hamiltonian system. Whether the endpoint  $(\faty_{\rm E},\fatp_{\rm E})$ of a leapfrog trajectory is accepted as a new sample therefore is decided by a usual Metropolis-Hastings comparison; the acceptance probability is
\begin{equation}
    P_{\rm accept} = {\rm min}\left\{1, \exp\left[-H( \faty_{\rm E},\fatp_{\rm E} ) + H(\faty_{\rm S},\fatp_{\rm S})\right]\right\},
\end{equation}
where $(\faty_{\rm S},\fatp_{\rm S})$ is the start of the trajectory. If the trajectory's end point is accepted then the sampler continues running from this point; otherwise the previous point is repeated in the chain, which continues running from the trajectory's starting point. New momenta are drawn in either case.

%%%%%%%%%%%%%%%%%%%%%%%%%%%%%%%%%%%%%%%%%%%%%%%%%%%%%%%%%%%%%%%%%%%%%%%%
%                         SAMPLER OPTIMISATION
%%%%%%%%%%%%%%%%%%%%%%%%%%%%%%%%%%%%%%%%%%%%%%%%%%%%%%%%%%%%%%%%%%%%%%%%
\section{Optimising the Almanac sampler}
\label{sec:HMCtuning}

HMC is a generic sampling algorithm. Generic samplers have the advantage of being able to sample from a wide range of probability distributions and the disadvantage, in the absence of tuning, of sampling many of these inefficiently.

Inefficiency in a sampler can manifest itself in several ways, including unacceptably large correlation length of the chain, slow exploration of the posterior, and avoiding (or alternatively getting stuck in) difficult regions of the posterior. The ultimate problem with an inefficient sampler is that it produces chains with a low effective sample size (see Section~\ref{ssec:ESS}). Observable corollaries of inefficient sampling are that it is time-consuming and wastes numerical efforts.

Once we have specified the data model (Sect.~\ref{sec:datamodel}) and sampler (Sect.~\ref{sec:HMCsampler}), there are still several choices that will affect the behaviour of the sampler in practice (and indeed the best choice will also depend on the character of the data, for example the signal-to-noise ratio in different regimes): choice of how to parameterise the probability space; choice of starting point for the sampler; choice of the mass matrix (which appears in the kinetic energy term of the Hamiltonian) as well as other parameters of our integration scheme.

\subsection{Parameterising Maps and Spectra}\label{sub:params}
\begin{figure}%
    \centering
    \includegraphics[width=0.45\textwidth]{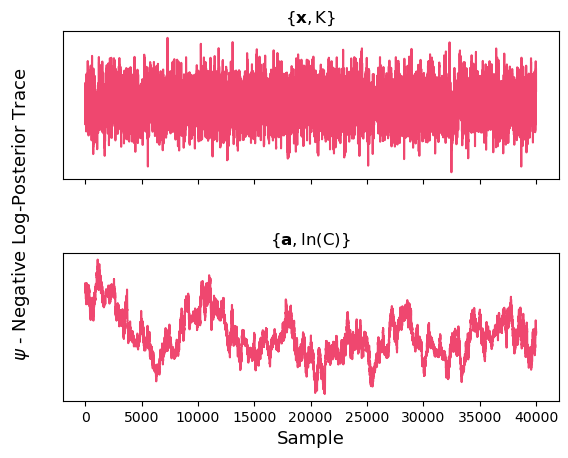}%
    \caption{Comparison between the negative log-posterior (potential $\nlp$) of two different coordinate systems: \textit{(top)} Cholesky parameterisation $\{\fatx, \sfK \}$ and the \textit{(bottom)} $\{\fata, \ln(\sfC) \}$ parameterisation applied for the same  spin-weight 2 lower dimensional test case ($\ell_{\rm max} = 32$). }%
    \label{Fig:PostTrac-NormXChol}%
    \vspace{0.3cm}
\end{figure}
Our data model (Sect.~\ref{sec:datamodel}) expresses the posterior probability distribution for the underlying signal maps and all of the related auto- and cross-power spectra. However, we are free to parameterise these maps and spectra in a way that might simplify the distribution. 

In particular, the matrix $\sfC$ must be positive definite at each $\ell$ and this requirement induces very strong nonlinear constraints on its entries, which in turn gives a very complicated boundary to the parameter space; naively sampling the elements of $\sfC$ will not guarantee positive-definiteness.

There are at least two solutions to this problem. One is to sample the elements of the matrix log $\sfG$ of the covariance matrix, defined so that $\sfC = \exp(\sfG)$. Another is to use the Cholesky decomposition $\sfC=\sfL\sfL^{\transpose{}}$, where  $\sfL$ is a lower triangular matrix with positive entries on the diagonal, and to sample the elements of the \textit{diagonal-log} $\sfK$ of $\sfL$ (here we take the logarithm of the diagonal entries of $\sfL$ and leave the off-diagonal entries unchanged -- see Eq.~\ref{Eq:DiagonalLog_L}). Using the formalism developed in Appendix~\ref{apx:DerivsOfMatrixExp} for derivatives of the exponential of symmetric matrices, the derivatives needed for the Hamiltonian sampler are derived in Appendix~\ref{apx:FormulaeForLogCoordinates} (for the log parameterisation) and Appendix~\ref{apx:FormulaeForCholeskyCoordinates} (for the Cholesky parameterisation). We have investigated both methods and compare results below.

Generally, the Cholesky decomposition gives shorter correlation lengths in our investigations. It is the matrix equivalent of taking the (positive) square root of $\sfC$; note that the diagonal element parameterisation is equivalent to a Jeffreys prior on $L_{ii}$, which is appropriate for a scale parameter. 

Problems such as those solved by \almanac{} in which we are simultaneously estimating a mean (the signal map) and variance (the spectra) exhibit a posterior geometry sometimes called a `funnel' or `stingray': there is a very long, narrow tail of high probability density and small gradient down toward very small positive values of the covariance. This can be very difficult to explore efficiently. With the Cholesky decomposition, we ameliorate this by decorrelating the modes and normalising them to unit variance. For a scalar, this means using $x_{\ell m}=a_{\ell m}/\sqrt{C_\ell}$ as our parameter; more generally we transform using the Cholesky decomposition: $\fatx=\sfL^{-1}\fata$. This has the effect that our new $\fatx$ are standard normal parameters.

The favoured parameterisation for \almanac{} is therefore the Cholesky one, using the decorrelated spherical harmonic coefficients $\fatx$ and the modified Cholesky decomposition $\sfK$. We use these parameters internally in the sampler, but we impose our prior (Eq.~\eqref{Eq:Prior}) on the power spectra and we use appropriate Jacobians to translate between the parameters (as described in Appendices~\ref{apx:FormulaeForLogCoordinates} and~\ref{apx:FormulaeForCholeskyCoordinates}). 

In Fig.~\ref{Fig:PostTrac-NormXChol}, we investigate the impact of the coordinate system choice on the trace of the posterior distribution for a spin-weight 2 toy example ($\ell_{\rm max} =32$). The $\{\fata, \sfG) \}$ coordinate system fails to explore the underlying target distribution efficiently. In contrast, the Cholesky coordinate system, $\{\fatx, \sfK \}$, successfully explores the target distribution. We further analyse the parameterisation choices in Sect.~\ref{sec:paramresults}, while Appendices~\ref{apx:FormulaeForLogCoordinates} and~\ref{apx:FormulaeForCholeskyCoordinates} give further mathematical detail. The companion paper \citep{2022AlmanacWL}{} presents a summary of results.

%%%%%%%%%%%%%%%%%%%%%%%%%%%%%%%%%%%%%%%%%%%%%%%%%%%%%%%%%%%%%%%%%%%%%%%%
%                           STARTING POINTS
%%%%%%%%%%%%%%%%%%%%%%%%%%%%%%%%%%%%%%%%%%%%%%%%%%%%%%%%%%%%%%%%%%%%%%%%
\subsection{Dispersed Starting Points}\label{SSec:StartingPoints}
We wish to choose a starting point that is not too far away from the typical region of our posterior as this also makes it reasonable to use this point as the location where we calculate the Hessian. However, we also need to maintain a certain randomness in our starting points in order to test whether chains from different starting points converge to the same target density. Hence, we implement the following steps to generate dispersed starting points:

\begin{figure*}%
    \centering
    \subfloat[\centering Spin-Weight 0]{{\includegraphics[width=.45\textwidth]{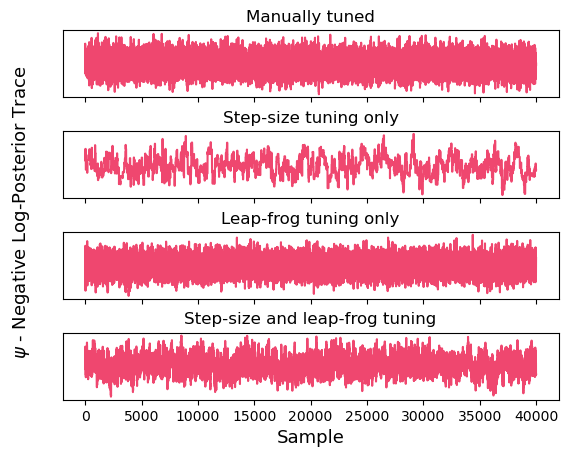} }}%
    \qquad
    \subfloat[\centering Spin-Weight 2]{{\includegraphics[width=0.45\textwidth]{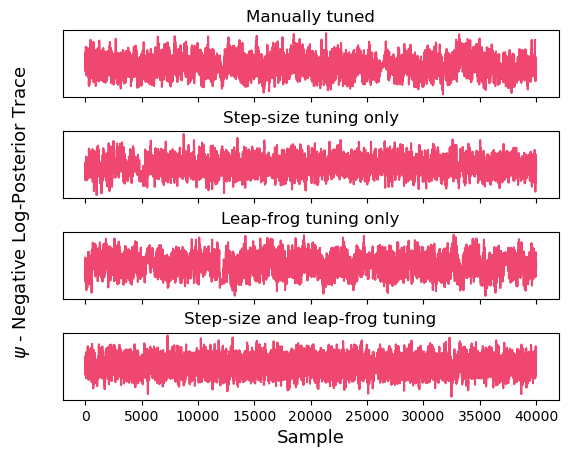} }}%
    \caption{Comparison of the impact of the different phases and methods of HMC tuning for \subf{(a)} a spin-weight 0 lower dimensional case and \subf{(b)} a spin-weight 2 lower dimensional case -- both with $\ell_{\rm max} = 32$. \textit{(First row)} shows the case for a manually tuned HMC, i.e., when the user finds an apparently optimal set of HMC parameters by trial and error. \textit{(Second row)} Only the step-size tuning is performed by using the standard deviation of the post burn-in samples as a proxy for the step-sizes. \textit{(Third row)} The Hessian is kept as the step-sizes and only tuning of the leap-frog parameters is performed, finding the factor $\tilde{g}$, which re-scales the step-sizes. \textit{(Fourth row)} All tuning stages are applied as described in Sect.~\ref{sec:HMCtuning}. }%
    \label{Fig:PostTracer_Tuning_Tests}%
    \vspace{8pt}
\end{figure*}

\begin{enumerate}
    \item Calculate the spherical harmonic coefficients of the data, $\fata_{\rm data}^i$, for each of the maps (treating masked regions as zero);
    \item Estimate pseudo-$\sfC_{\ell}$ power spectra of the data maps
        \begin{equation}
            \hat\sfC_{\ell} \equiv \frac{1}{2\ell +1} \sum_m \fata_{\rm data}^{\phantom{\dagger}} \fata_{\rm data}^{\dagger};
            \label{Eq:PCL}
        \end{equation}
    \item Randomly sample a new set of spherical harmonics by drawing from a gaussian
        \begin{equation}
            \tilde{\fata}_{\rm data} \sim \calG(0, \hat\sfC_{\ell})\, ,
        \end{equation}
        and use these as the starting points for the $\fata$ dimensions in the parameter space;
    \item Estimate a new set of pseudo-$\sfC_{\ell}$ from the new spherical harmonics $\tilde{\fata}_{\rm data}$ and use these as the starting point for the $\sfC$ dimensions in the parameter space.
\end{enumerate}

This gives a starting point in $\{\fata, \sfC\}$ space, which may then be transformed to the desired coordinate system for use in the sampler.

%%%%%%%%%%%%%%%%%%%%%%%%%%%%%%%%%%%%%%%%%%%%%%%%%%%%%%%%%%%%%%%%%%%%%%%%
%                               TUNING
%%%%%%%%%%%%%%%%%%%%%%%%%%%%%%%%%%%%%%%%%%%%%%%%%%%%%%%%%%%%%%%%%%%%%%%%

\subsection{Tuning}

\begin{figure}%
    \centering
    \subfloat[\centering]{{\includegraphics[width=.45\textwidth]{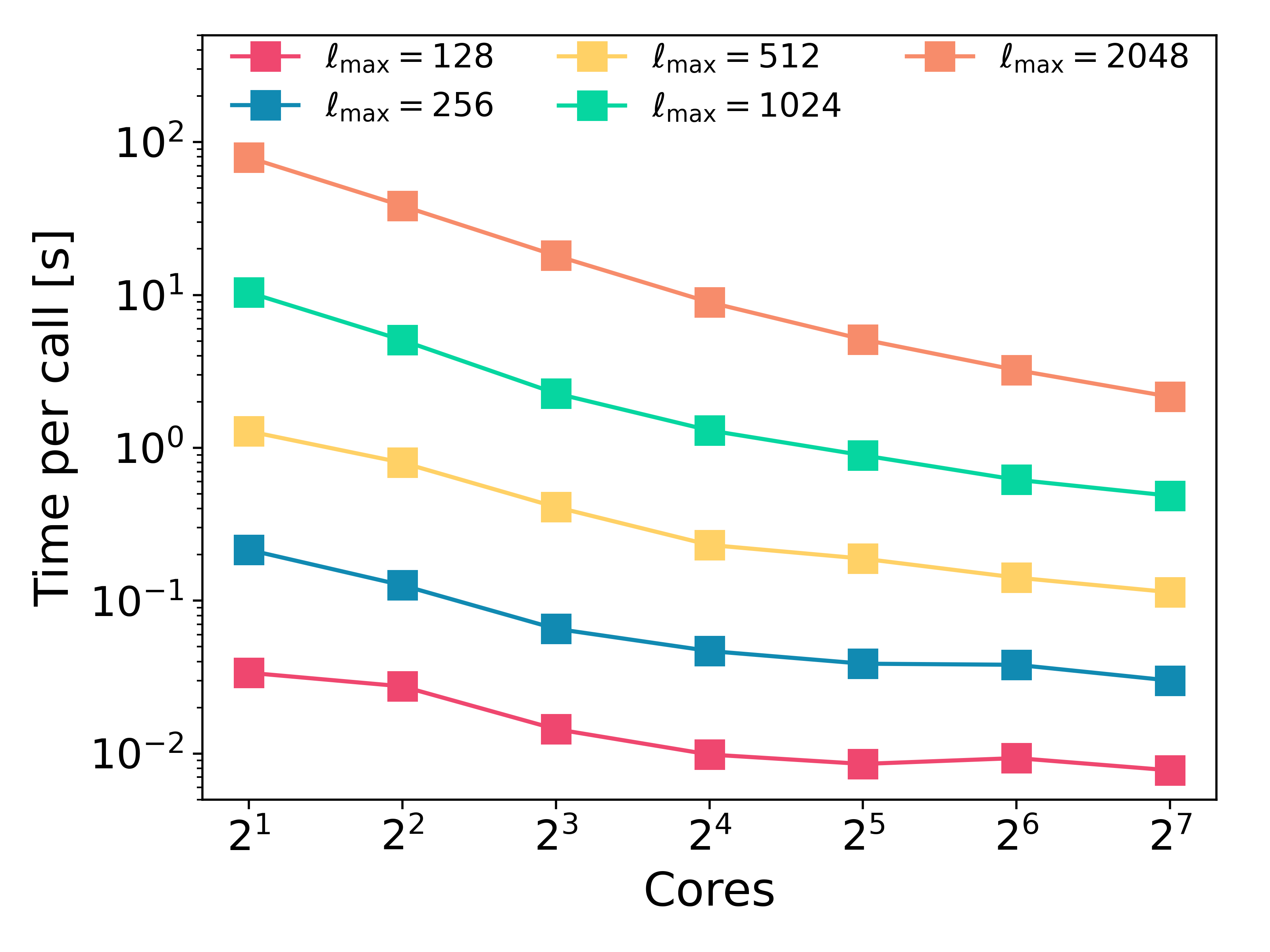}}\label{Fig:time_scale_left}}%
    \qquad
    \subfloat[\centering]{{\includegraphics[width=0.45\textwidth]{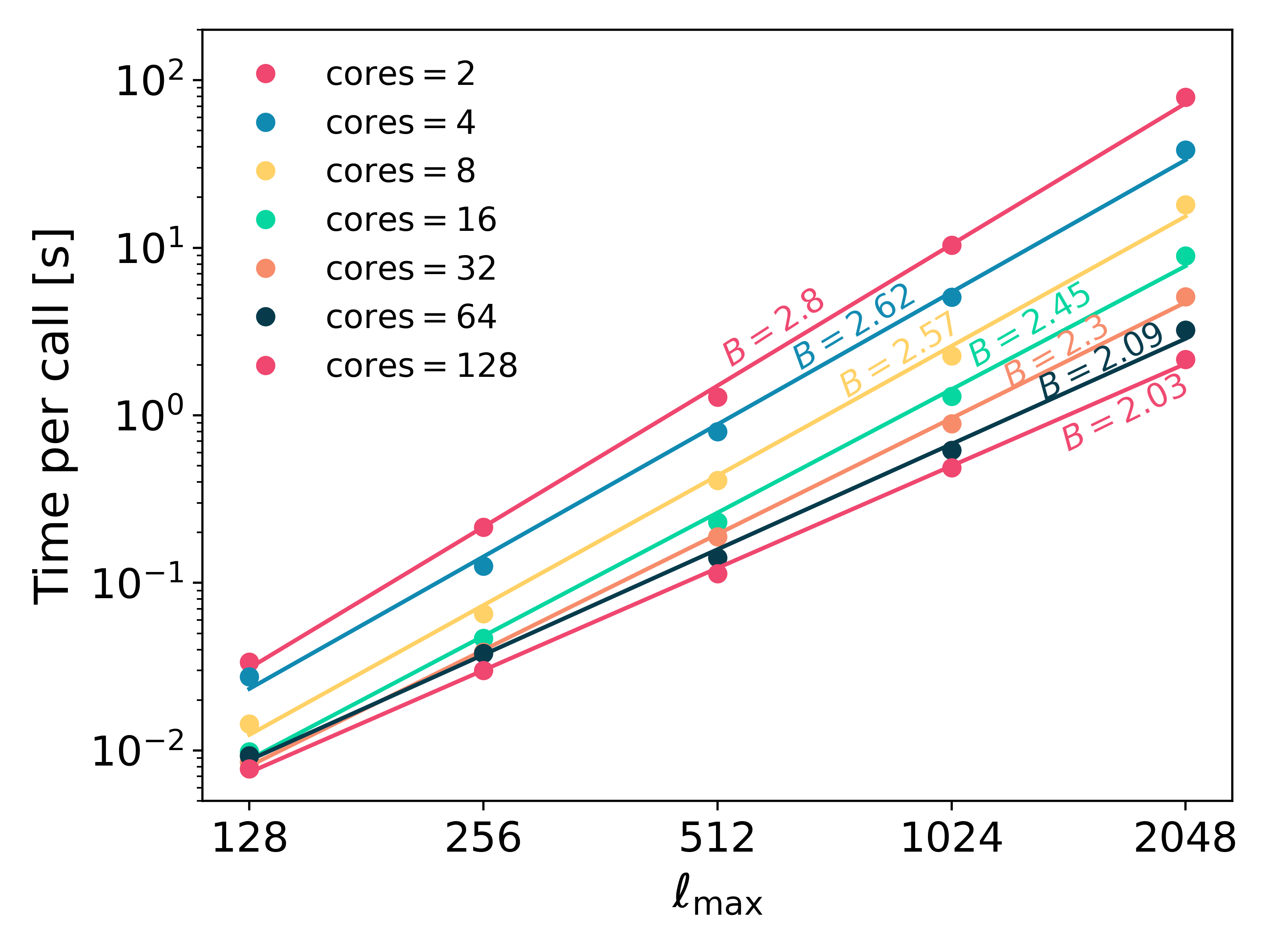}}\label{Fig:time_scale_right}}%
    \caption{Performance of the \almanac{} code when evaluating the negative log-likelihood (Eq.~\eqref{Eq:Likelihood}). \subf{(a)} Scaling of the average time per call with the number of cores for different resolutions ($\ell_{\rm max}$). \subf{(b)} As before, but now with $\ell_{\rm max}$ as the independent variable; we also show a power-law fit $t=A\,\ell_{\rm max}^B$. We report the $B$ exponent, which changes with the number of cores; we obtain the value closest to the ideal ($B=2$) with 128 cores ($B=2.03$).}%
    \label{Fig:time_scaling}%
    \vspace{8pt}
\end{figure}

The \almanac{} algorithm proceeds in four stages. Each successive stage begins from the final sample point of the previous stage.

\begin{enumerate}
\item {\textit{Burn-in}: At startup, \almanac{} sets the start point using the method of Sect.~\ref{SSec:StartingPoints} and sets the step sizes to be the inverse square root of the diagonal of the Hessian (see discussion in Sect.~\ref{sec:HMCsampler}), calculated at the start point.\footnote{If the data are unconstraining, the gradients will start to vanish. No matter how big a step size is then chosen, the sampler will get stuck: it is confronted with a huge prior range and must traverse it with tiny steps due to vanishing gradients. This is the failure mode of the HMC sampler and hence causes extremely long correlation lengths. A remedy can be to draw from the prior if the gradient vanishes.} We then calculate $n_{\textrm{burn-in}}$ samples (usually a few thousand suffice) so as to move from the start point to the typical region.}
\item {\textit{First tuning stage}: This stage continues to use the Hessian-derived step sizes. We take $n_{\rm tune}$ samples from the typical region, and at the end of the stage we reset the step sizes, $\mathbf{s}$, to be the standard deviations of these samples.}
\item {\textit{Second tuning stage}: This stage of $n_{\rm lf-tune}$ samples uses the standard-deviation-derived step sizes, but now all multiplied by an additional overall scale factor $g$. During this stage we vary $g$, keeping track of the consequential variation in the HMC acceptance ratio. At the end of the stage we may therefore infer which value of $g$ (call it $\tilde{g}$) yields an acceptance ratio matching a user-specified target, and we fix the step sizes to be the standard-deviation-derived step sizes scaled by this $\tilde{g}$.
Smaller step sizes lead to a smaller energy error in the leapfrog step and hence a higher acceptance ratio. Extreme step sizes (both small and large) lead to inefficient sampling; there is therefore an optimal step size and hence an optimal acceptance rate. We have arbitrarily used a target acceptance rate of 70\%, close to the 65\% value found to be optimal with simpler posteriors \citep{2010arXiv1001.4460B}; we leave for future work the determination of which target acceptance rate leads to the most efficient sampling. We have chosen here to optimise acceptance ratio simply because it is easy to measure.}
\item {\textit{Main sampling stage}: At the start of this final stage we discard all samples taken so far. We then take $n_{\rm samples}$ samples, using the standard-deviation-derived step sizes, all scaled by $\tilde{g}$. This sampling will be from the typical region, and will achieve (close to) the desired acceptance ratio.}
\end{enumerate}

Other hyperparameters for the HMC algorithm are not subject to such tuning. For example, the number $r$ of position updates for a single trajectory is simply randomly chosen from the range $[1,10)$. We use a non-fixed $r$ to avoid resonant trajectories that hinder ergodic sampling; see \cite{Taylor} and \cite{2011-Neal}. A proposed enhancement to the algorithm is to tune this range to optimise performance.

In Fig.~\ref{Fig:PostTracer_Tuning_Tests}, we show the effects of the different stages of tuning, applied in combination and separately, in the negative log-posterior trace for a spin-weight 0 and spin-weight 2 lower dimension case ($\ell_{\rm max} = 32$). We compare these to a case where the step sizes are manually tuned by visually inspecting the chain post-burn-in via its acceptance ratio and other convergence statistics (see Sect.~\ref{sec:paramresults}). If the manual convergence inspection shows poor convergence, the chains are re-started with a different step size until the diagnostics indicate a better convergence. Note that there is no guarantee that inspecting convergence immediately after burn-in for a few thousand samples means we will have an optimal sampling. While manually tuning the chains is possible for a lower-dimensional case, this becomes infeasible as we approach realistic circumstances. Fig.~\ref{Fig:PostTracer_Tuning_Tests} illustrates that the full tuning sets our sampler in the proper optimisation path, indicating that the leap-frog tuning is the most crucial stage.

\subsection{Performance}\label{Sec:Performance}
To test the performance of \almanac{}, we measure the average calculation time for a configuration based on spin-weight 2 maps for two bins with different resolutions and $\ell_{\rm max}= 2 n_{\rm side}$ (similar to the example presented in \citealt{2022AlmanacWL}). Fig.~\ref{Fig:time_scale_left} shows the time per call for the negative log-likelihood (as this dominates the overall calculation time) as a function of the number of cores in one node, comparing $2,4,8,16,32,64$ and $128$ cores. The plot shows how the code is robust in the scaling (although these graphs are not linear, as is expected since some parts of the code are not parallelisable). We note, especially for $\ell_{\rm max}=128$, that performance plateaus once we use more than $16$ cores. 
Similarly, Fig.~\ref{Fig:time_scale_right} shows how time per call for the negative log-likelihood scales with the maximum resolution of the map ($\ell_{\rm max}$); we also show a power-law fit $t=A\,\ell_{\rm max}^B$, and the resultant $B$ exponents. We see that the performance scales slightly worse than the ideal $\ell_{\rm max}^2$ (i.e., proportional to the number of pixels).

%%%%%%%%%%%%%%%%%%%%%%%%%%%%%%%%%%%%%%%%%%%%%%%%%%%%%%%%%%%%%%%%%%%%%%%%
%                                RESULTS
%%%%%%%%%%%%%%%%%%%%%%%%%%%%%%%%%%%%%%%%%%%%%%%%%%%%%%%%%%%%%%%%%%%%%%%%
\section{Applications to Simulated CMB data}\label{sec:results}

\begin{figure}%
    \centering
    \subfloat[\centering CMB Temperature]{{\includegraphics[width=.45\textwidth]{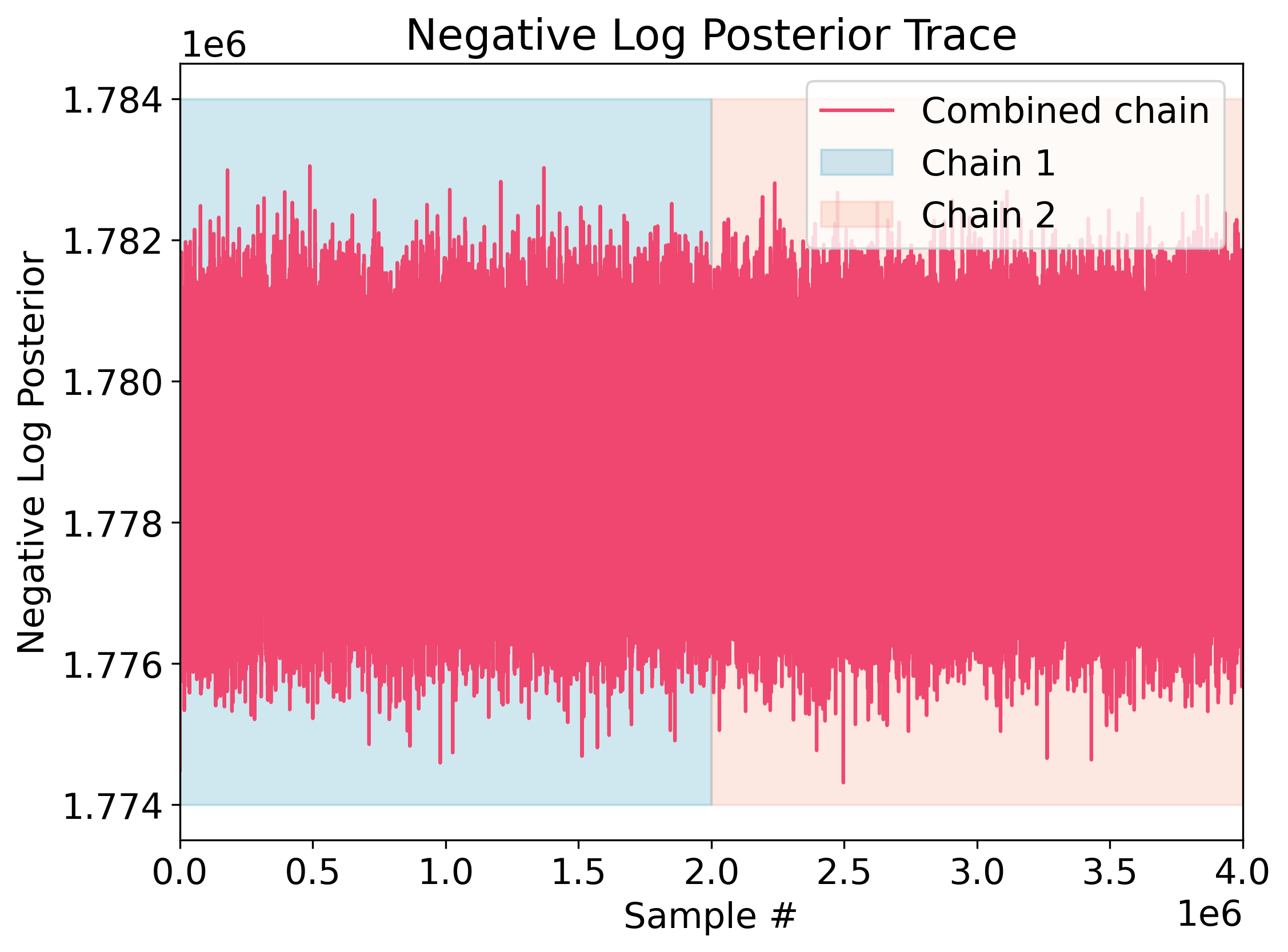} }}%
    \qquad
    \subfloat[\centering CMB Polarization]{{\includegraphics[width=0.45\textwidth]{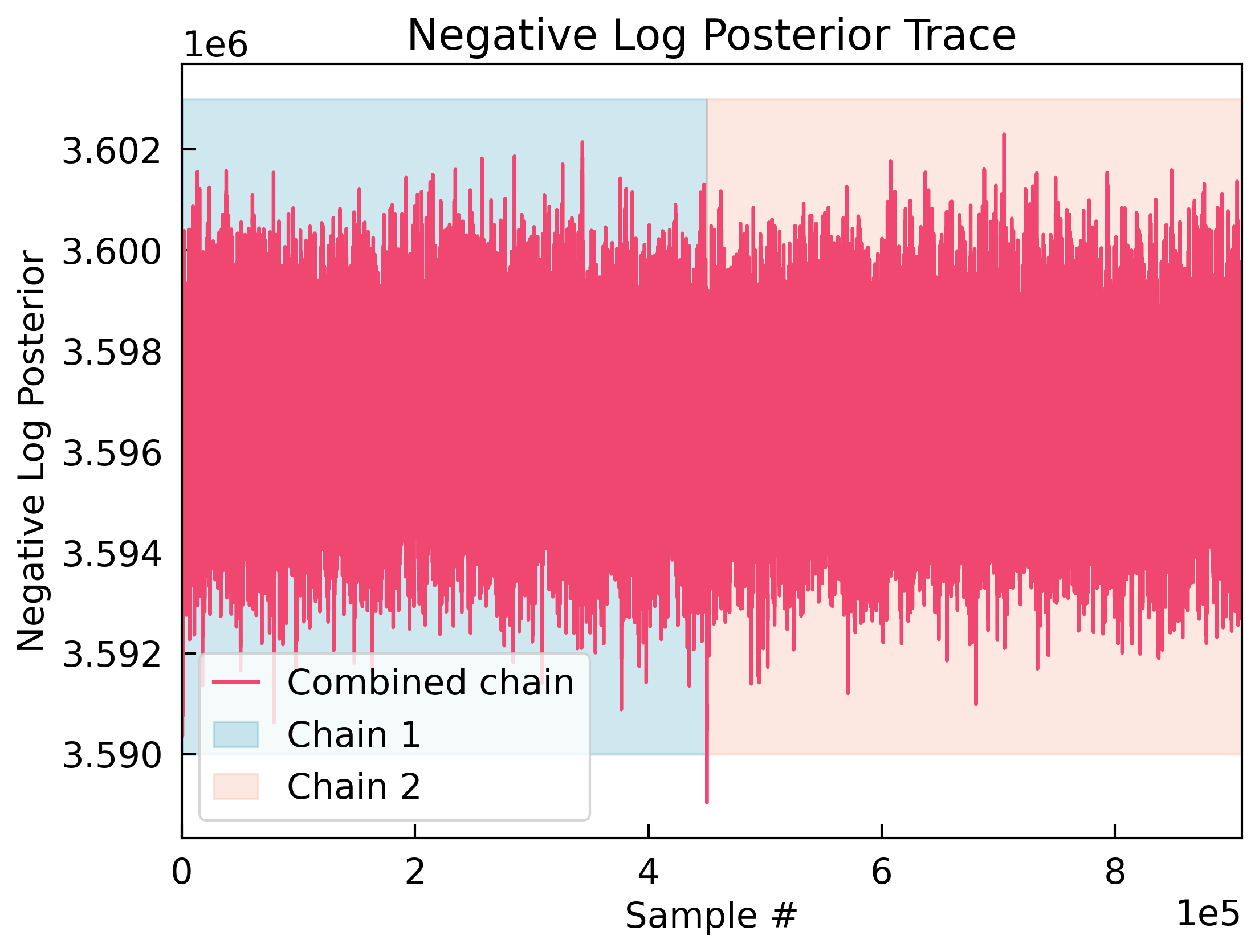} }}%
    \caption{The negative log-posterior trace for the concatenated chains used in analysis for both the \subf{(a)} CMB temperature and \subf{(b)} CMB polarization tests.}%
    \label{Fig:PostTracer_CMB_Tests}%
    \vspace{8pt}
\end{figure}

\begin{figure}%
    \centering
    \subfloat[\centering CMB Temperature]{{\includegraphics[width=.45\textwidth]{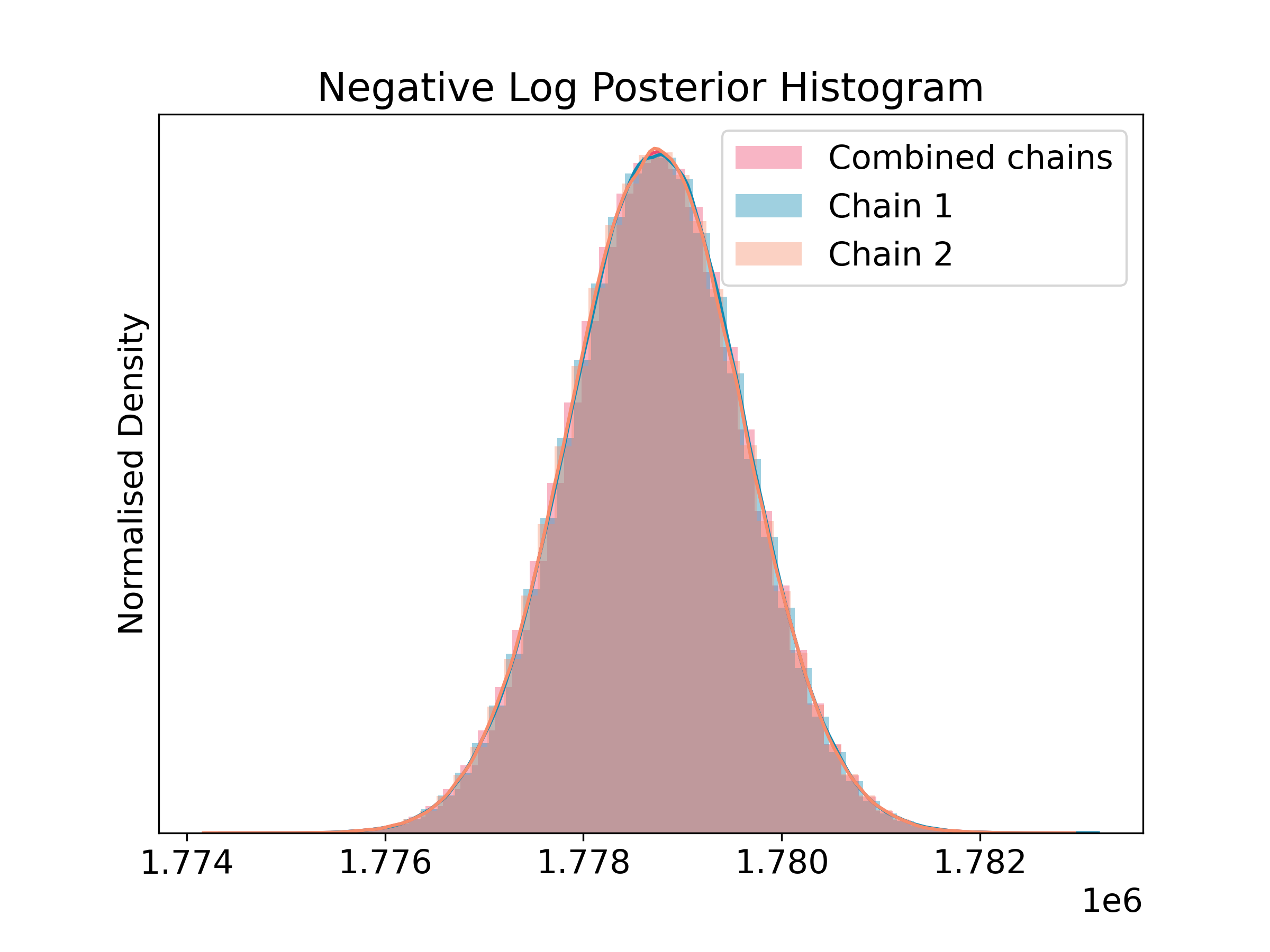} }}%
    \qquad
    \subfloat[\centering CMB Polarization]{{\includegraphics[width=0.45\textwidth]{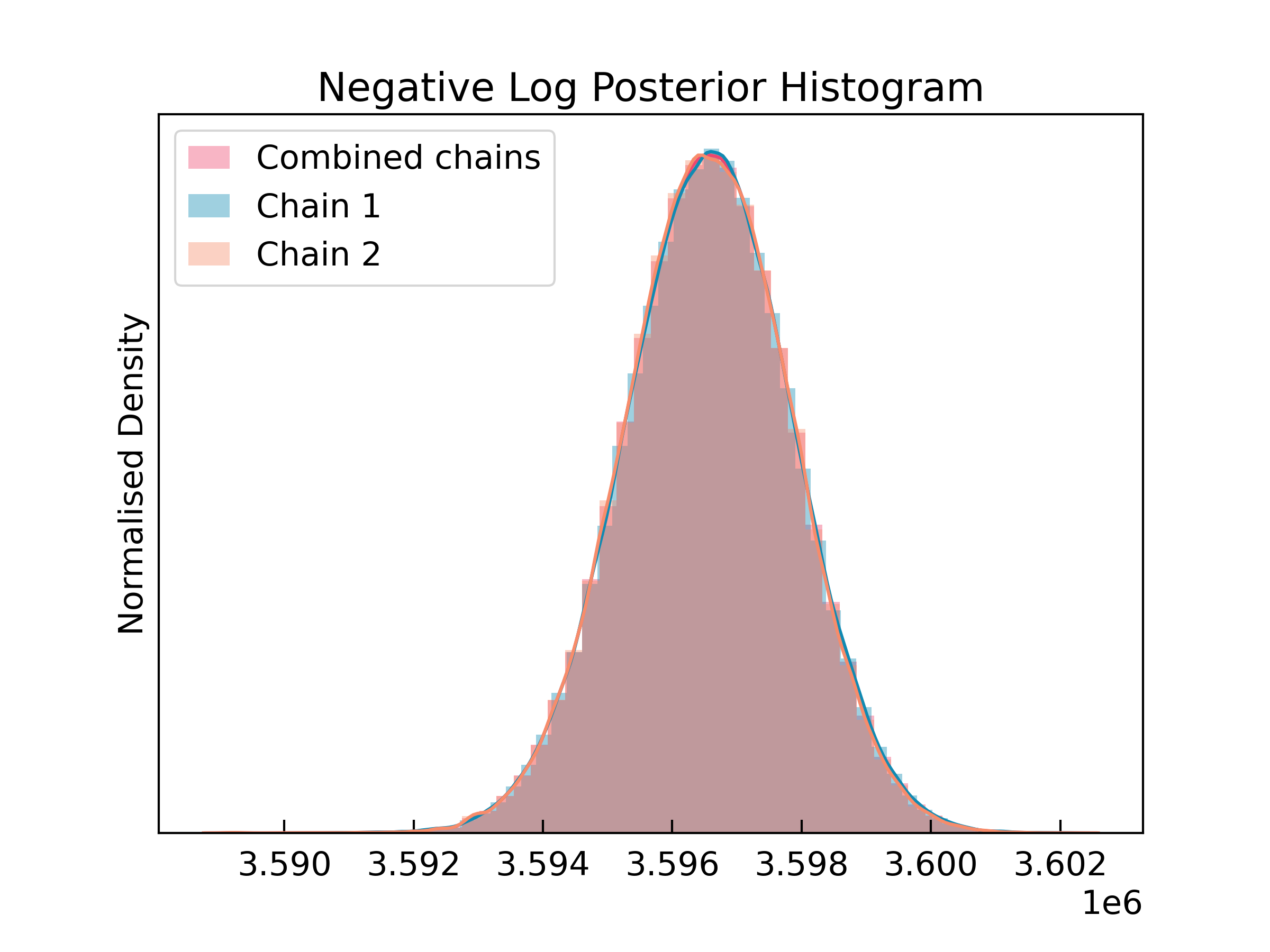} }}%
    \caption{Histogram for the individual chains and combined chains for the analysis of \subf{(a)} CMB temperature and \subf{(b)} CMB polarization in Sect.~\ref{sec:results}.}%
    \label{Fig:PostHist_CMB_Tests}%
    \vspace{8pt}
\end{figure}

\begin{figure*}%
    \centering
    \subfloat[\vspace{0.2cm}\centering Spin Weight 0]{{\includegraphics[width=.40\textwidth]{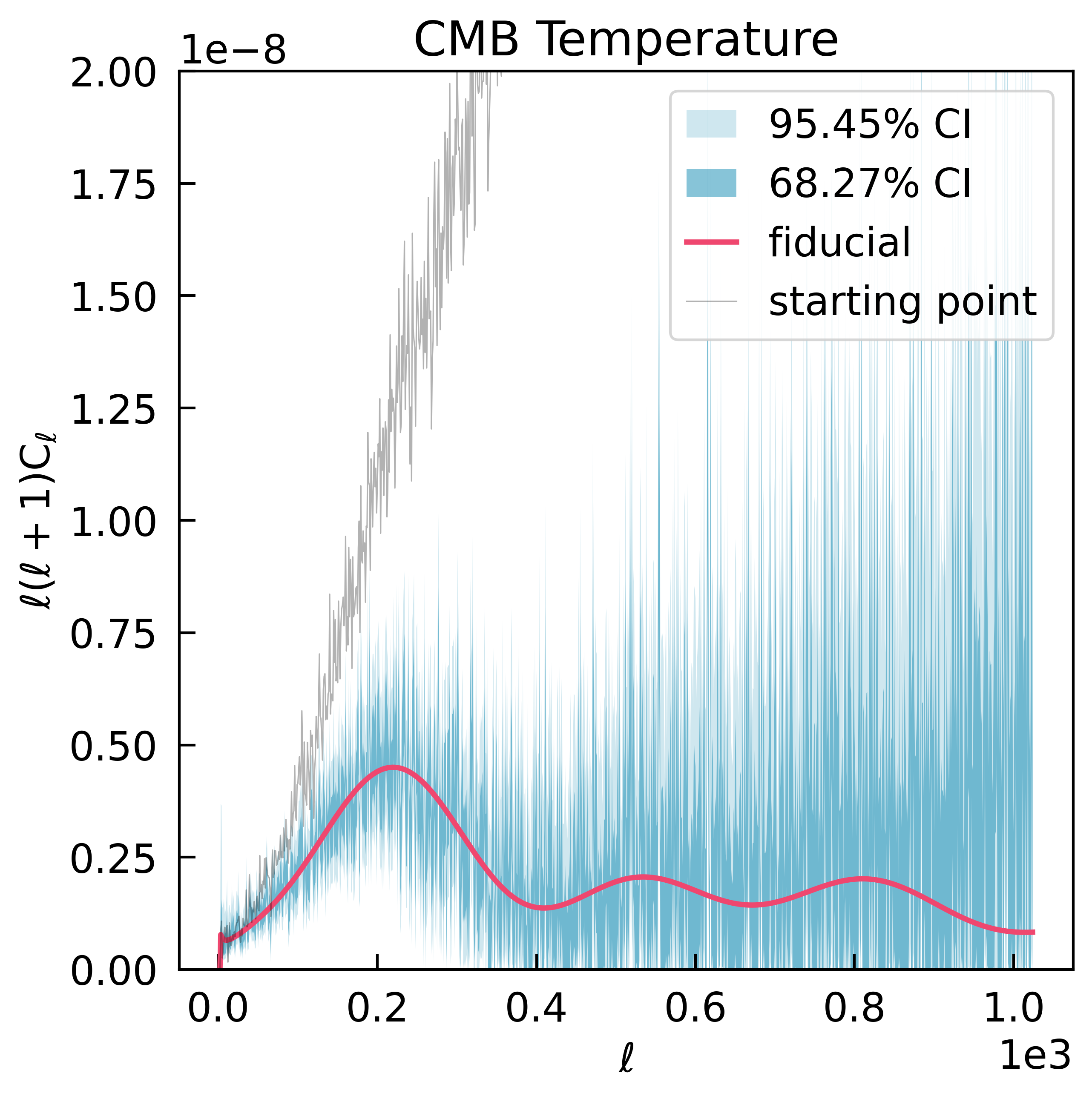} } \label{Fig:Results_Cls_T}}%
    \qquad
    \subfloat[\centering Spin Weight 2]{{\includegraphics[width=0.75\textwidth]{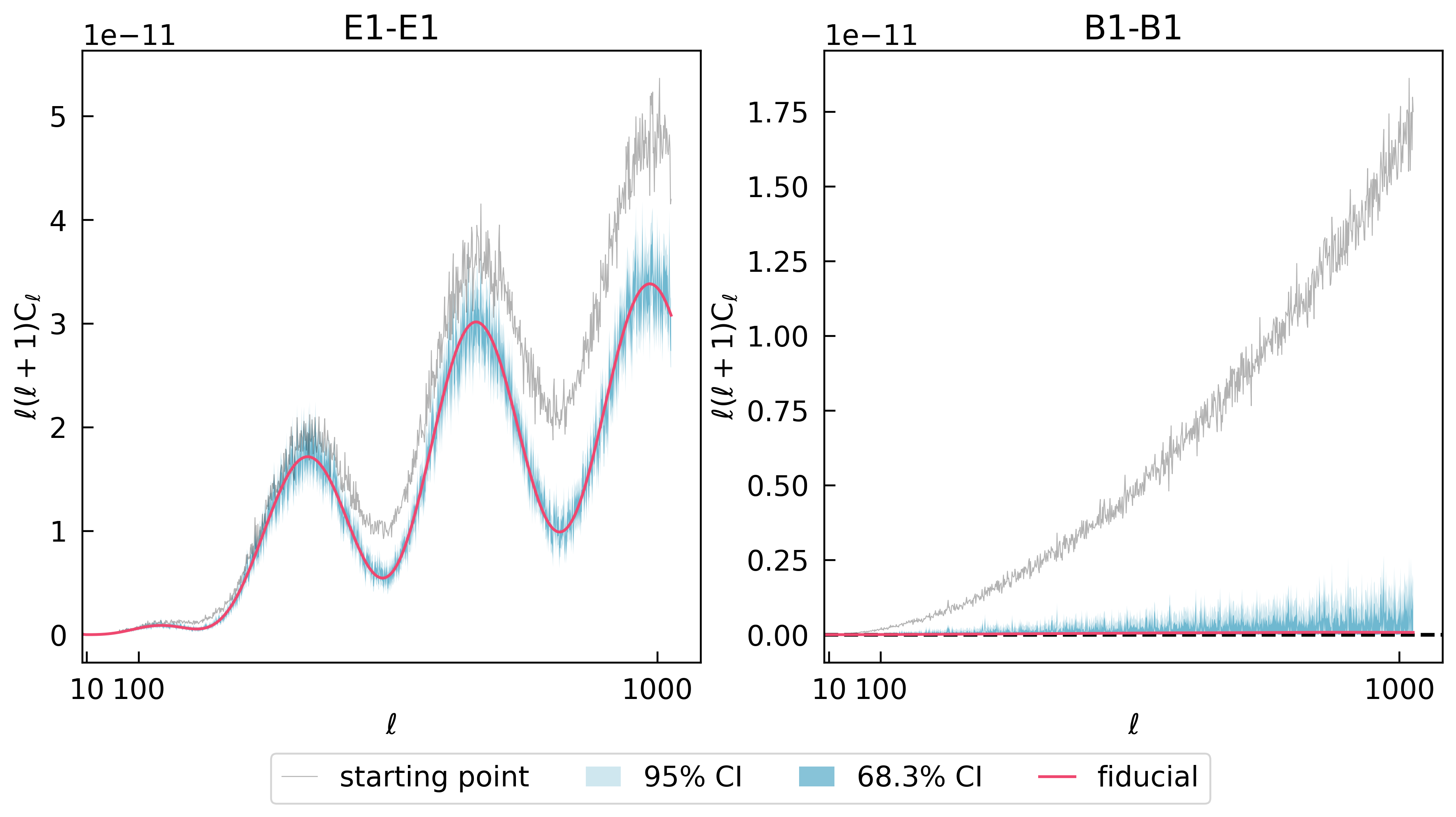}}\label{Fig:Results_Cls_EB}}%
    \caption{The marginalised one dimensional constraints on the inferred \subf{(a)} spin-weight 0 CMB temperature and \subf{(b)} spin-weight 2 CMB polarization multipoles. The dark (light) blue contours show the 68.27$\%$ (95.45$\%$) C.I. The starting point for one of the chains, calculated according to Sect.~\ref{SSec:StartingPoints}, is shown by the gray line. Due to the low signal-to-noise ratio in this case, the chain starts very far off from the typical region of this posterior. The red line represents the fiducial value, used to generate the simulation realisation analysed by \almanac{}.}%
    \label{Fig:Results_Cls}%
    \vspace{8pt}
\end{figure*}

\begin{figure*}%
    \centering
    \includegraphics[width=\textwidth]{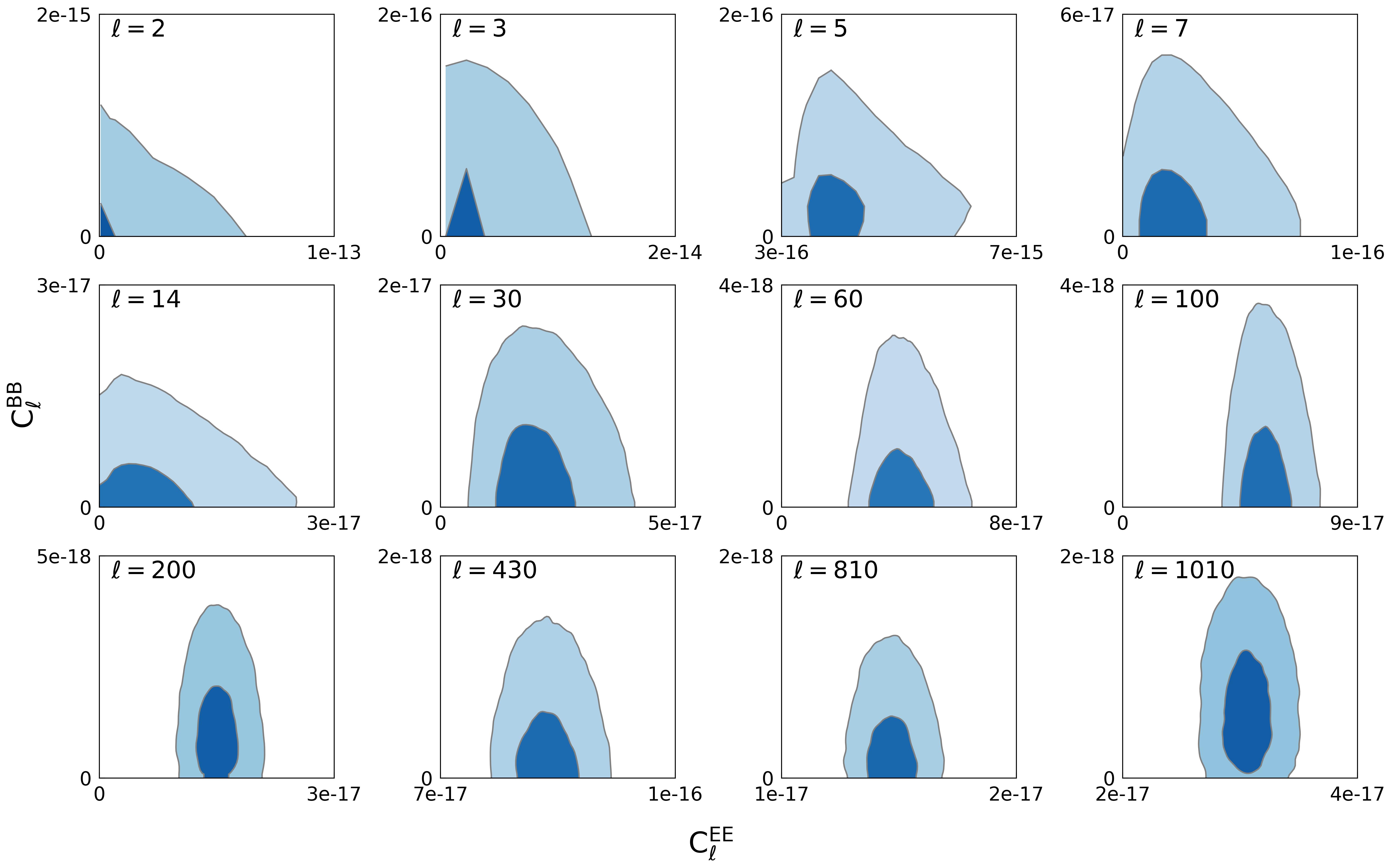}%
    \caption{Plots of marginal joint distributions of $E$ and $B$ power spectra for different multipoles showing how reducing the posteriors to point estimates could lead to leakage of $E$ to $B$. However, as also shown in Fig.~\ref{Fig:EBleakage1D}, no significant correlations between $E$ and $B$ modes were found for a WMAP-mask simulation. }%
    \label{Fig:EBleakage2D}%
\end{figure*}

\begin{figure}%
    \centering
    \includegraphics[width=0.45\textwidth]{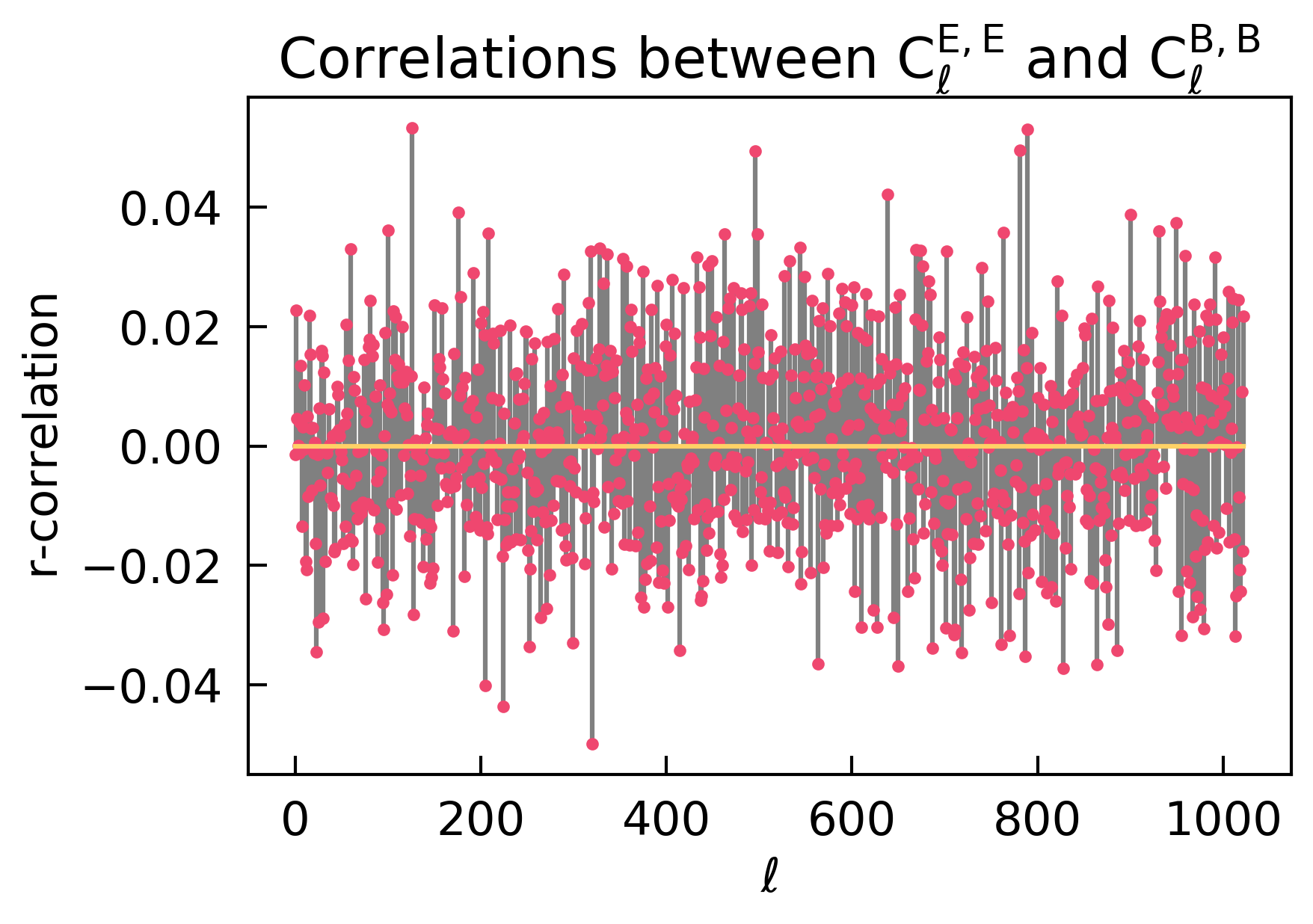}%
    \caption{Correlation between $E$ and $B$ modes for the same multipoles in the CMB polarization example. We note here that the WMAP mask with the isotropic noise used in our spin-weight 2 test case does not introduce any spurious correlations between modes. A few detailed two dimensional marginalised constraints are shown in Fig.~\ref{Fig:EBleakage2D}.}%
    \label{Fig:EBleakage1D}%
\end{figure}

In this section, we present applications of \almanac{} to CMB simulated data for temperature (spin-weight 0) and, separately, polarization (spin-weight 2). We present the inferred angular power spectra for each case, marginalised individual multipoles, and a set of recovered maps from the posterior distributions of the inferred fields.

For CMB temperature only or polarization only, the covariance matrix of the true map coefficients is diagonal and the reparametrization essentially scales ${\bf a}_{\ell m} \rightarrow {\bf x}_{\ell m} = {\bf a}_{\ell m}/\sqrt{C_\ell}$.  When temperature and polarization are both included, the covariance matrix is no longer diagonal; instead it is block diagonal with small blocks, and either the $\ln{\sfC}$ parametrization or the Cholesky decomposition parametrization can be used. The latter involves more complex derivatives, and we leave this to the more extensive investigation of correlated multiple fields (with a focus on tomographic cosmic shear) in the companion paper \citep{2022AlmanacWL}{}.

For the studies we present here, we simulate CMB maps with $n_{\rm side}=512$ and $\ell_{\rm max}=1024$, with a noise level of $\qty{2}{\micro \kelvin}$ per pixel for a temperature-only experiment, and separately for a polarization experiment (with the same noise level for each of $Q,U$). This noise level corresponds to low signal-to-noise for the CMB temperature case and, for the CMB polarization case, a combination of high signal-to-noise for the $E$-modes and very low signal-to-noise for the $B$-modes. We generate the simulations using a fiducial power spectrum from \textsc{CAMB} \citep{CAMB} and produce a gaussian realisation of this signal. We apply a gaussian noise (with the noise level we previously specified) and a WMAP mask \citep{2009-WMAP-Cls}. This procedure is the same for both temperature and polarization simulations. For the polarization experiment, the $B$-mode power is close to zero in the dataset. It is allowed to vary in the inference; however, we force to zero the parity violating modes $C^{EB}_{\ell}$.

We run two independent chains for each case (spin-weight 0 and spin-weight 2) using dispersed starting points as outlined in Sect.~\ref{SSec:StartingPoints}. The tuning phase (as outlined in Sect.~\ref{sec:HMCtuning}) used $10 000$ samples for burn-in, $5 000$ samples for step-size tuning, and $5 000$ samples for leap-frog tuning. Following tuning, we generate $2\times 10^6$ samples for each chain for the low signal-to-noise CMB temperature case, and $4.5\times10^5$ samples for each chain for the CMB polarization case. Upon confirmation that the chains are converged (see Sect.~\ref{sec:paramresults}), we merge them to obtain the results exhibited in this section. Fig.~\ref{Fig:PostTracer_CMB_Tests} depicts the trace of the log posterior for the combined chains, while Fig.~\ref{Fig:PostHist_CMB_Tests} shows the histogram for this quantity for the individual and joint chains. This rapid diagnostic indicates that both chains, starting at different starting points, converge to the same underlying target distribution.

Following the same procedure outlined in~\cite{2022AlmanacWL}, we extract the 68.3$\%$ and 95.5$\%$ credible intervals (C.I.) from the resulting concatenated chains. We exhibit the results in Fig.~\ref{Fig:Results_Cls_T}, for the spin-weight 0 CMB temperature, and in Fig.~\ref{Fig:Results_Cls_EB}, for the spin-weight 2 CMB polarization (with $C_{\ell}^{EB}$ fixed to zero).
 
The spin-weight 0 case in Fig.~\ref{Fig:Results_Cls_T} has a low signal-to-noise ratio, leading to a higher uncertainty on the small scales. This low signal-to-noise ratio is apparent when looking at the starting point in Fig.~\ref{Fig:Results_Cls_T}; this is a dispersed pseudo-angular power spectrum containing not only the partial-sky signal but also noise (see Sect.~\ref{SSec:StartingPoints} for details). Nevertheless, \almanac{} can recover the fiducial angular power spectra within its 1- and 2-$\sigma$ C.I.
    
By comparison, the spin-weight 2 case in Fig.~\ref{Fig:Results_Cls_EB} exhibits a combination of high and low signal-to-noise. The high signal-to-noise $E$-modes are accurately inferred from the chains, while the low signal-to-noise $B$-modes are consistent with zero. This example demonstrates \almanac{}'s ability handle a mixture of signal-to-noise ratios.

In the spin-weight 2 case we consider samples of $C^{EE}_{\ell}$ and $C^{BB}_{\ell}$; these samples exhibit covariance between those quantities at a given $\ell$, due to anisotropic noise and/or the geometry of the mask. This is fully described by the posterior distribution in our Bayesian setting, but can translate into so-called `$E-B$ leakage' between point estimators of the spectrum at that $\ell$.
By looking at marginalised two-dimensional contours, as pictured in Fig.~\ref{Fig:EBleakage2D}, one can better understand the relationship between the $EE$ and $BB$ spectra for a given multipole $\ell$. The shape of the contours for $\ell\la20$ indicates that the total power is better constrained than either individual spectrum; samples with excursions to high power in one quantity can only occur when the other quantity exhibits low power. However, because the posterior is cut off at $C_\ell=0$, the actual correlation coefficients between  $C^{EE}_{\ell}$ and $C^{BB}_{\ell}$ are low (mostly less than $0.02$ in absolute value) even at low $\ell$, as shown in Fig.~\ref{Fig:EBleakage1D} over a range of multipoles.
As anticipated, the $B$-mode is generally consistent with zero; although, as highlighted above and in \citealt{2022AlmanacWL}, our log-prior excludes zero from the parameter space. 

\begin{figure*}%
    \centering
    \includegraphics[width=0.8\textwidth]{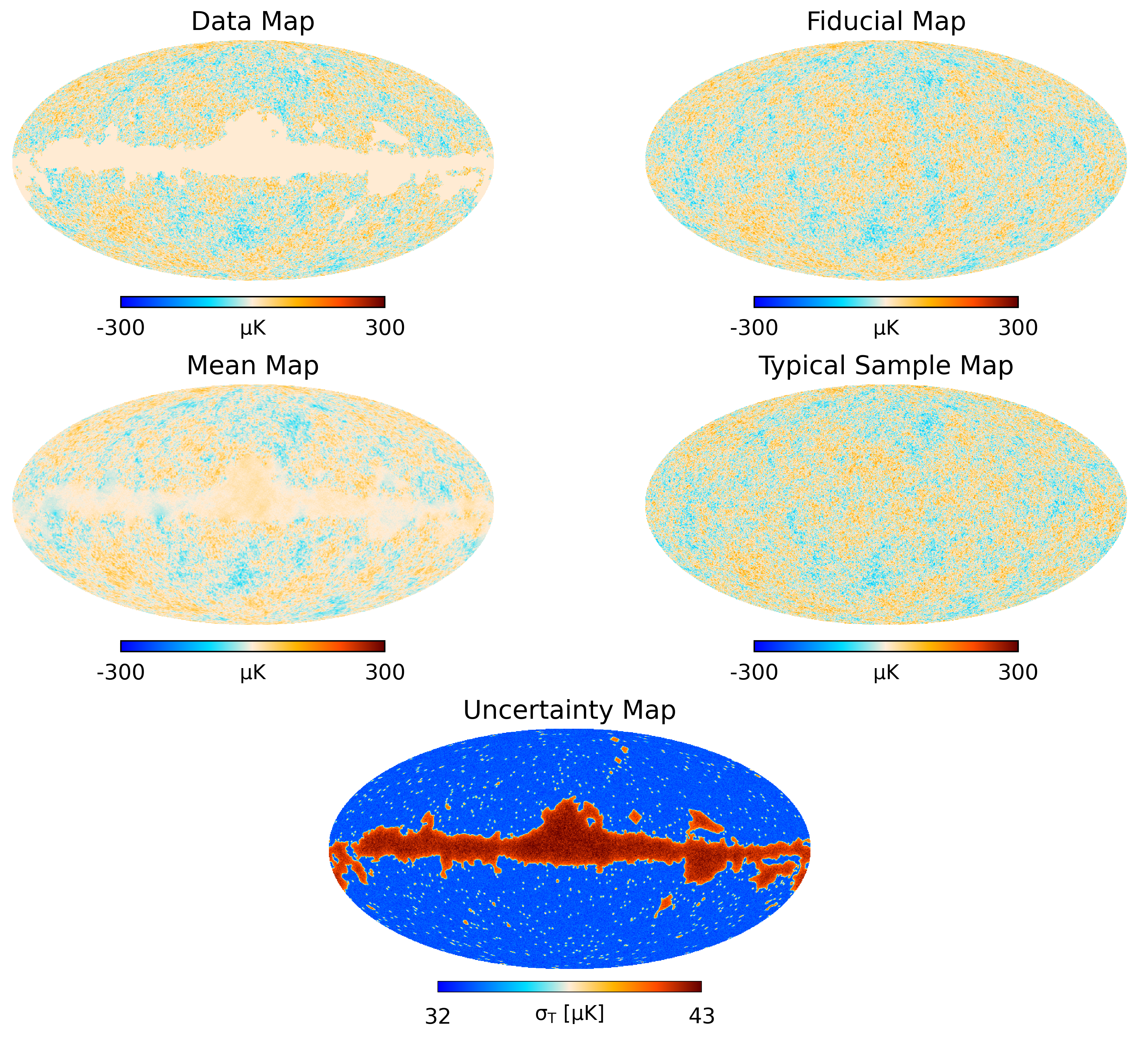}%
    \caption{Mean, Standard Deviation, and Typical Sample maps from the \almanac{} posterior compared to the input data, with a WMAP mask, and the fiducial noise-less and full sky map for a CMB temperature case (spin-weight 0). All maps have a resolution of $N_{\rm side}=512$. For more details, see the text.}%
    \label{Fig:Results_T}%
\end{figure*}

\begin{figure*}%
    \centering
    \includegraphics[width=0.8\textwidth]{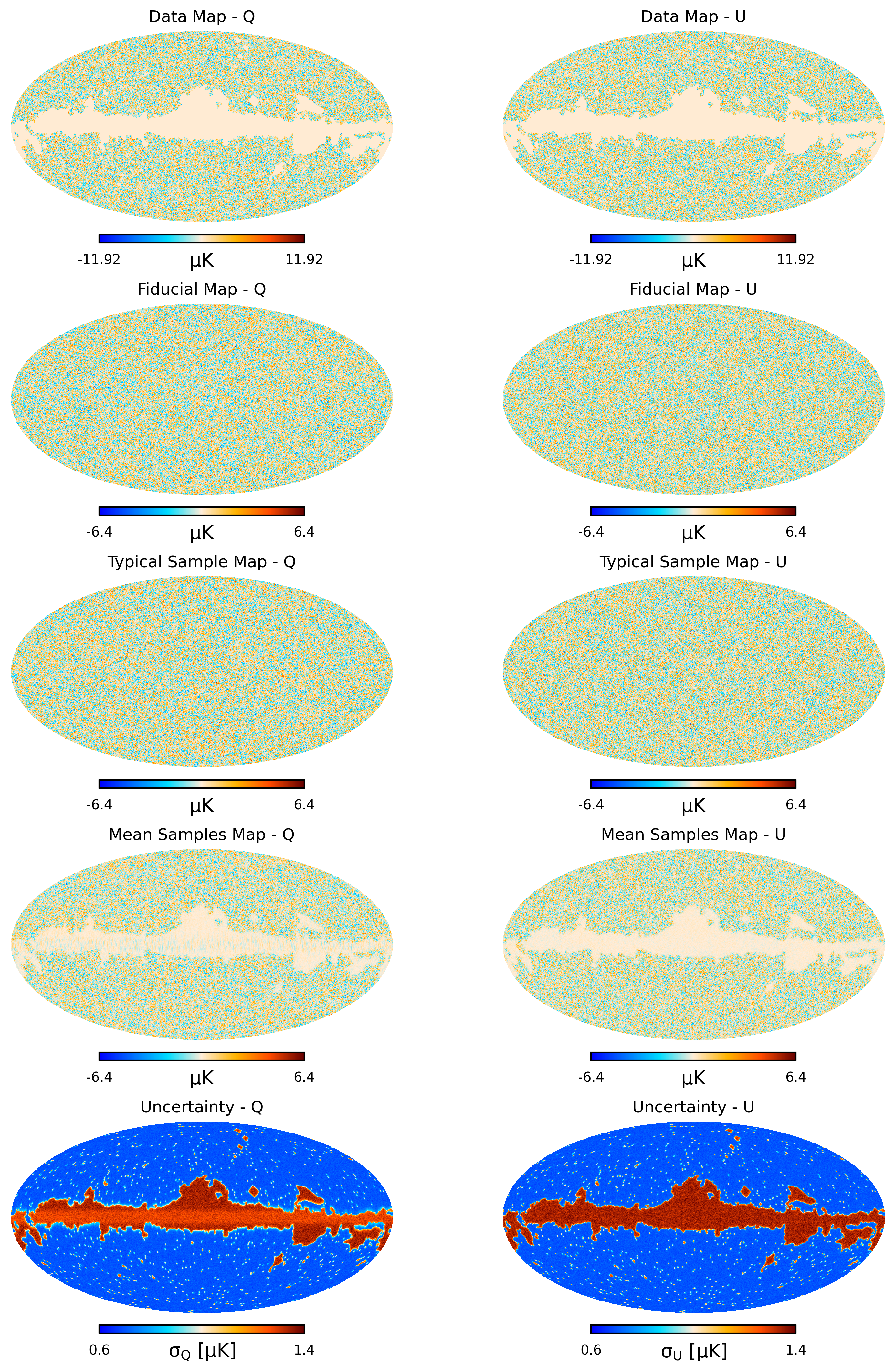}%
    \caption{Similar to Fig.~\ref{Fig:Results_T} but for a CMB polarization case (spin-weight 2) showing the $Q$, $U$ (or $E$, $B$) component maps.}%
    \label{Fig:Results_EB}%
\end{figure*}

Finally, from the $\fata$ samples, we can (using spherical harmonic transformations) reconstruct a distribution of maps. Figs.~\ref{Fig:Results_T} and~\ref{Fig:Results_EB} show the input data, the fiducial realisation (noiseless and full sky), the mean posterior maps and associated variances, as well as a random typical realisation map from the \almanac{} posterior. The mean maps show some smoothing, similar to the Wiener filter smoothing expected for the mean posterior of a gaussian field of a given power spectrum and gaussian noise. Here the power spectrum is not fixed but sampled, but this smoothing feature is still expected when taking the average of the unobserved pixels. We also note that, as expected, the variance is much larger in regions where data is missing. For example, the typical sample in Fig.~\ref{Fig:Results_T} displays a structure indistinguishable from the fiducial simulation. Naturally, small-scale information in the masked regions will differ as expected since these are primarily random draws from the prior.

%%%%%%%%%%%%%%%%%%%%%%%%%%%%%%%%%%%%%%%%%%%%%%%%%%%%%%%%%%%%%%%%%%%%%%%%
%                       CONVERGENCE DIAGNOSTICS
%%%%%%%%%%%%%%%%%%%%%%%%%%%%%%%%%%%%%%%%%%%%%%%%%%%%%%%%%%%%%%%%%%%%%%%%
\section{Monitoring convergence in high dimensions}
\label{sec:paramresults}

\begin{figure}%
    \centering
    \subfloat[\centering CMB Temperature]{{\includegraphics[width=.45\textwidth]{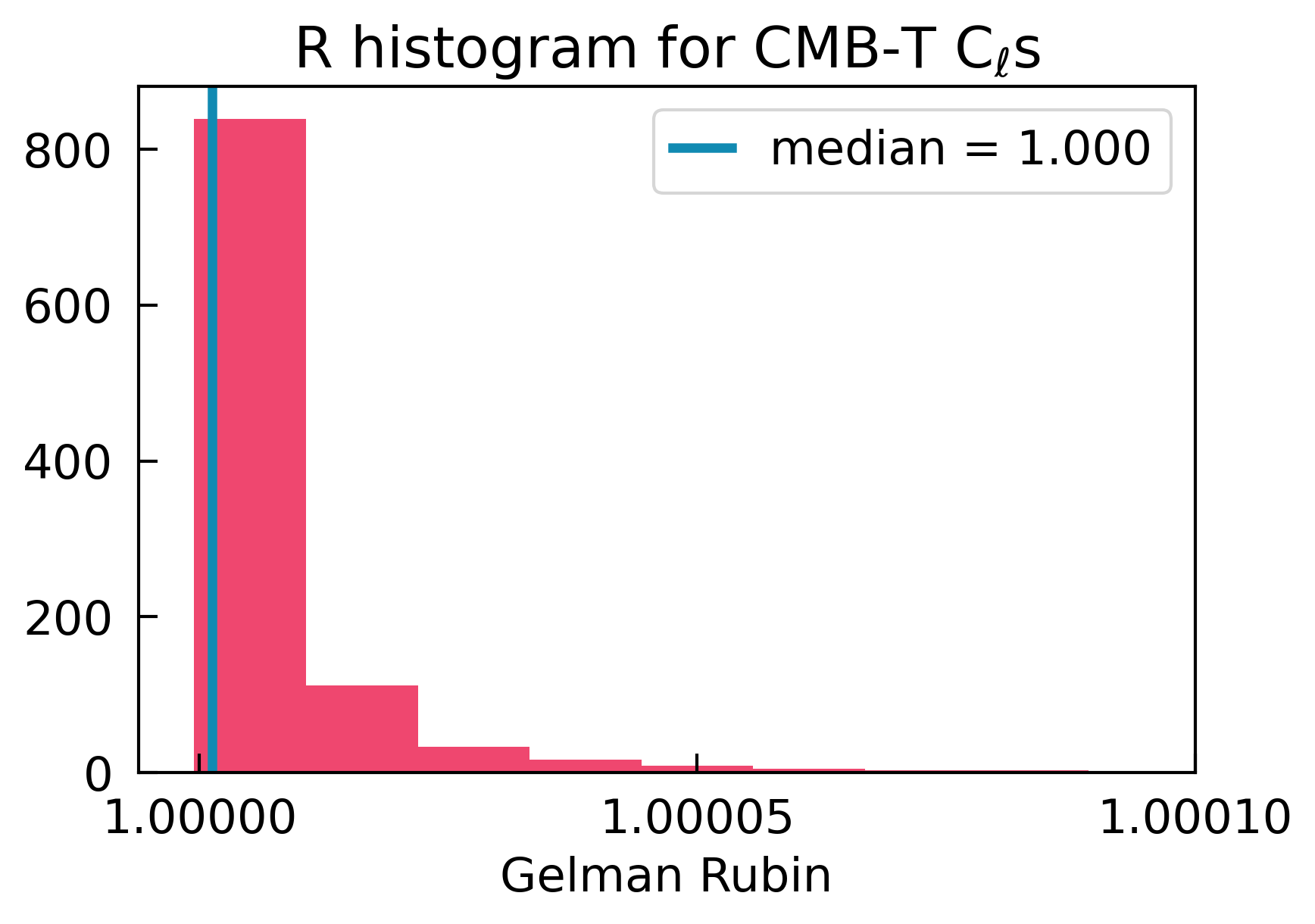} }}%
    \qquad
    \subfloat[\centering CMB Polarization]{{\includegraphics[width=.45\textwidth]{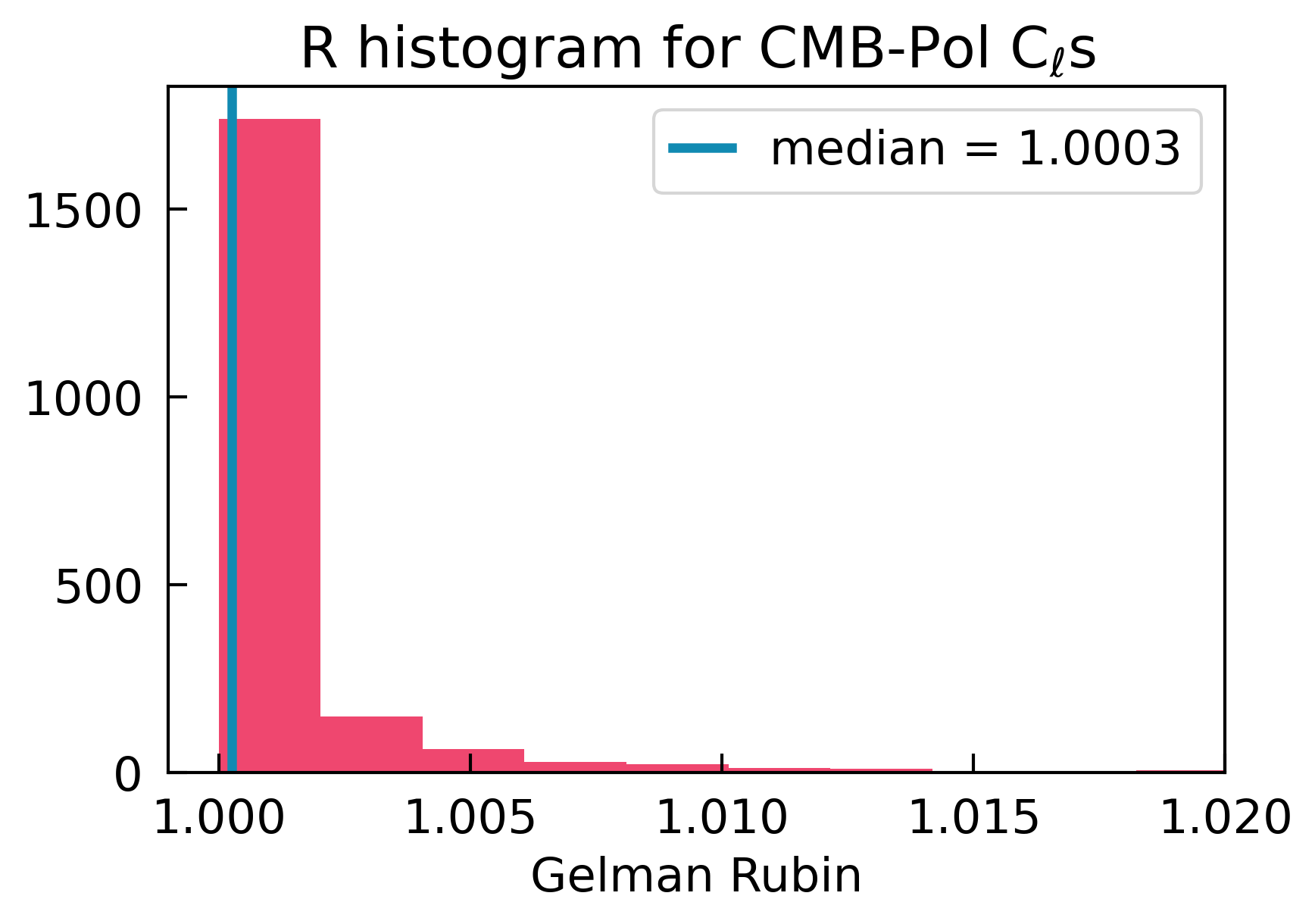} }}%
    \caption{Histograms of the Gelman-Rubin parameter, $R$, for the \subf{(a)} spin-weight 0 CMB temperature and \subf{(b)} spin-weight 2 CMB polarization angular power spectra presented in Fig.~\ref{Fig:Results_Cls}. }%
    \label{Fig:GelRub_CMB_Tests}%
    \vspace{8pt}
\end{figure}

Achieving and monitoring chain convergence is particularly difficult in high dimensions; this section describes the statistics that we examined to monitor convergence and the overall performance of our algorithm.

We use four convergence and performance diagnostics: the Gelman-Rubin test, the Fraction of Missing Information\footnote{This statistic has also been called the Bayesian Fraction of Missing Information (BFMI); however, the statistic is not specifically Bayesian and hence the shorter name is preferred.} (FMI), Hanson's statistic, and the effective sample size; each test has a different goal. The Gelman-Rubin test \citep{GelmanRubin} works with the one-dimensional marginal distributions and monitors whether several chains are converged with respect to each other; the FMI test monitors whether a chain is converged with respect to the underlying target distribution; Hanson's statistic works with the one-dimensional marginal distributions and monitors whether each dimension is converged with respect to the underlying target marginal distribution; the effective sample size (closely related to the correlation length) approximates the number of independent samples that are available to determine the statistical properties of a given parameter.

In practice, we found the FMI to be a valuable convergence diagnostic, in particular because it can be evaluated `on the fly', i.e., while a chain is running: a badly performing chain (due, e.g., to poorly chosen hyperparameters) can be spotted and stopped early.

Several chains that are individually and jointly converged are combined into one large set of posterior samples from which our research results are then derived. The well-known Gelman-Rubin test, shown in Fig.~\ref{Fig:GelRub_CMB_Tests}, is the ratio of a) the variance within a single chain and b) the variance between chains; it should be close to one (ideally within 5--10\%) for a converged chain. 

The following gives details of the other diagnostics, which are generally less well-known.

%%%%%%%%%%%%%%%%%%%%%%%%%%%%%%%%%%%%%%%%%%%%%%%%%%%%%%%%%%%%%%%%%%%%%%%%
%                       FRACTION OF MISSING INFO
%%%%%%%%%%%%%%%%%%%%%%%%%%%%%%%%%%%%%%%%%%%%%%%%%%%%%%%%%%%%%%%%%%%%%%%%
\subsection{Fraction of Missing Information}
\begin{figure*}%
    \centering
    \subfloat[\centering $\{\fata, \ln(\sfC_{\ell})\}$]{{\includegraphics[width=.45\textwidth]{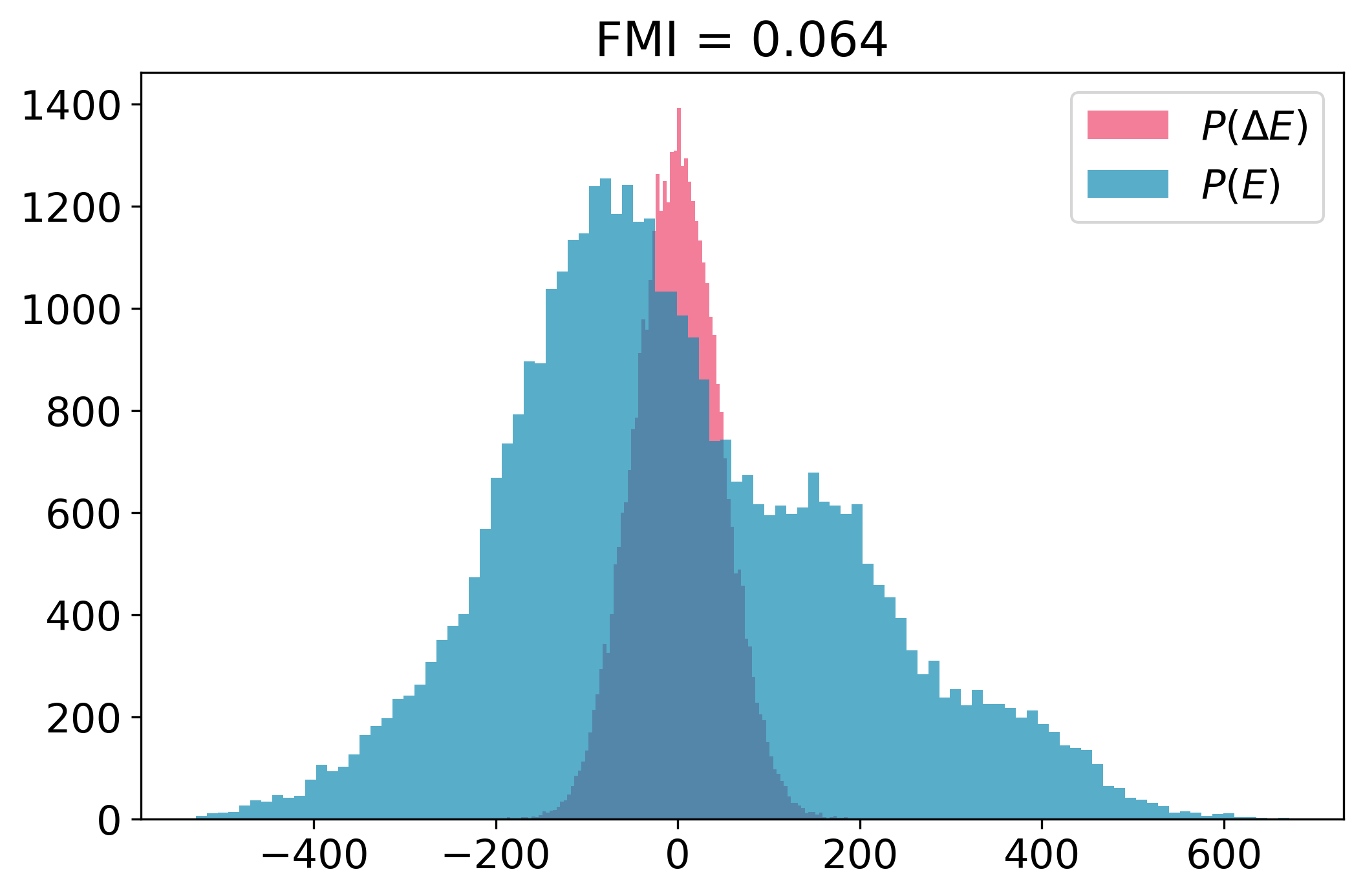} } \label{Fig:FMI_LogC}}%
    \qquad
    \subfloat[\centering $\{\fatx, \sfK \}$]{{\includegraphics[width=.45\textwidth]{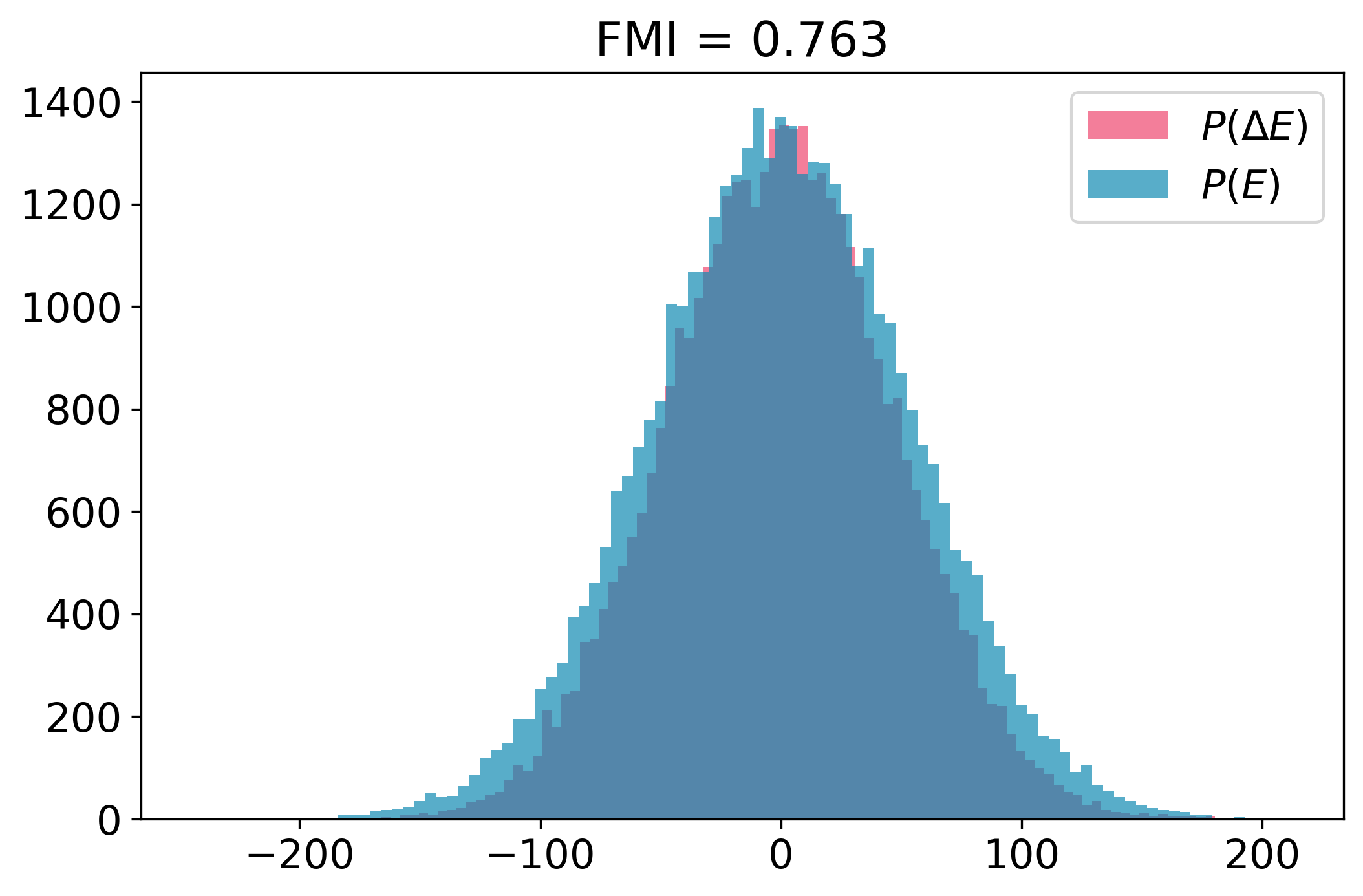} }\label{Fig:FMI_Chol}}%
    \caption{Fraction of Missing Information (FMI) for sampling the \almanac{} posterior using different parametrizations. Blue is the distribution of (mean-subtracted) energies, while pink is the distribution of energy transitions (as caused by momentum resampling). If the two distributions have similar variances then the sampler will easily reach all energy levels of phase space, and hence will not avoid the tails or other posterior regions that are difficult to access. Plot \subf{(a)} depicts sampling in the $\{\fata, \ln(\sfC)\}$ parameterisation; here the energy transition distribution (pink) is too narrow, implying the sampler struggles to reach the posterior tails. Plot \subf{(b)} depicts sampling in the $\{ \fatx, \sf{K} \}$-parameterisation; here the energy transition distribution matches the posterior's marginal energy distribution well, and the sampler has reached most posterior regions. The FMI quantifies the comparison of histograms in a single number given by Eq.~\eqref{eq:FMI}. This figure analyzes one spin-weight 2 field up to $\ell_{\rm max} = 64$. }%
    \label{Fig:FMI_Compare}%
    \vspace{8pt}
\end{figure*}
\begin{figure*}%
    \centering
    \subfloat[\centering CMB Temperature]{{\includegraphics[width=.45\textwidth]{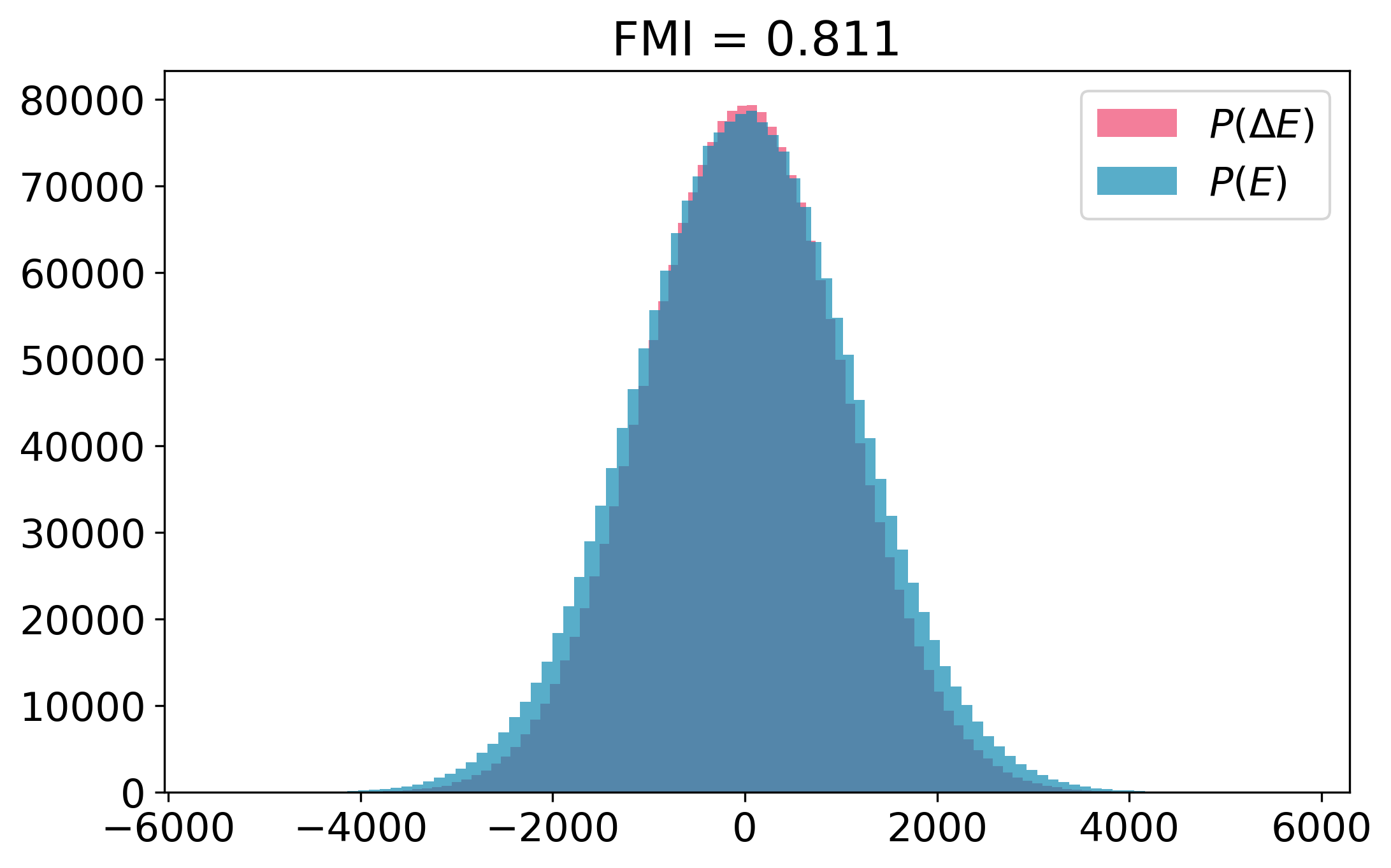} }}%
    \qquad
    \subfloat[\centering CMB Polarization]{{\includegraphics[width=.45\textwidth]{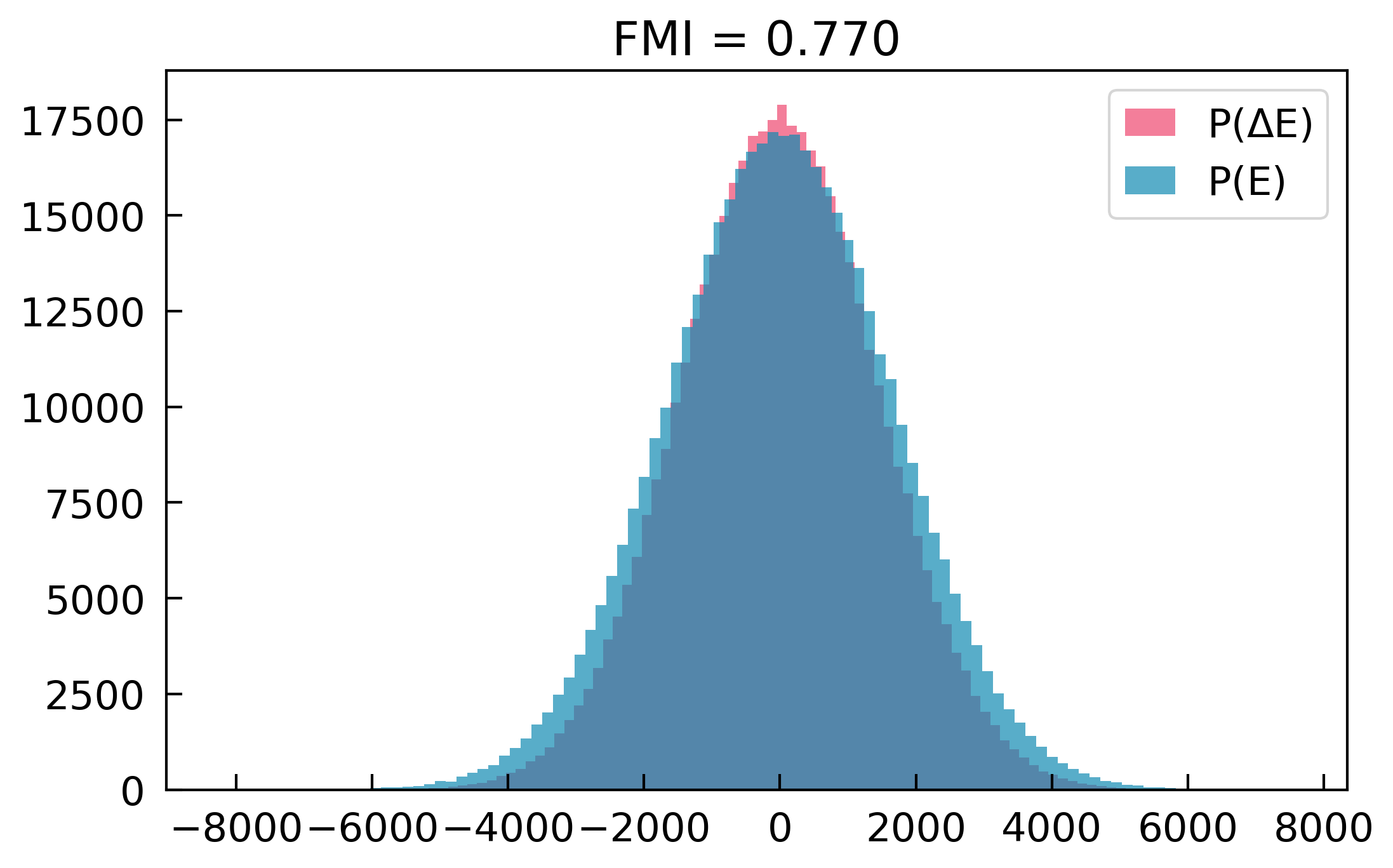} }}%
    \caption{Fraction of Missing Information (FMI) diagnostics for the\subf{(a)} spin-weight 0 CMB temperature case and \subf{(b)} spin-weight 2 CMB polarization case described in Sect.~\ref{sec:results}. We display only one of the chains (the other chain is similar).}%
    \label{Fig:FMI_CMB_Tests}%
    \vspace{8pt}
\end{figure*}

Let $\faty$ be the position variables of the sampler (i.e., the parameters to be inferred) and let $E=H(\faty,\fatp)$ [Eq.~\eqref{eq:Hamiltonian}] be the energy of the sampler. The Fraction of Missing Information (FMI) is defined to be
\begin{equation}
    \mathrm{FMI} = \frac{\mathbb{E} \left[ \Var (E|{\faty})\right] }{\Var(E) } \, ,
\end{equation}
where $\mathbb{E}$ and $\Var$ are the expectation value and variance using the posterior probabilities. The numerator is hence the expected variance of the energy at a \emph{fixed} value of the parameters, $\faty$, then averaged over all parameters; the denominator is the unconstrained variance of the energy. A larger ratio indicates an ability to explore the full distribution as we vary the kinetic energy (see \cite{2016-Betancourt-FMI-1,Betancourt} for further discussion).

If $N$ is the number of samples in the chain, $k$ the index of a sample in the chain (written as a superscript), and $\bar{E}$ the mean energy, then the FMI can be estimated from the chain via \citep{2016-Betancourt-FMI-1,Betancourt}
\begin{equation}
    \widehat{\mathrm{FMI}} =  \frac{ \sum_{k=1}^N (E^k -E^{k-1})^2}{\sum_{k=0}^N (E^k - \bar{E})^2 }.
    \label{eq:FMI}
\end{equation}
We see that the estimated FMI is the ratio of two variances: the numerator is the variance of the energy difference between adjacent samples while the denominator is the variance of the energy across the entire chain. If this fraction is close to zero, then each time that new momenta are drawn, the sampler has only a small chance of jumping to an energy level that is independent of the previous energy level, and therefore will not quickly explore all the energy levels of the Hamiltonian; instead, it will stay stuck on an energy level and will have difficulty transiting from the tails to the core of the distribution. Hence FMI not near zero is desired; FMI levels below which problems have been reported include $0.3$ \citep{Betancourt} and $0.7$ \citep{2022AlmanacWL}{}. Large correlation lengths in the chain are expected when the FMI is too low. 

We can visualize the FMI calculation by comparing histograms of the quantities whose variances form its numerator and denominator. At step $k$, the energy level of the posterior is
\begin{equation}
    E = E^k - \bar{E}
\end{equation}
(the removal of the mean energy is optional), while the transition between energy levels is
\begin{equation}
    \Delta E = E^k - E^{k-1}.
\end{equation}
By plotting simultaneously the histograms of $E$ and $\Delta E$ we can judge whether the sampler transits quickly between energy levels. If the distribution of $\Delta E$ is much narrower than that of $E$, then FMI will be close to zero and the sampler will require many iterations to transit between the typical region and the tails; if the two histograms are similar, then the FMI will be significantly non-zero and this problem is not expected to arise.

To illustrate this behaviour, Fig.~\ref{Fig:FMI_Compare} compares the energy histograms for the two coordinate systems discussed in Sect.~\ref{sub:params}. Fig.~\ref{Fig:FMI_LogC} uses the na\"ive parameterisation $\{\fata, \ln(\sfC)\}$; here the two histograms are quite different (FMI $ = 0.064$, very low). This parameterisation is highly sub-optimal when there are non-diagonal correlations in the $\sfC_{\ell}$ matrix (see Sect.~\ref{sub:params}). Meanwhile, for the same data but using the Cholesky coordinate system $\{\fatx, \sfK \}$, Fig.~\ref{Fig:FMI_Chol} shows a much better agreement between the two histograms (FMI $ = 0.763$, acceptable), indicating that this is a more suitable coordinate system for this case. Other MCMC diagnostics can reach a similar conclusion, but our experience is that FMI highlights problems in the sampling process much earlier than other diagnostics. For example, again comparing the na\"ive parameterisation to the Cholesky parameterisation (for a lower dimensional $\ell_{\textrm{max}} = 32$ spin-weight 2 example), we found FMI values after 1,000, 10,000, and 100,000 samples of 0.290, 0.051, and 0.067 respectively for the naive parameterisation versus 0.772, 0.807 and 0.782 for the Cholesky parameterisation. Here we see that the FMI is already after only 1,000 samples able to distinguish between a chain that converged slowly (FMI $ < 0.3$ after 1,000 samples) and a chain that converged well (FMI $ > 0.7$ after 1,000 samples).

Fig.~\ref{Fig:FMI_CMB_Tests} shows the energy histograms for the results presented in Sect.~\ref{sec:results} (for one of the chains; the other chain is similar). We find FMI $\approx 0.80$ for the spin-weight 0 CMB temperature case and FMI $\approx 0.77$ for the spin-weight 2 CMB polarization case. Both these results are considered acceptable. Our experience is that bad convergence is always associated with FMI $< 0.3$ and good convergence usually has FMI $> 0.7$. Even with good convergence, however, FMI seldom reaches values greater than $0.85$.

%%%%%%%%%%%%%%%%%%%%%%%%%%%%%%%%%%%%%%%%%%%%%%%%%%%%%%%%%%%%%%%%%%%%%%%%
%                        HANSON STATISTICS
%%%%%%%%%%%%%%%%%%%%%%%%%%%%%%%%%%%%%%%%%%%%%%%%%%%%%%%%%%%%%%%%%%%%%%%%
\subsection{Hanson's Statistic}
\begin{figure}%
    \centering
    \subfloat[\centering CMB Temperature]{{\includegraphics[width=.45\textwidth]{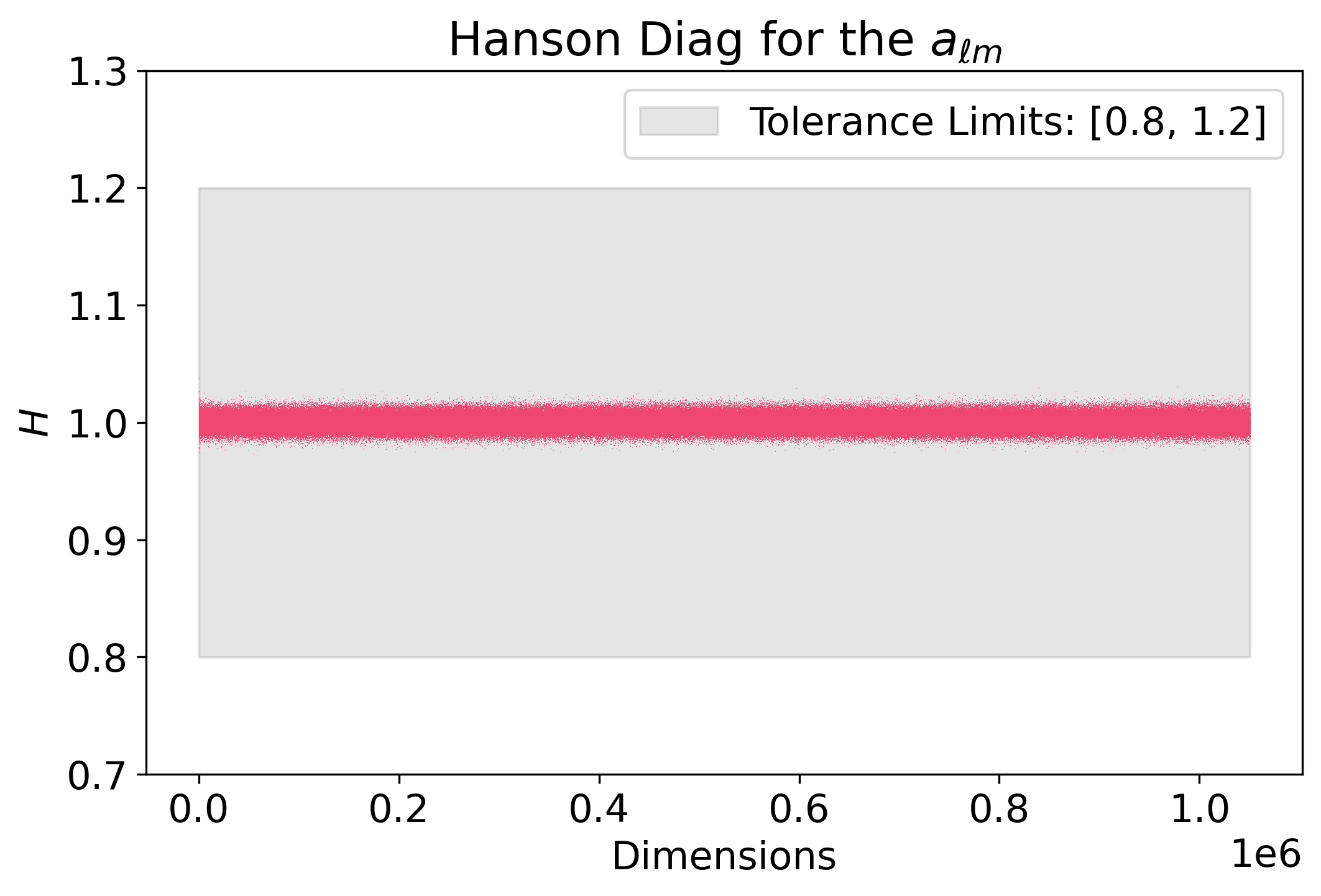} }}%
    \qquad
    \subfloat[\centering CMB Polarization]{{\includegraphics[width=.45\textwidth]{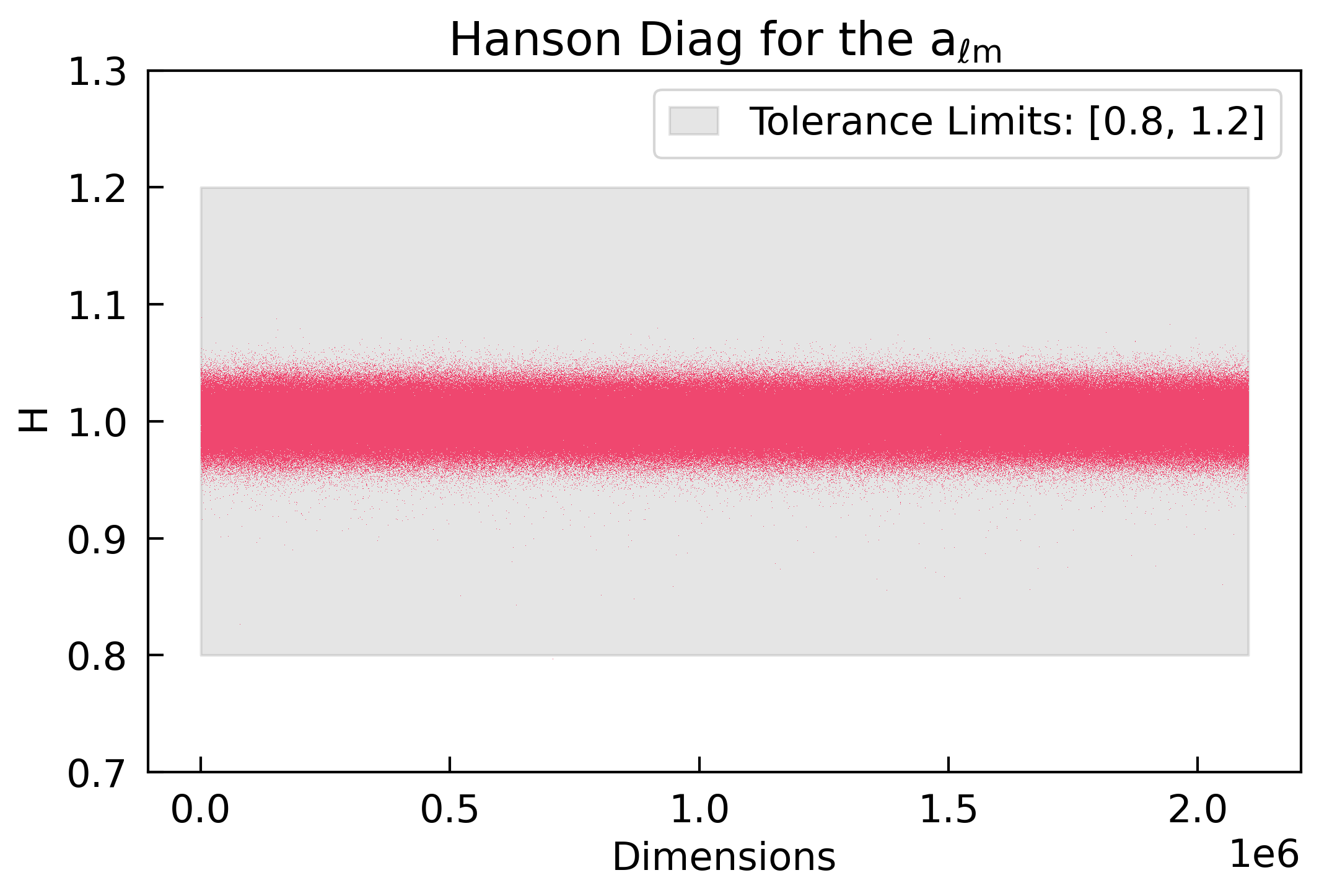} }}%
    \caption{Hanson diagnostics for the field dimensions, $\fata$, in the \almanac{} BHM chains from Sect.~\ref{sec:results}, demonstrating the convergence for the maps presented in Figs.~\ref{Fig:Results_T} and~\ref{Fig:Results_EB}. We display only one of the chains (the other chain is similar).}%
    \label{Fig:Hanson_CMB_Tests}%
    \vspace{8pt}
\end{figure}

Sampling in multiple millions of dimensions benefits from convergence diagnostics that can be computed immediately after each sample has been taken, and that resolve the convergence per component. An analysis of the convergence per component is important as components might converge at different rates. \citet{2001-Hanson-HMC} proposes such a component-wise convergence diagnostic by comparing two expressions for the sample-estimated variance.

Recalling the definition
\begin{equation}
    \nlp{}(\faty) = -\ln \calP(\faty)\, ,
\end{equation}
\citet{2001-Hanson-HMC} points out that the expression for the variance per dimension can be integrated by parts, yielding

\begin{equation}
\begin{aligned}
    \textrm{Var}(y_i) & \equiv \int (y_i - \bar{y}_i)^2\calP(\faty) \mathd \faty \\
    & = \frac{1}{3} \int (y_i - \bar{y}_i)^3 \calP(\faty) \frac{\partial}{\partial y_i} \nlp{}(\faty) \mathd \faty \\
    & \quad + \frac{1}{3} \int (y_i - \bar{y}_i)^3 \calP(\faty)\big\vert_{y_i=-\infty}^{y_i=+\infty} \mathd \faty_{\textrm{exclude }i} \, .
    \end{aligned}
\end{equation}
The first term in the final expression includes the derivative of the negative of the logarithm of the posterior density, a quantity already computed as it is required by the HMC sampler. The second term (the \textit{boundary} term) in the final expression includes an integration over all components of $\faty$ except $y_i$. \citet{2001-Hanson-HMC} points out that for continuous distributions with exponential tails (and more generally for distributions whose tails decay faster than $|y|^{-3}$), the boundary term can be neglected. We obtain a convergence diagnostic by replacing the above integrals with Monte Carlo estimators based on the chain, and then computing the ratio
\begin{equation}
    {\cal H}(y_i) = \frac{\sum_k (y_i^k - \bar{y}_i)^3 \frac{\partial}{\partial y_i} \nlp{}(\faty)\big\vert_{\faty^k}} {3 \sum_k(y_i^k - \bar{y}_i^{k})^2}
\end{equation}
of these two estimators for the variance (recall that a superscript index denote sample number). We expect that well-converged chains sampling continuous distributions with steeply decreasing tails should display values of ${\cal H}(y_i) \approx 1$ for all $i$; \cite{Taylor} suggests a tolerance range from $0.8$ to $1.2$.

For gaussian (exponential) tails, the derivative $ \partial y_i/ \partial \nlp{}(\faty)$ is proportional to $y_i - \bar{y}_i$. The numerator will hence dominate if a sample falls far away from the mean, yielding ${\cal H} > 1$ for those outer samples. In consequence, at least for exponential tails, values of $ {\cal H}(y_i) < 1$ indicate a bad sampling of the tails. In summary, while ${\cal H}(y_i)\approx 1$ indicates good convergence, a deviation of ${\cal H}(y_i)$ can indicate bad convergence, the existence of fat tails, or the hitting of a prior boundary (which would lead to a discontinuity in the posterior).

This quantity is ideal for diagnosing convergence as it can be calculated easily for all the dimensions at each step in the chain, without requiring the post-processing of a huge file. Fig.~\ref{Fig:Hanson_CMB_Tests} displays the Hanson diagnostics of the spherical harmonic dimensions ($\fata$ or $\fatx$) for the maps shown in Figs.~\ref{Fig:Results_T} and~\ref{Fig:Results_EB} for one of the chains (the other chain displays similar results); we find excellent convergence according to this diagnostic.

%%%%%%%%%%%%%%%%%%%%%%%%%%%%%%%%%%%%%%%%%%%%%%%%%%%%%%%%%%%%%%%%%%%%%%%%
%                       CORRELATION LENGTH
%%%%%%%%%%%%%%%%%%%%%%%%%%%%%%%%%%%%%%%%%%%%%%%%%%%%%%%%%%%%%%%%%%%%%%%%
\subsection{Autocorrelation, Correlation Length, and Effective Sample Size}
\label{ssec:ESS}
\begin{figure}%
    \centering
    \subfloat[\centering CMB Temperature]{{\includegraphics[width=.45\textwidth]{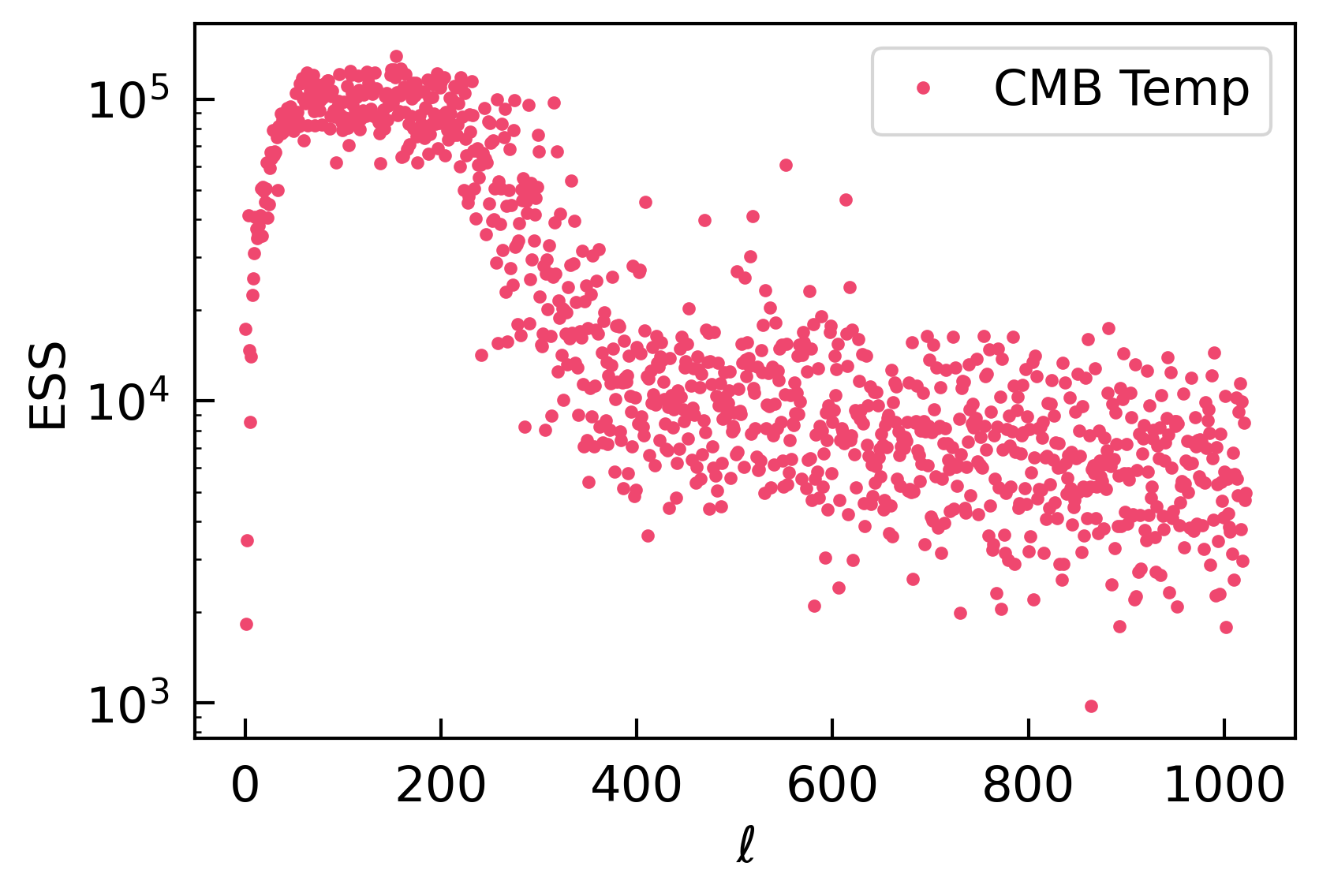} }}%
    \qquad
    \subfloat[\centering CMB Polarization]{{\includegraphics[width=.45\textwidth]{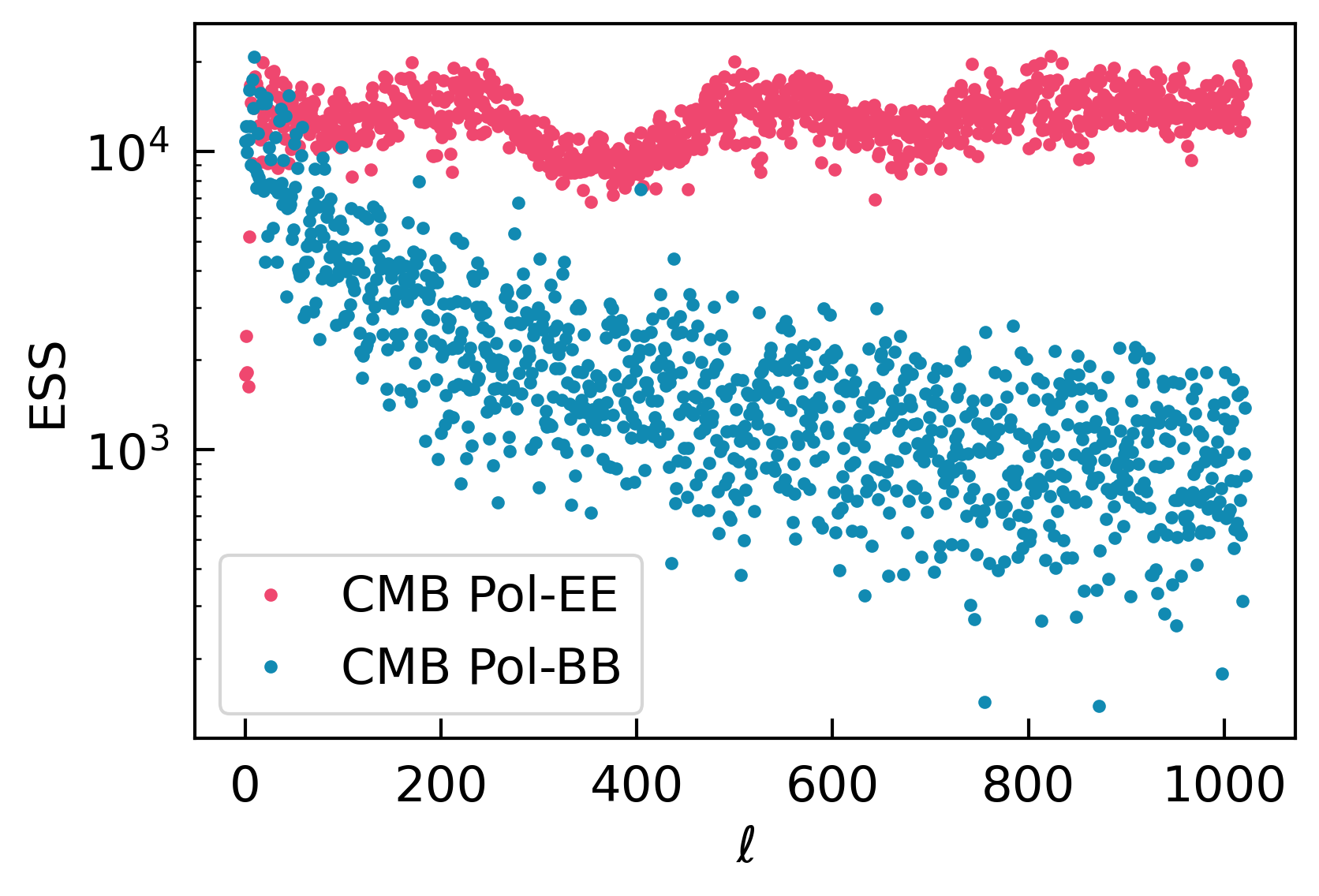} }}%
    \caption{Effective Sample Size (bulk ESS) for the combined chains from \almanac{} used in the angular power spectra results shown in Sect.~\ref{sec:results}. Here, we show the bulk ESS per multipole per probe. We note the apparent structure due to the variable signal-to-noise ratio across different scales.}%
    \label{Fig:ESS}%
    \vspace{8pt}
\end{figure}
	
Each new step in a Monte Carlo Markov Chain depends on the previous step, and this creates intrinsic correlations between the samples. This effect can be quantified by the autocovariance function ${\hat c}_i(t)$, defined for any parameter $y_i$ as a function of the lag $t$:
\begin{equation}
    {\hat c}_i(t) = \frac{1}{N-t} \sum_{k=1}^{N-t} (y_i^k - \bar{y_i})(y_i^{k+t} - \bar{y_i}) \, .
    \label{Eq:Autocorr}
\end{equation}
This can be calculated efficiently using the Fast Fourier Transform \citep[e.g.,][]{NumerialRecepies,arviz_2019}. For $t=0$ the autocovariance is simply the variance of the samples: $\sigma_i^2={\hat c}_i(0)$. The autocorrelation function is the autocovariance function normalized by this variance: ${\hat\zeta}_i(t) = {\hat c}_i(t)/{\hat c}_i(0)$. Correlations at non-zero lags imply that nearby samples are not independent and hence a good MCMC chain has an autocorrelation function that rapidly approaches zero as $t$ increases. 

If we have $S$ \emph{independent} samples from a chain, then the variance of the estimate of the mean of parameter $y_i$, ${\bar y_i} \equiv \sum_k y_i^k/N$, is given by $\Var(\bar{y_i})=\sigma_i^2/S$; this is also known as the Monte Carlo standard error (MCSE). For a chain derived from a stationary random process with underlying correlation function $\zeta_i(t)$, we have (for $N\to\infty$)
\begin{equation}
	\Var(\bar{y_i})=\frac{\sigma_i^2}{N}\sum_{t=-\infty}^{+\infty}\zeta_i(t)\, ,
\end{equation}
and hence the effective sample size (ESS) for parameter $i$ is $S_i=N/\sum_t\zeta_i(t)$. This implicitly defines the correlation length to be $\sum_t\zeta_i(t)$. Because $\zeta_i(t)$ is only estimated as ${\hat\zeta}_i(t)$ from finite samples at finite lags, the correlation length must itself be estimated, usually by truncating the sum at some appropriate lag \citep{Geyer2011,arviz_2019}. In practice \citep{Vehtari2021}, better estimates of the effective sample size and correlation length are found using the correlation function of the rank-normalised score for half-splits of each chain (the `bulk ESS'); moreover, the effective sample size of specified quantiles of the data can monitor convergence in the tails of the distribution as well (the `tail ESS').

For any parameter, an effective sample size much greater than one indicates that our chain has enough independent samples such that the variance due to the Markov process does not dominate our estimate of the variance of the target distribution; still larger effective sample sizes are typically needed for accurate measurements of quantities related to the tails of the distribution.

We show in Figure~\ref{Fig:ESS} the estimated bulk ESS for our spectral parameters. Even for the low signal-to-noise polarization $B$ mode, effective sample sizes of hundreds or thousands are typical; for the better-detected temperature and $E$ mode, our chains reach effective sample sizes of $10^4$--$10^5$. As a further indication that our chains are exploring the parameter space, the tail ESS for these parameters is of a comparable order of magnitude. Hence, correlation lengths of 50--100 are typical for our $T$ and $E$ spectra, and are at least an order of magnitude larger for the $B$ mode. All of these numbers depend upon the details of the data --- pixel resolution, signal-to-noise ratio, shape of the mask on the sky, etc. --- but are in a range that indicates our ability to sample efficiently with realistic data.

%%%%%%%%%%%%%%%%%%%%%%%%%%%%%%%%%%%%%%%%%%%%%%%%%%%%%%%%%%%%%%%%%%%%%%%%
%                           DISCUSSION
%%%%%%%%%%%%%%%%%%%%%%%%%%%%%%%%%%%%%%%%%%%%%%%%%%%%%%%%%%%%%%%%%%%%%%%%
\section{Discussion}
\label{sec:discussion}

This paper describes the \almanac{} sampler for field-level inference of cosmic fields on the sky. The sampler can in principle be used on any statistically isotropic fields, such as the cosmic microwave background, galaxy distributions, cosmic shear, or 21cm intensity.  The Hamiltonian Monte Carlo (HMC) sampler can analyse multiple spin-weight 0 and spin-weight 2 fields, inferring the fields (on the entire sky) and their auto- and cross-power spectra, with $E$- and $B$-mode decompositions when applied to spin-weight 2 fields. In this paper, we focus on single fields, and investigate different parametrizations in harmonic space, which are constructed to ensure that the samples of the covariance matrix $\sfC$ are always positive definite.  The two parametrizations correspond to a) the matrix log $\sfG$ of the covariance matrix, so that $\sfC = \exp(\sfG)$, and b) the Cholesky decomposition $\sfC = \sfL \sfL^{\transpose{}}$, where $\sfL$ is a lower-triangular matrix with positive diagonal elements.  In the examples here, the covariance matrix is diagonal, and the Cholesky decomposition reduces to a rescaling of the spherical harmonic coefficients ${\bf a}$ to ${\bf x} = {\bf a}/\sqrt{\sfC_\ell}$. We explore in the companion paper \cite{2022AlmanacWL}{} the Cholesky decomposition in more detail in the context of multiple correlated fields, such as in cosmic shear tomography, where the covariance matrices are far from diagonal. We find that the Cholesky decomposition performs better than the matrix log approach.

The algorithm, being intrinsically all-sky, is more suitable for surveys that map a significant fraction of the full celestial sphere; it infers the fields across the whole sky, including the masked (i.e., unobserved) regions, albeit with larger posterior variance.  It avoids any boundary effects that are inherent in Fourier-transform methods. It also avoids issues of ambiguous modes (neither pure $E$- nor pure $B$-mode), since the samples are of the underlying modes that are well-defined on the full sky. For further discussion of these issues, see, for example \citet{Ghosh2021} and references therein.

Following the results shown in Sect.~\ref{Sec:Performance}, computation time for the posterior calculations scales only marginally worse than the ideal quadratic scaling, with time $\propto\ell_{\rm max}^{2.03}$ for 128 cores. For $\ell_{\rm max}=2048$, which is beyond the scales currently investigated by Stage-III photometric surveys \citep{2022-Camacho-DES-Cls,2021-Loureiro-Kids-PCL,2023-Dalal-HSC-WL}, the algorithm requires only a few seconds per sample on a 64-core computing node when studying two redshift bins.  This takes a few minutes for each effectively independent sample at $\ell_{\rm max}=2048$.  While being slow in comparison with the analysis of more traditional summary statistics, it is not prohibitively so, and the improved inference merits the extra computational expense. Scaling the analysis for a more realistic case for a Stage-IV survey, with $\ell_{\rm max} = 5000$ and 10 or more redshift tomographic bins, would be challenging but feasible since our posterior calculation, which currently relies on OpenMP \citep{chandra2001parallel}, can be further parallelised for taking advantage of other high-performance computing architectures.

The main aim of \almanac{} is to infer the two-point statistical properties of the fields involved, and to infer the underlying, all-sky maps. The inference has the advantage of being independent of the cosmological model, beyond minimal assumptions of isotropy and (in the case of tomography) the presence of an effective distance indicator.  This motivates our choice of gaussian prior for the spherical harmonic coefficients (for given power spectrum).  This choice is of course well-motivated for the CMB, but it is also well justified for late-time fields such as the galaxy density field or cosmic shear fields. It is the least informative (in the sense of maximum entropy) for a field of given mean and covariance \citep{jaynes03}, and hence is the optimal choice if we want to avoid making further assumptions about the physics or cosmology. It does not lead to zero-mean gaussian posterior maps (since they are conditioned on the data), nor does it assume that the underlying fields themselves are gaussian. An alternative approach for late-time fields is to include gravity in the data model, sample from the initial field, which is assumed to be gaussian, and to evolve it and make the comparison of the data with the gravitationally-evolved fields.  This is the approach of the \textsc{BORG} programme, which can either infer fields for a given cosmology \citep[e.g.,][]{2013-Jasche-BORG1,2019-Jasche-BORG2} or which (for lensing) can infer the cosmological parameters simultaneously \citep{2021-Porqueres-WL-1,2022-Porqueres-2,Porqueres2023}, exploiting all the field-level data (including information of higher order than two-point).  On the other hand, the approach of \almanac{} allows investigation of $B$ modes (which may be useful both for fundamental physics and for the investigation of systematic errors) and potentially parity-violating $EB$ modes; it produces samples that are not dependent on the cosmological model assumed.  The two approaches are therefore quite complementary, and there is clear merit in applying both approaches.

\section*{Acknowledgements}
We thank the support staff of Leiden University's ALICE High Performance Computing infrastructure and UCL's Hypatia cluster, Martin Reinecke for assistance with \textsc{libsharp2} \citep{2013-Libsharp,2018-Libsharp2}, and Justin Alsing for useful discussions. The work presented here was made possible thanks to the following software: \textsc{HEALPix} \citep{1999-Healpix}, \textsc{libsharp2} \citep{2013-Libsharp,2018-Libsharp2}, \textsc{Arviz} \citep{arviz_2019}, \textsc{GetDist} \citep{2019-Lewis-GetDist}, \textsc{CAMB} \citep{CAMB}, \textsc{matplotlib} \citep{Hunter:2007-matplotlib}, and \textsc{NumPy} \citep{harris2020numpy}. This work was supported by the Science and Technology Facilities Council [grant numbers ST/N000838/1, ST/S000372/1]. This work used computing equipment funded by the Research Capital Investment Fund (RCIF) provided by UKRI, and was partially funded by the UCL Cosmoparticle Initiative.

\section*{Data Availability}
The simulated data and data-products, including the MCMC chains, discussed in this article are available upon reasonable request. Please contact the corresponding author for access to the requested materials.

\bibliographystyle{mnras}
\bibliography{Almanac.bib}

\appendix

\section{The geometry of spin-weight functions}
\label{apx:SpinWeightFunctions}

We describe the geometry of spin-weight functions, following Appendix A in \cite{2005-Castro} but using coordinate-free language and adding detail. We also make various mathematical comments, distinguished from the main text by being enclosed in square brackets. 

Let $s$ be a positive integer and consider a field of `$s$-fold symmetric shapes' defined on the sphere $S^2$ --- this means that at each point on the sphere we have (in the tangent plane at that point) a shape which, due to its symmetry, could be described by any one of $s$ different vectors; these vectors have equal length and in angle are equally spaced around the circle (adjacent vectors are separated by an angle of $2 \pi / s$). A prototypical example with $s=2$ is a field of ellipses (as would describe polarization or weak lensing shear on the full sky); here the two vectors point in either of the two opposite directions determined by the major axis while the length of the vectors is set by the eccentricity. (An additional scalar field is then needed to carry information about the area of the ellipses). Gravitational wave physics provides additional examples with $s=4$. We seek a mathematical description of such fields. For this we need to deal both with the non-trivial topology of the sphere and with the symmetry of the shapes; we accomplish this with a series of `tricks' (thus named because they cannot be generalised to higher dimensions).

What follows will work on an arbitrary two-dimensional Riemannian manifold $M$. Let $T_p$ denote the tangent plane at a point $p \in M$. If we have an ordered basis $\calB = (\hat{e}_1, \hat{e}_2)$ for $T_p$ that is orthonormal (relative to the Riemannian inner product), then we can identify $T_p$ with the complex plane $\mathbb{C}$ by mapping $\hat{e}_1$ to $1$ and $\hat{e}_2$ to $\imagunit$, and we may then use this mapping to convert the $s$-fold symmetric shape vectors at $p$ into $s$ complex numbers $z_j=re^{\imagunit\theta+2\pi \imagunit j/s}$ (here $j=1, \ldots, s$ and $\theta = \theta(p)$ is some angle). Denote by $S$ the operation of multiplying the angle of a complex number by $s$, i.e., $S (r e^{\imagunit \theta}) = r e^{\imagunit s \theta}$; we see that $S(z_1) = \ldots = S(z_s)=re^{\imagunit\theta s}$, and we denote this common value by $f(p, \calB)$. This process is invertible; i.e., given $f(p, \calB)$ we can reconstruct the $s$-fold symmetric shape vectors.

[The point of $S$ is that it eliminates the ambiguity caused by the shape symmetry. Mathematically $S$ works because $S^1$ modulo $\mathbb{Z}_s$ is topologically the same as $S^1$ (so that it can continue to be parameterised by an angle). Note that this is not true in higher dimensions. For example, $S^2$ modulo $\mathbb{Z}_2$ (which arises in neutrino experiments that cannot distinguish in which direction a neutrino was moving along a certain line) is not $S^2$; it is not even orientable.]

Consider rotating the orthonormal basis $\calB$ through an angle of $\alpha$ to create a new basis $\Rot_{\alpha}(\calB)$. (To make precise the direction of this rotation, we remark that if $\alpha = \pi/2$ then $\hat{e}_1$ will rotate to the original position of $\hat{e}_2$.) We see that such a rotation multiplies the $z_i$ by $\exp(- \imagunit \alpha)$ (the minus sign arising because it is the coordinate system that we are rotating, not the vectors); the effect of this rotation on $f$ is therefore
\begin{equation}
    \label{Eq:apx:spinWeightFunctionDefRotation}
    f(p, \Rot_{\alpha}(\calB)) = \exp(- \imagunit s \alpha) f(p, \calB).
\end{equation}
Alternatively, consider the operation $\Flip$ which acts on the basis $\calB$ via $(\hat{e}_1, \hat{e}_2) \mapsto (\hat{e}_1, -\hat{e}_2)$. Such a flip will act as complex conjugation on the $z_i$, and therefore on $f$ as well:
\begin{equation}
    \label{Eq:apx:spinWeightFunctionDefFlip}
    f(p, \Flip(\calB)) = (f(p, \calB))^*.
\end{equation}

These equations allow us to define the abstract notion of a spin-weight function: an arbitrary complex-valued function of position and orthonormal basis that satisfies Eqs.~\eqref{Eq:apx:spinWeightFunctionDefRotation} and~\eqref{Eq:apx:spinWeightFunctionDefFlip} will be said to be of \textit{spin-weight $s$}. Such functions are often denoted ${}_{s}f$.

In this abstract definition we may allow $s$ to be an arbitrary integer (not necessarily positive). By taking the complex conjugate of Eq.~\eqref{Eq:apx:spinWeightFunctionDefRotation} we see that the complex conjugate function $f^*$ will be of spin-weight $-s$. Note that $f^*$ is often denoted ${}_{-s}f$, but this notation is not ideal: it hides that a complex conjugation has been done, and it is ambiguous when $s=0$.

The symmetry count $s$ for the symmetric shapes will always be positive, and generating a spin-weight function from a field of symmetric shapes by following the prescription given above will yield a function of positive spin-weight. For example, if on the sphere we apply this prescription to a field of ellipses that describe weak lensing shear, then we will get a spin-weight +2 shear field. However, it is also common to use a slightly amended prescription (in which $\alpha$ is defined to be a rotation in the `other' direction) that instead creates a function of spin-weight $-s$. Such a divergence of conventions arises naturally (essentially because on the sphere there is no canonical choice of right-handedness; the geophysicist looking down at the Earth from above will have a different notion of right-handedness from the astronomer looking at the celestial sphere from the inside). Users of this other convention will create a weak lensing shear field of spin-weight $-2$. The difference is not critical, in that one such field can be converted to the other by taking the complex conjugate, but it is crucial to be know when interpreting data which spin-weight sign (or, equivalently, choice of handedness) was chosen.

[An alternative, more common, abstract definition of a spin-weight function instead requires $f$ only to be defined for right-handed $\calB$; in this case condition~\eqref{Eq:apx:spinWeightFunctionDefFlip} is not required. However this requires a choice of which bases are right-handed (as noted above, this choice is not canonical), and this in turn would require $M$ to be orientable (which would restrict our choice of $M$). The definition given here ensures that the spin-weight function responds appropriately to all inner-product-preserving transformations of the tangent plane (i.e., all of $O(2)$) and not just rotations (i.e., $SO(2)$); for example, with this definition a spin-weight function defined on the celestial sphere will respond appropriately to using `left ascension and declination' as coordinates instead of `right ascension and declination'.]

We define spin-weight $s=0$ functions to be complex-valued functions on $M$ (overriding Eq.~\eqref{Eq:apx:spinWeightFunctionDefFlip}, which would suggest that such functions should be real-valued). The need for complex values is related to complexification, as described below.

[We can ask `what is the geometry of the space $\tilde{M}$ on which $f$ is defined?' This is discussed in detail in \cite{2016-Boyle}. We see that $\tilde{M}$ will be a fibre bundle over $M$; it is known as the \textit{orthonormal frame bundle} of $M$. The fibre over a point $p \in M$ is the set of orthonormal bases for $T_p$, which (in our case) is topologically two circles (one for bases of each handedness). If $M$ is connected and orientable then $\tilde{M}$ will have two connected components but if $M$ is connected and not orientable (e.g., if $M$ is a M{\"{o}}bius strip) then it will have only one connected component. In the case that $M = S^2$ and that $f$ is defined only for right-handed bases, \cite{2016-Boyle} shows that $\tilde{M} \cong S^3$ ($\cong$ unit quaternions $\cong$ Hopf bundle).]

We next describe the complexification $T_p \otimes \mathbb{C}$ of the tangent plane as this space gives us a convenient way to parameterise the orthonormal bases $\calB$. Recall that for a real vector space $V$, $V \otimes \mathbb{C}$ is the complex vector space consisting of linear combinations of elements of $V$ with complex coefficients. Since $\{1, \imagunit\}$ is a basis for $\mathbb{C}$ over $\mathbb{R}$, elements $\alpha$ of $V \otimes \mathbb{C}$ may be uniquely represented as $\alpha = v \otimes 1 + w \otimes \imagunit$ for $v, w \in V$, which we abbreviate to $\alpha = v + \imagunit w$. Define $\alpha^* = v - \imagunit w$. The inner product 
on $T_p$ may be extended to yield a bilinear form $\langle,\rangle$ on $T_p \otimes \mathbb{C}$ via $\langle v + \imagunit w , \tilde{v} + \imagunit \tilde{w} \rangle = (\langle v , \tilde{v} \rangle - \langle w , \tilde{w} \rangle)  + \imagunit (\langle v , \tilde{w} \rangle + \langle w, \tilde{v} \rangle)$; this form is not an inner product. Define a mapping $C$ from the set of ordered orthonormal bases to $T_p \otimes \mathbb{C}$ via
\begin{equation}
C(\calB) = (\hat{e}_1 + \imagunit \hat{e}_2) / \sqrt{2} \quad \textrm{where} \quad \calB = (\hat{e}_1, \hat{e}_2).
\end{equation}
A simple calculation shows that the image of $C$ in $T_p \otimes \mathbb{C}$ is the set of all $\alpha$ satisfying
\begin{equation}
    \label{Eq:apx:imageofc}
    \langle \alpha, \alpha \rangle\ = 0 \quad \textrm{and} \quad \langle \alpha, \alpha^* \rangle\ = 1
\end{equation}
[topologically this set is $O(2)$, i.e., is the union of two circles] and that the inverse of $C$ is $v+\imagunit w \mapsto (\sqrt{2} v, \sqrt{2} w)$. [This trick does not work in high dimensions; there is no natural embedding of $O(n)$ in $\mathbb{R}^n \otimes \mathbb{C}$.] Simple calculations show that $C$ interacts appropriately with $\Rot$ and $\Flip$:
\begin{equation}
    \label{Eq:apx:rotationofcomplexifiedtangentplane}
    C(\Rot_{\alpha}(\calB)) = \exp(- \imagunit \alpha) C(\calB) \qquad \textrm{and} \qquad C(\Flip(\calB)) = (C(\calB))^* \, .
\end{equation}

We now describe \citep[following][]{2005-Castro} the relationship between spin-weight functions and tensors. Restrict to the case of non-negative $s$ and consider a $(0,s)$ tensor $F$ (i.e., a fully covariant tensor) on $M$. In coordinate-free language $F$ is, at each point $p \in M$, a multilinear (i.e., linear in each argument) function $F_p$ mapping $s$ tangent vectors (i.e., $s$ elements of $T_p$) to a real scalar. Such a function immediately extends to be a multilinear mapping of $s$ elements of $T_p \otimes \mathbb{C}$ to a complex scalar. The point of $F$ is that we may use it to define a complex-valued function $f$ acting on ordered bases via
\begin{equation}
    \label{Eq:apx:FnFromTensor}
    f(p, \calB) = F_{p}(C(\calB), \ldots, C(\calB))\, ,
\end{equation}
i.e., we put the complexified tangent vector that represents $\calB$ simultaneously into each of the $s$ input slots of $F_p$. By Eq.~\ref{Eq:apx:rotationofcomplexifiedtangentplane} and using the linearity and real-valuedness of $F_p$ we see that $f$ satisfies:
\begin{equation}
\begin{split}
    \label{Eq:apx:TensorGivesSpinWeightFn}
    f(p, \Rot_{\alpha}(\calB)) &= F_{p}(C(\Rot_{\alpha}(\calB)), \ldots, C(\Rot_{\alpha}(\calB))) \\
    &= F_{p}(\exp(- \imagunit \alpha)C(\calB), \ldots, \exp(- \imagunit \alpha)C(\calB) = \exp(- \imagunit s \alpha) f(p, \calB) \\
    \textrm{and} \qquad f(p, \Flip(\calB)) &= (f(p, \calB))^*.
\end{split}
\end{equation}
Thus $f$ defined in this way is a spin-weight $s$ function.

Conversely given a spin-weight $s$ function $f$ we may define a $(0,s)$ tensor $F$ by setting
\begin{equation}
    \label{Eq:apx:TensorFromFn}
    F_p(v_1, \ldots, v_s) = f(p, \calB_0) \langle v_1, (C(\calB_0))^* \rangle \ldots \langle v_s, (C(\calB_0))^* \rangle + (f(p, \calB_0))^* \langle v_1, C(\calB_0) \rangle \ldots \langle v_s, C(\calB_0) \rangle
\end{equation}
for arbitrary input tangent vectors $v_1, \ldots , v_s$. Here $\calB_0$ is some fixed but arbitrary orthonormal basis; $F$ is easily seen to be independent of the choice of $\calB_0$ (note Eq.~\eqref{Eq:apx:spinWeightFunctionDefFlip} is needed for this result). Eq.~\ref{Eq:apx:imageofc} shows that converting a spin-weight function $f$ to a tensor and then converting back again returns us to $f$. The opposite is not true (tensor to function to tensor) as many tensors can represent the same spin-weight function. However, the tensor created in Eq.~\ref{Eq:apx:TensorFromFn} is both symmetric (unchanged under a permutation of the inputs) and traceless (meaning in this context that $F(\hat{e}_1, \hat{e}_1, v_3, \ldots, v_s) = -F(\hat{e}_2, \hat{e}_2, v_3, \ldots, v_s)$ where $(\hat{e}_1, \hat{e}_2)$ is any orthonormal basis for $T_p$) and it is straightforward to show that Eq.~\ref{Eq:apx:TensorFromFn} gives a \textit{unique} tensor representative for $f$ among tensors with these properties.

Example: a Riemannian manifold $M$ has a canonical (0, 2) tensor (namely the metric); to what spin-weight 2 function $f$ does this tensor correspond? We see via Eqs.~\ref{Eq:apx:imageofc} and~\ref{Eq:apx:FnFromTensor} that $f=0$ (which makes sense, as no non-trivial function could be canonical).

So far we have examined properties at a specific point in $M$; we next consider how a spin-weight function varies from point to point (and we assume that the functions under consideration are sufficiently smooth so that what follows is well-defined). Let $s$ be non-negative. If $F$ is a $(0,s)$ tensor (and hence corresponds to a spin-weight $s$ function $f$) then its covariant derivative $\nabla F$ is a $(0,s+1)$ tensor (and hence corresponds to a spin-weight $s+1$ function); we define $\eth f$ to be the spin-weight $s+1$ function corresponding to $-\sqrt{2} \nabla F$. A calculation shows that this quantity is independent of the tensor used to represent $f$. The $-\sqrt{2}$ prefactor is historical, and makes later equations simpler. Alternatively we may create a function $\bar{\eth} f$ by mapping $\calB$ to $-\sqrt{2} \nabla F(C(\calB), \ldots, C(\calB), (C(\calB))^*)$; a calculation similar to Eq.~\eqref{Eq:apx:TensorGivesSpinWeightFn} shows that $\bar{\eth} f$ is a spin-weight $s-1$ function. We may extend $\eth$ and $\bar{\eth}$ to all $s$ by setting
\begin{equation}
    (\bar{\eth}f)^* = \eth(f^*) \, .
\end{equation}

We now move to the specific case of the sphere, i.e., we set $M = S^2$. We use standard `physics' spherical coordinates colatitude $\theta$ and longitude (= azimuthal angle) $\phi$. Let $U$ be the open subset of $S^2$ given by $0 < \theta < \pi$ and $0 < \phi < 2 \pi $; this is the patch on which $\theta, \phi$ provide a non-degenerate coordinate system. Note that $U$ is dense in $S^2$, i.e., a continuous function on $S^2$ is completely determined by its behaviour on $U$. [Note also that the bad coordinate behaviour at the `date line' $\phi = 0 = 2\pi$ is easily removable but the bad coordinate behaviour at the poles $\theta = 0$ or $\theta = \pi$ is not.] The standard metric on $U$ is then $d\theta \otimes d\theta + \sin^2{\theta} \, d\phi \otimes d\phi$, with Christoffel symbols $\Gamma^{\theta}_{\phi \phi} = -\sin{\theta} \cos{\theta}$ and $\Gamma^{\phi}_{\theta \phi} = \Gamma^{\phi}_{\phi \theta} = \cot{\theta}$ (all others vanishing). We see that $\calB_{\theta \phi} = (\partial_{\theta}, \csc{\theta} \, \partial_{\phi})$ is an orthonormal ordered basis for $TS^2$ on $U$ (these vectors point south and east, respectively). If ${}_{s}f(p, \calB)$ is a spin-weight $s$ function then we may define a complex-valued function $f$ on $U$ via $f(p) = {}_{s}f(p, \calB_{\theta \phi})$. Such a function $f$ must have special behaviour near the poles (it will either vanish, or behave as a multiple of $\exp(- \imagunit s \phi)$ (respectively $\exp(\imagunit s \phi)$) near the north (respectively south) pole); this is a consequence of the behaviour of $\calB_{\theta \phi}$ near the poles. Note that $f$ might not be defined outside of $U$ (e.g., might not be defined at the poles).

Conversely any complex function on $U$ with such polar behaviour will lead to a spin-weight $s$ function. A calculation shows that $\eth$ and $\bar{\eth}$ behave on such functions as given by Eq.~\ref{eqn:eth}.

\section{Notational preliminaries for HMC sampler formulae}
\label{apx:Preliminaries}
We start with the notation for the indexing of the vector $\fata$ of spherical harmonic coefficients. We treat $\fata$ as a vector of \textit{real} numbers, splitting each complex value into its real and imaginary parts. In every case we can ignore coefficients with $m<0$ as they can be deduced from the $m \geq 0$ coefficients; furthermore, coefficients with $m=0$ will have no imaginary part. (In the spin-weight 0 case this is a consequence of the data being real, whereas in the spin-weight 2 case this is a consequence of Eq.~\ref{eq:eandb}). For a fixed $\ell$, the vector $\fata$ will therefore have $2\ell+1$ real parts, which we will label using the index $\bar{m}$ with range $0 \leq \bar{m} \leq 2\ell$ (referring, in order, to $\Re(\fata_{\ell,m=0}), \Re(\fata_{\ell,m=1}), \Im(\fata_{\ell,m=1}), \ldots, \Re(\fata_{\ell,m=\ell}), \Im(\fata_{\ell,m=\ell})$). As we have multiple correlated fields, there is an additional field index $i$ with $1 \leq i \leq n$; in this Appendix the field index is written as a subscript (rather than as a superscript as in the main text). Thus a particular single entry in $\fata$ will be denoted $\fata_{\ell \bar{m} i}$, while $\fata_{\ell \bar{m}}$ (with the field index unspecified) will denote an $n$-vector. When using Cholesky coordinates we have a transformed variable $\fatx$, for which we use similar indexing.

The \textit{complex} $\fata_{\ell m i}$ values are draws from a distribution with certain variance; the \textit{real} components will have half this variance (except when $m=0$ (and hence $\bar{m}=0$), where there is no corresponding imaginary part). Therefore define
\begin{equation}
    \epsilon_{\bar{m}} = 
    \begin{cases}
        1 & \textrm{if } \bar{m} = 0 \\
        1/\sqrt{2} & \textrm{otherwise},
    \end{cases}
\end{equation}
so that
\begin{equation}
\left< \fata_{\ell \bar{m}}^{\phantom{\transpose{}}} \fata_{\ell \bar{m}}^{\transpose{}} \right> = \epsilon_{\bar{m}}^2 \sfC_{\ell}^{}.
\end{equation}

In $\{\fata, \sfC\}$ coordinates the negative logarithm $\nlp{}$ of the posterior density Eq.~\eqref{Eq:Cond_Posterior} is (ignoring irrelevant additive constants)
\begin{equation}
\label{Eq:apx:nlpaC}
\begin{split}
    \nlp{}(\fata, \sfC) &= 
    \half (\fatd - \sfR\sfY\fata)^{\transpose{}}\sfN^{-1}(\fatd - \sfR\sfY\fata)
    + \half \sum_{\ell} \sum_{\bar{m}} \left( \ln{\determinant{\sfC_{\ell}}} + \epsilon_{\bar{m}}^{-2} \fata_{\ell \bar{m}}^{\transpose{}} \sfC_{\ell}^{-1} \fata_{\ell \bar{m}}^{\phantom{\transpose{}}} \right)
    - q \sum_{\ell} \ln{\determinant{\sfC_{\ell}}} \\
    &= \half (\fatd - \sfR\sfY\fata)^{\transpose{}}\sfN^{-1}(\fatd - \sfR\sfY\fata)
    + \sum_{\ell} ( \ell + \half - q) \ln{\determinant{\sfC_{\ell}}}
    + \half \sum_{\ell} \sum_{\bar{m}} \epsilon_{\bar{m}}^{-2} \fata_{\ell \bar{m}}^{\transpose{}} \sfC_{\ell}^{-1} \fata_{\ell \bar{m}}^{\phantom{\transpose{}}} \, .
\end{split}
\end{equation}

\section{Differentiating the matrix exponential and a related trace w.r.t. matrix elements}
\label{apx:DerivsOfMatrixExp}

For use in Appendix~\ref{apx:FormulaeForLogCoordinates}, we consider the derivative of the exponential of a symmetric matrix with respect to the matrix elements. Our calculations follow from and extend \cite{2001-Ortiz}, clarifying some of the indices and derivations, and extending to the trace calculations we require. We also calculate the derivatives of the Jacobian of the matrix exponential.

\subsection{First and second derivatives of the matrix exponential}
Let $\sfA$ be symmetric, with spectral decomposition $\sfA = \sfU \diag(\lambda_{\alpha})\sfU^{\transpose{}}$ (here $\{\lambda_{\alpha}\}$ are the eigenvalues and the columns of $\sfU$ are the corresponding eigenvectors ($\alpha=1,...,n$); $\sfU$ is orthogonal: $\sfU^{-1} = \sfU^{\transpose{}}$). Then $\sfA_{ij} = \sum_{\alpha=1}^n\lambda_{\alpha}\sfU_{i\alpha}\sfU_{j\alpha}$.

Let $\exp(\tau \sfA) = \sum_{k=0}^{\infty} \tau^k \sfA^k/k!$ (definition valid even for non-symmetric $\sfA$); then $\exp(\tau\sfA) = \sfU \diag(\exp(\tau \lambda_{\alpha}))\sfU^{\transpose{}}$ and so $[\exp(\tau\sfA)]_{ij} = \sum_{\alpha=1}^n\exp(\tau\lambda_{\alpha})\sfU_{i\alpha}\sfU_{j\alpha}$.

The first derivatives are given by
\begin{equation}
    \label{Eq:apx:firstDeriv1}
    D_{ijkl} \equiv
    \frac{\partial \left[ \exp(\sfA)\right]_{ij}}{\partial \sfA_{kl}} =
    \int_0^1\left\{\exp\left[(1-\tau)\sfA\right]\right\}_{ik} \left[\exp(\tau\sfA)\right]_{lj} d\tau\ =
    \sum_{\alpha,\beta=1}^n f(\lambda_{\alpha}, \lambda_{\beta})\sfU_{i\alpha }\sfU_{j \beta}\sfU_{k \alpha}\sfU_{l \beta}\, ,
\end{equation}
where the representation as an integral is Equation (15) in \cite{2001-Ortiz} (with indices clarified) and the final equality is simply the evaluation of the integral. Here
\begin{equation}
   f(x, y) = 
   \begin{cases}
        [\exp(y)-\exp(x)]/(y-x) & \textrm{if } x \neq y \\
        \exp(x) & \textrm{otherwise}.
    \end{cases}
\end{equation}

The perturbation of $\sfA$ described in Eq.~\eqref{Eq:apx:firstDeriv1} results in a non-symmetric $\sfA$. More typically we will be interested in perturbations (denoted $|_S$) that keep $\sfA$ symmetric:
\begin{equation}
    \left. \frac{\partial \left[ \exp(\sfA)\right]_{ij}}{\partial \sfA_{kl}} \right|_S = 
    \begin{cases}
    D_{ijkl} + D_{ijlk} & \textrm{if } k < l \\
    D_{ijkl} & \textrm{if } k = l.
    \end{cases}
    \label{Eq:apx:firstDerivSymmetrised}
\end{equation}

For the second derivative again following \cite{2001-Ortiz} we differentiate the integral representation from Eq.~\eqref{Eq:apx:firstDeriv1}, apply the end result from Eq.~\eqref{Eq:apx:firstDeriv1}, and then finally evaluate the resulting integral:
\begin{equation}
\begin{split}
    D_{ijklmn} \equiv
    \frac{\partial^2 \exp(\sfA_{ij})}{\partial \sfA_{kl} \partial \sfA_{mn}} & =
    \int_0^1 \left( \frac{\partial \left\{\exp\left[(1-\tau)\sfA\right]\right\}_{ik}}{\partial \sfA_{mn}} \left[\exp(\tau\sfA)\right]_{lj} + \left\{\exp\left[(1-\tau)\sfA\right]\right\}_{ik} \frac{\partial \left[\exp(\tau\sfA)\right]_{lj}}{\partial \sfA_{mn}} \right) d\tau \\
    & = \sum_{\alpha,\beta,\gamma=1}^n g(\lambda_{\alpha}, \lambda_{\beta}, \lambda_{\gamma}) \sfU_{m \alpha} \sfU_{n \beta} \left( \sfU_{i\alpha} \sfU_{j \gamma} \sfU_{k \beta} \sfU_{l \gamma} + \sfU_{i \gamma} \sfU_{j \beta}  \sfU_{k \gamma} \sfU_{l \alpha} \right) \, .
\end{split}
\end{equation}
Here
\begin{equation}
\begin{split}
    g(x,y,z) & = \frac{x[\exp(y)-\exp(z)] + y[\exp(z)-\exp(x)] + z[\exp(x)-\exp(y)]}{(x-y)(y-z)(z-x)} \textrm{ if $x,y,z$ are distinct;} \\
    g(y,y,z) & = \frac{\exp(y)}{y-z} - \frac{\exp(y) - \exp(z)}{(y-z)^2} \textrm{ if $y \neq z$ (and similarly for other such cases);}\\
    g(z,z,z) & = \frac{\exp(z)}{2}.\\
\end{split}
\end{equation}

For symmetric derivatives we have
\begin{equation}
    \label{Eq:apx:secondderivsymmetric}
    \left. \frac{\partial^2 \exp(\sfA_{ij})}{\partial \sfA_{kl} \partial \sfA_{mn}} \right|_S =
    \begin{cases}
    D_{ijklmn} + D_{ijlkmn} + D_{ijklnm} + D_{ijlknm} \textrm{ if $k < l$ and $m < n$} \\
    D_{ijklmn} + D_{ijklnm} \textrm{ if $k = l$ and $m < n$} \\
    D_{ijklmn} + D_{ijlkmn} \textrm{ if $k < l$ and $m = n$} \\
    D_{ijklmn} \textrm{ if $k = l$ and $m = n$}. \\
    \end{cases}
\end{equation}

\subsection{Derivative of a certain trace}
Let $\sfB$ be symmetric; we will need the derivative of $\Tr[\exp(-\sfA)\sfB]$ with respect to the elements of symmetric matrix $\sfA$. Let $\Phi_{\alpha \beta} = -f(-\lambda_{\alpha}, -\lambda_{\beta})$. Then (using Eq.~\eqref{Eq:apx:firstDeriv1} and the Hadamard product $(\sfA \circ \sfB)_{ij} \equiv \sfA_{ij}\sfB_{ij}$):

\begin{equation}
    \frac{\partial \Tr( \exp(-\sfA) \sfB)}{\partial \sfA_{kl}}
    = \sum_{i,j=1}^n \frac{\partial [\exp(-\sfA)]_{ij} \sfB_{ij}}{\partial \sfA_{kl}}
    = \sum_{i,j,\alpha,\beta=1}^n \Phi_{\alpha \beta}\sfU_{i\alpha }\sfU_{j \beta}\sfU_{k \alpha}\sfU_{l \beta} \sfB_{ij}
    = [\sfU (\Phi \circ \sfU^{\transpose{}} \sfB \sfU) \sfU^{\transpose{}} ]_{kl} \, .
\end{equation}

Now $\sfU (\Phi \circ \sfU^{\transpose{}} \sfB \sfU) \sfU^{\transpose{}}$ is symmetric, and so (using the Kronecker delta $\delta^{\rm K}$):
\begin{equation}
    \label{Eq:apx:dTrdAsymmetric}
    \left. \frac{\partial \Tr[\exp(-\sfA) \sfB]}{\partial \sfA_{kl}} \right|_S
    = (2 - \delta^{\rm K}_{kl}) [\sfU (\Phi \circ \sfU^{\transpose{}} \sfB \sfU) \sfU^{\transpose{}} ]_{kl} \, .
\end{equation}

\subsection{Jacobian of matrix exp and its derivative}

For symmetric $\sfA$ let $\sfJ = \partial \exp(\sfA) / \partial \sfA$ (here treating $\sfA$ and $\exp(\sfA)$ as symmetric matrices, so that $\sfJ$ has $n(n+1)/2$ rows and columns). The entries of $\sfJ$ may be evaluated via Eq.~\eqref{Eq:apx:firstDeriv1}, and from these the value of $\determinant{\sfJ}$ may be calculated. Note that $\sfJ$ is invertible but is generally not symmetric.

We will also need $\partial \ln \determinant{\sfJ} / \partial \sfA_{kl}$. Let $p,q$ be row-and-column indices for $\sfJ$ ($1 \leq p, q \leq n(n+1)/2$), so that
\begin{equation}
    \sfJ_{pq} = \left. \frac{\partial [\exp(\sfA)]_{r(p),c(p)}}{\partial \sfA_{r(q),c(q)}} \right|_S \, .
\end{equation}
Jacobi's formula for invertible $\sfJ$ is
\begin{equation}
    \frac{\partial \determinant{\sfJ}}{\partial \sfJ_{pq}} = \determinant{\sfJ} (\sfJ^{-1})_{qp}.
\end{equation}
Applying the chain rule gives
\begin{equation}
    \label{Eq:apx:derivofjacobianofmatrixexp}
    \frac{\partial \ln \determinant{\sfJ}}{\partial \sfA_{kl}} =
    \frac{1}{\determinant{\sfJ}} \sum_{pq} \frac{\partial \determinant{\sfJ}}{\partial \sfJ_{pq}} \frac{\partial \sfJ_{pq}}{\partial \sfA_{kl }} =
    \sum_{pq} (\sfJ^{-1})_{qp} \left. \frac{\partial^2 [\exp(\sfA)]_{r(p),c(p)}}{\partial \sfA_{r(q),c(q)}\partial \sfA_{kl}} \right|_S,
\end{equation}
which may be calculated using Eq.~\eqref{Eq:apx:secondderivsymmetric}.

\section{Formulae for implementing the HMC sampler using log coordinates}
\label{apx:FormulaeForLogCoordinates}

We supply the formulae necessary for implementing the HMC sampler using log coordinates $\{\fata,\sfG\}$ where $\sfG = \ln(\sfC)$.

\subsection{Jacobian}

Reparameterising from $\{\fata,\sfC\}$ to $\{\fata,\sfG\}$ modifies the posterior density by a Jacobian factor $\determinant{\sfJ}$ where $\sfJ = \partial \sfC / \partial \sfG = \partial \exp(\sfG) / \partial \sfG$. The entries of $\sfJ$ may be evaluated using Eq.~\eqref{Eq:apx:firstDeriv1}, from which the determinant may then be calculated.

\subsection{Negative Logarithm of the Posterior Density}

In log coordinates the negative logarithm $\nlp{}$ of the posterior density Eq.~\eqref{Eq:apx:nlpaC} is

\begin{equation}
    \label{Eq:apx:nlpclassic}
    \nlp{}(\fata, \sfG) =
    \half (\fatd - \sfR\sfY\fata)^{\transpose{}}\sfN^{-1}(\fatd - \sfR\sfY\fata)
    + \sum_{\ell} \left( \ell + \half - q \right) \Tr(\sfG_{\ell})
    + \sum_{\ell} \left( \ell + \half \right)\Tr[\exp(-\sfG_{\ell}) \sfZ_{\ell}]
    - \ln \determinant{\sfJ}.
\end{equation}
Here we have included the Jacobian term and we have used the identities $\ln \determinant{\sfA} = \Tr \ln(\sfA)$, $\Tr(\sfA \sfB) = \Tr(\sfB \sfA)$, and $\Tr(\sfA) = \sfA$ if $\sfA$ is a singleton. The $n \times n$ signal covariance matrix is defined to be
\begin{equation}
    \sfZ_{\ell} = \frac{1}{2 \ell + 1} \sum_{\bar{m}} \epsilon_{\bar{m}}^{-2} \fata_{\ell \bar{m}}^{} \fata_{\ell \bar{m}}^{\transpose{}}.
\end{equation}

\subsection{Gradients}

The HMC sampler requires the derivatives of $\nlp{}$ with respect to both $\fata$ and $\sfG$ parameters. We calculate
\begin{equation}
    \frac{\partial \nlp}{\partial \fata_{\ell \bar{m} i}} = -[\sfY^{\transpose{}} \sfR^{\transpose{}} \sfN^{-1}(\fatd - \sfR \sfY \fata)]_{\ell \bar{m} i}
    + \epsilon_{\bar{m}}^{-2} [\exp(-\sfG_{\ell}) \fata_{\ell \bar{m}}]_{i}. 
\end{equation}
The two terms arise from the first and third terms, respectively, on the right hand side of Eq.~\eqref{Eq:apx:nlpclassic}.

The derivative $\partial \nlp / \partial \sfG$ has three terms. The first arises from the second term in Eq.~\eqref{Eq:apx:nlpclassic}:
\begin{equation}
    \frac{\partial}{\partial \sfG_{\ell ij}} \sum_{\ell'} \left( \ell' + \half - q \right) \Tr(\sfG_{\ell'}) =
    \left(\ell + \half - q \right) \frac{\partial \Tr(\sfG_{\ell}) }{\partial \sfG_{\ell ij}} =
    \left(\ell + \half - q \right) \sum_k \frac{\partial \sfG_{\ell kk} }{\partial \sfG_{\ell ij}} = 
    \left(\ell + \half - q \right) \delta^{\rm K}_{ij} \, .
\end{equation}
The two remaining terms arise from the third and fourth terms in Eq.~\eqref{Eq:apx:nlpclassic} and may be evaluated using Eq.~\eqref{Eq:apx:dTrdAsymmetric} (with $\sfA = \sfG_{\ell}$ and $\sfB = \sfZ_{\ell \bar{m}}$) and Eq.~\eqref{Eq:apx:derivofjacobianofmatrixexp} (respectively).

\subsection{Hessian}

For the initial step sizes of the HMC sampler we require the diagonal of the Hessian matrix (see Sect.~\ref{sec:HMCsampler}). We calculate
\begin{equation}
    \frac{\partial^2 \nlp}{\partial \fata_{\ell \bar{m} i} \partial \fata_{\ell \bar{m} i}} = 
    [\sfY^{\transpose{}} \sfR^{\transpose{}} \sfN^{-1} \sfR \sfY]_{\ell \bar{m} i, \ell \bar{m} i} + \epsilon_{\bar{m}}^{-2} \exp(-\sfG_{\ell})_{ii} \, .
\end{equation}
We may estimate $(\sfY^{\transpose{}} \sfR^{\transpose{}} \sfN^{-1} \sfR \sfY)_{\alpha \alpha}$ as follows: let $\faty$ be a multi-variate normal random variable with zero mean and covariance $ \langle \faty \faty^{\transpose{}} \rangle = \sfN^{-1}$. Then observe that
\begin{equation}
(\sfY^{\transpose{}} \sfR^{\transpose{}} \sfN^{-1} \sfR \sfY)_{\alpha \alpha} =
\langle [(\faty^{\transpose{}} \sfR \sfY)_{\alpha}]^2\rangle \, .
\end{equation}
The expectation on the right hand side may be estimated using Monte-Carlo style draws of $\faty$.

The formulae necessary for analytically calculating
\begin{equation}
    \frac{\partial^2 \nlp}{\partial \sfG_{\ell ij} \partial \sfG_{\ell ij}}
\end{equation}
could be derived using the techniques of Appendix~\ref{apx:DerivsOfMatrixExp}; however the computation need not be fast (it is only done once) so for simplicity we have calculated it using numerical shocking of the first derivative.

\section{Formulae for implementing the HMC sampler using Cholesky coordinates}
\label{apx:FormulaeForCholeskyCoordinates}

We supply the formulae necessary for implementing the HMC sampler using Cholesky coordinates $\{\fatx,\sfK\}$ as described in Sect.~\ref{sub:params}, adding detail to the presentation in the companion paper \citep{2022AlmanacWL}{}.

For positive definite $\sfC_{\ell}$ the Cholesky factor $\sfL_{\ell}$ is the unique lower-triangular matrix with positive elements on the diagonal satisfying $\sfC_{\ell} = \sfL_{\ell}\sfL_{\ell}^{\transpose{}}$.

Define the \textit{diagonal-log} $\sfK_{\ell}$ of $\sfL_{\ell}$ to be 
\begin{equation}
    \sfK_{\ell ij} = 
    \begin{cases}
        \ln(\sfL_{\ell ij}) & \textrm{ if } i=j\, , \\
        \sfL_{\ell ij} & \mbox{ otherwise;}
    \end{cases}
    \label{Eq:DiagonalLog_L}
\end{equation}
differentiating gives
\begin{equation}
    \sfD_{\ell ij} \equiv \frac{\partial \sfL_{\ell ij}}{\partial \sfK_{\ell ij}} = 
    \begin{cases}
        \sfL_{\ell ij} & \textrm{ if } i=j\, , \\
        1  & \mbox{ otherwise.}
    \end{cases}
    \label{Eq:DiagonalLog_LDeriv}
\end{equation}
There is a one-to-one correspondence between lower-triangular $\sfK$ and positive-definite $\sfC$.

Given $\fata$ define $\fatx$ implicitly via
\begin{equation}
    \epsilon_{\bar{m}} \sfL_{\ell} \fatx_{\ell \bar{m}} = \fata_{\ell \bar{m}} \, ;
    \label{Eq:Scale_fata}
\end{equation}
then $\{ \fatx, \sfK \}$ provides an alternative parameterisation of $\{ \fata, \sfC \}$.

\subsection{Jacobian}

Reparameterising from $\{ \fata, \sfC \}$ to $\{ \fatx, \sfK \}$ modifies the posterior density by a Jacobian factor. Now
\begin{equation}
     p(\fatx,\sfK) \ \mathd \fatx \ \mathd \sfK = p(\fata,\sfC) \ \mathd \fata \ \mathd \sfC 
\end{equation}
and so the Jacobian factor is
\begin{equation}
    \determinant{\sfJ} = \begin{vmatrix} \partial \fata/\partial \fatx & \partial \fata/\partial \sfK\\
    \partial \sfC/\partial \fatx & \partial \sfC/\partial \sfK\\
    \end{vmatrix} = \determinant{ \partial \fata/\partial \fatx } \ \determinant{ \partial \sfC/\partial \sfK };
\end{equation}
the second equality follows from $ {\partial \sfC} / {\partial \fatx} = 0 $ (since it is evaluated at constant $\sfK$).
We have $\fata_{\ell \bar{m}} \propto \sfL_{\ell} \fatx_{\ell \bar{m}}$ and so
\begin{equation}
    \label{Eq:dadx}
    \determinant{ \partial \fata/\partial \fatx } =
    \prod_{\ell} \prod_{\bar{m}} \determinant{\partial \fata_{\ell \bar{m}} / \partial \fatx_{\ell \bar{m}}} \propto
    \prod_{\ell} \prod_{\bar{m}} \determinant{\sfL_{\ell}} = 
    \prod_{\ell} \left( \prod_{i=1}^{n} \sfL_{\ell ii}^{2\ell+1} \right) \, .
\end{equation}
The determinant of the Cholesky transformation is \citep{RedMatrixBook}:
\begin{equation}
    \determinant{\partial \sfC/\partial \sfL} =
    \prod_{\ell} \left( 2^n\prod_{i=1}^n \sfL_{\ell ii}^{n+1-i} \right) \, ,
\end{equation}
while (using Eq.~\ref{Eq:DiagonalLog_LDeriv})
\begin{equation}
    \determinant{\partial \sfL/\partial \sfK } =
    \prod_{\ell} \left( \prod_{i=1}^{n} \sfD_{\ell ii} \right) = \prod_{\ell} \left( \prod_{i=1}^{n} \sfL_{\ell ii} \right) \, .
\end{equation}
Combining these results yields
\begin{equation}
    \label{Eq:Jax}
    \determinant{\sfJ} \propto \exp \left[ \sum_{\ell} \sum_{i=1}^n (n+3+2\ell-i)\sfK_{\ell ii} \right] \ .
\end{equation}

\subsection{Negative Logarithm of the Posterior Density}
\label{SSec:nlp}

In Cholesky coordinates the negative logarithm $\nlp{}$ of the posterior density Eq.~\eqref{Eq:apx:nlpaC} is
\begin{equation}
    \nlp{}(\fatx, \sfK) = \half (\fatd - \sfR \sfY \fata)^{\transpose{}}\sfN^{-1}(\fatd - \sfR \sfY \fata) +
    \half \sum_{\ell} \sum_{\bar{m}} \fatx_{\ell \bar{m}}^{\transpose{}} \fatx_{\ell \bar{m}}^{\phantom{\transpose{}}} + \sum_{\ell} \sum_{i=1}^n (i - 2 - n - 2q) \sfK_{\ell ii} \ .
\end{equation}
Here we have dropped further irrelevant additive constants, we have included the Jacobian term Eq.~\eqref{Eq:Jax}, and we have used $\determinant{\sfC_{\ell}} = \determinant{\sfL_{\ell}}^2$. Note that $\fata$ is as defined in Eq.~\ref{Eq:Scale_fata}.

The negative log posterior has a shape in this coordinate system that can be explored more efficiently by the HMC sampler than the shape in $\{\fata, \ln(\sfC)\}$ coordinates.

\subsection{Gradients}
\label{SSec:gradients}

The HMC sampler requires the derivatives of $\nlp{}$ with respect to both $\fatx$ and $\sfK$ parameters. We calculate
\begin{equation}
    \frac{\partial\nlp{}}{\partial \fatx_{\ell \bar{m} i}}  = - \epsilon_{\bar{m}} \sum_{j=i}^{n} \sfL_{\ell j i} [\sfY^{\transpose{}} \sfR^{\transpose{}} \sfN^{-1} (\fatd - \sfR \sfY \fata)]_{\ell \bar{m} j} +\fatx_{\ell \bar{m} i} \,
\end{equation}
and
\begin{equation}
    \label{Eq:dpsidK}
    \begin{split}
        \frac{\partial\nlp{}}{\partial \sfK_{\ell ij}} & =
        - \sfD_{\ell ij} \sum_{\bar{m}} \epsilon_{\bar{m}} \left[\sfY^{\transpose{}} \sfR^{\transpose{}} (\fatd - \sfR \sfY \fata)\right]_{\ell \bar{m} i} \fatx_{\ell \bar{m} j} + \delta_{ij}^K (i - 2 - n - 2q) \, .
    \end{split}
\end{equation}
Here $\sfD$ is as defined in Eq.~\ref{Eq:DiagonalLog_LDeriv}.

\subsection{Hessian}

For the initial step sizes of the HMC sampler we require the diagonal of the Hessian matrix (see Sect.~\ref{sec:HMCsampler}). We calculate

\begin{equation} \label{Eq:hessian_x}
    \frac{\partial^2\nlp{}}{\partial \fatx_{\ell \bar{m} i} \fatx_{\ell \bar{m} i}} = \epsilon_{\bar{m}}^2 \sum_{j = i}^{n} \sum_{k = i}^{n} \sfL_{\ell ji} \sfL_{\ell ki} \left[ \sfY^{\transpose{}} \sfR^{\transpose{}} \sfN^{-1} \sfR \sfY \right]_{\ell \bar{m} j, \ell \bar{m} k} + 1
\end{equation}
and

\begin{equation}
    \label{Eq:hessian_K}
    \begin{split}
        \frac{\partial^2\nlp{}}{\partial \sfK_{\ell ij} \partial \sfK_{\ell ij}} = & 
        \sfD_{\ell ij}^2 \sum_{\bar{m}} \sum_{\bar{m}'} \epsilon_{\bar{m}} \epsilon_{\bar{m}'}[\sfY^{\transpose{}} \sfR^{\transpose{}} \sfN^{-1} \sfR \sfY]_{\ell \bar{m} i, \ell \bar{m}' i} \, \fatx_{\ell \bar{m} j} \fatx_{\ell \bar{m}' j}  \\
        & \,  - \delta_{ij}^K \sfD_{\ell ij} \sum_{\bar{m}} \epsilon_{\bar{m}} [\sfY^{\transpose{}} \sfR^{\transpose{}} \sfN^{-1}(\fatd-\sfR\sfY\sfL\fatx)]_{\ell \bar{m} i} \fatx_{\ell \bar{m} j} \, .
    \end{split}
\end{equation}

We may estimate the diagonal terms of $\sfL^{\transpose{}} \sfY^{\transpose{}} \sfR^{\transpose{}} \sfN^{-1} \sfR \sfY \sfL$ (where $\sfL$ is a block diagonal matrix in which each $\sfL_{\ell}$ is repeated $2\ell + 1$ times) using Monte Carlo integration as described in  Appendix~\ref{apx:FormulaeForLogCoordinates}, and this technique allows us to calculate Eq.~\eqref{Eq:hessian_x}. The double sum in Eq.~\eqref{Eq:hessian_K} makes this expression more challenging to calculate efficiently and we therefore estimate it by numerical differentiation.

\end{document}